\documentclass[twocolumn, twocolappendix]{aastex63}

\usepackage{graphicx}
\usepackage{amssymb}
\usepackage{amsmath}
\usepackage{fnpct}

\newcommand{\oii}{[\ion{O}{2}]}
\newcommand{\oiii}{[\ion{O}{3}]}
\newcommand{\hb}{H$\beta$} 
\newcommand{\ha}{H$\alpha$}
\newcommand{\lya}{Ly$\alpha$} 
\newcommand{\sii}{[\ion{S}{2}]}
\newcommand{\nii}{[\ion{N}{2}]}

\newcommand{\hg}{H$\gamma$} 
\newcommand{\hd}{H$\delta$}

\usepackage{amsmath} 
\usepackage{hyperref}

\begin{document} 
\title{The mass-metallicity relation at $z\sim 1-2$ and its dependence on star formation rate} 
\author{Alaina Henry} 
\affiliation{Space Telescope Science Institute; 3700 San Martin Drive, Baltimore, MD,  21218, USA} 
\affiliation{Department of Physics \& Astronomy, Johns Hopkins University, Baltimore, MD 21218, USA}
\author{Marc Rafelski} 
\affiliation{Space Telescope Science Institute; 3700 San Martin Drive, Baltimore, MD,  21218, USA} 
\affiliation{Department of Physics \& Astronomy, Johns Hopkins University, Baltimore, MD 21218, USA}
\author{Ben Sunnquist} 
\affiliation{Space Telescope Science Institute; 3700 San Martin Drive, Baltimore, MD,  21218, USA} 
\author{Norbert Pirzkal} 
\affiliation{Space Telescope Science Institute; 3700 San Martin Drive, Baltimore, MD,  21218, USA}
\author{Camilla Pacifici} 
\affiliation{Space Telescope Science Institute; 3700 San Martin Drive, Baltimore, MD,  21218, USA}
\author{Hakim Atek} 
\affiliation{Institut d'astrophysique de Paris, CNRS UMR7095, Sorbonne Universit\'e, 98bis Boulevard Arago, F-75014 Paris, France } 
\author{Micaela Bagley} 
\affiliation{Department of Astronomy, The University of Texas at Austin, Austin, TX, 78712, USA}
\author{Ivano Baronchelli}
\affiliation{Dipartmento di Fisica e Astronomia, Universit\'a di Padova, Vicolo dell'Osservatorio, I-3 31522, Padova, Italy}  
\author{Guillermo Barro} 
\affiliation{Department  of  Physics,  University  of  the  Pacific,  3601 Pacific  Ave, Stockton, CA, 95211, USA} 
\author{Andrew J. Bunker} 
\affiliation{Sub-department of Astrophysics, Department of Physics, University of Oxford, Denys Wilkinson Building, Keble Road, Oxford OX1 3RH, UK}
\affiliation{Kavli Institute for the Physics and Mathematics of the Universe (WPI), The University of Tokyo Institutes for Advanced Study, The University of Tokyo, Kashiwa, Chiba 277-8583, Japan}
\author{James Colbert} 
\affiliation{Infrared Processing and analysis Center, California Institute of Technology, Pasadena, CA 91125, USA} 
\author{Y.Sophia Dai}
\affil{Chinese Academy of Sciences South America Center for Astronomy (CASSACA), NAOC, 20A Datun Road, Beijing, 100101, China}
\author{Bruce G. Elmegreen} 
\affiliation{IBM Research Division, T.J. Watson Research Center, 1101 Kitchawan Road, Yorktown Heights, NY10598, USA}
\author{Debra Meloy Elmegreen} 
\affiliation{Department of Physics  \& Astronomy, Vassar College, Poughkeepsie, NY 12604, USA}   
\author{Steven Finkelstein} 
\affiliation{Department of Astronomy, The University of Texas at Austin, Austin, TX, 78712, USA}
\author{Dale Kocevski} 
\affiliation{Department  of  Physics  and  Astronomy,  Colby  College,  Waterville, ME 04901, USA} 
\author{Anton Koekemoer} 
\affiliation{Space Telescope Science Institute; 3700 San Martin Drive, Baltimore, MD,  21218, USA}
\author{Matthew Malkan}
\affiliation{Department of Physics and Astronomy, UCLA, Los Angeles, CA 90095-1547, USA} 
\author{Crystal L. Martin} 
\affiliation{Department of Physics, University of California, Santa Barbara, CA 93106, USA} 
\author{Vihang Mehta} 
\affiliation{Minnesota Institute for Astrophysics, University of Minnesota, 116 Church Street SE, Minneapolis, MN 55455, USA}
\author{Anthony Pahl} 
\affiliation{Department of Physics and Astronomy, UCLA, Los Angeles, CA 90095-1547, USA} 
\author{Casey Papovich} 
\affiliation{Department of Physics and Astronomy, Texas A\&M University, College Station, TX 77843-4343, USA}
\author{Michael Rutkowski}
\affiliation{Department of Physics and Astronomy, Minnesota State University, Mankato, MN, 56001, USA}
\author{Jorge S\'anchez Almeida} 
\affiliation{Instituto de Astrof\'isica de Canarias, La Laguna, Tenerife 38200, Spain}
\author{Claudia Scarlata} 
\affiliation{Minnesota Institute for Astrophysics, University of Minnesota, 116 Church Street SE, Minneapolis, MN 55455, USA}
\author{Gregory Snyder} 
\affiliation{Space Telescope Science Institute; 3700 San Martin Drive, Baltimore, MD,  21218, USA}
\author{Harry Teplitz} 
\affiliation{Infrared Processing and analysis Center, California Institute of Technology, Pasadena, CA 91125, USA}

\begin{abstract} 

We present a new measurement of the gas-phase mass-metallicity relation (MZR), and its dependence on star formation rates (SFRs) at $1.3<z<2.3$.    Our sample comprises 1056 galaxies with a mean redshift of $z=1.9$, identified from the {\it Hubble Space Telescope} Wide Field Camera 3 (WFC3) grism spectroscopy in the Cosmic Assembly Near-Infrared Deep Extragalactic Survey (CANDELS) and the 
WFC3 Infrared Spectroscopic Parallel Survey (WISP). This sample is four times larger than previous metallicity surveys at $z\sim2$, and reaches an order of magnitude lower in stellar mass ($10^8$ M$_{\sun}$).  Using stacked spectra, we find that the MZR evolves by 0.3 dex relative to $z\sim 0.1$.    
 Additionally, we identify a subset of 49 galaxies with high signal-to-noise (SNR) spectra and redshifts between $1.3<z< 1.5$, where \ha\ emission is observed along with \oiii\ and \oii.   With accurate measurements of SFR in these objects, we confirm the existence of a mass-metallicity-SFR (M-Z-SFR) relation at high redshifts. 
  These galaxies show systematic differences from the local M-Z-SFR relation, which vary depending on the adopted measurement of the local relation. 
  However, it remains difficult to ascertain whether these differences could be due to redshift evolution, as the local M-Z-SFR relation is poorly constrained at the masses and SFRs of  our sample.   
 Lastly, we reproduced our sample selection in the IllustrisTNG hydrodynamical simulation, demonstrating that our line flux limit lowers the normalization of the simulated MZR by 0.2 dex.   We show that the M-Z-SFR relation in IllustrisTNG has an SFR dependence that is too steep by a factor of around three.

\end{abstract}

\section{Introduction}
The role of the baryon cycle in galaxy evolution is one of the most critical questions facing studies of galaxy evolution. 
 The growth of galaxies-- and the properties that we observe today-- are determined by gas accretion from the intergalactic medium (IGM) and circumgalactic medium (CGM), as well as star-formation feedback that heats and removes gas from the interstellar medium (ISM).   Theoretical studies of galaxy formation, both from hydrodynamical simulations and semi-analytic models, must balance these processes to produce realistic properties of galaxies over cosmic history \citep{SD15}.

One of the key constraints on galaxy formation models is the 
correlation between stellar mass and galaxy metallicities (both stellar and gas-phase\footnote{We hereafter define ``metallicity'' to mean gas-phase oxygen abundance throughout the paper, unless otherwise noted.}), known as the mass-metallicity relation (MZR).   Since metals are produced by star formation, their production is also regulated by feedback.   A large number of models, from analytical approaches \citep{Dalcanton07, Erb08, Peeples11, Dave12, Dayal13, Lilly13, Yabe15b}, to semi-analytical models \citep{Lu14, Croton16} and hydrodynamical simulations \citep{Finlator08, Ma16, DeRossi17, Dave17,Dave19, Torrey18, Torrey19}, have all modeled the MZR.  What is clear is that a large number of physical processes may contribute to the shape, normalization, and evolution of the MZR.  Supernova-driven outflows are usually considered a key ingredient, setting the slope of the MZR at low masses \citep{Tremonti04, Henry2013_clm, Henry2013_wisp, DeRossi17}.  These outflows may be mixed with the ISM, or metal-enriched from the sites of star formation and supernovae (e.g. \citealt{Chisholm18}).    Other studies have argued that the evolution of the MZR is largely tied to increasing gas fractions in lower mass and higher redshift systems \citep{Ma16, Torrey19}.   Feedback from active galactic nuclei (AGN) in simulations has also been shown to play an important role in shaping the MZR at high masses \citep{DeRossi17}.  Complicating all of these effects, gas is likely recycled through the CGM;  ejected gas may not leave the halo and re-accreted gas may be metal-deficient or metal-enriched \citep{SA14}. 

Nonetheless, evidence of an interplay between inflows, star formation, metal production, and feedback is observed.  The scatter in the MZR is shown to correlate with star-formation rates (SFRs) and (in some cases) gas fractions in galaxies out to $z\sim 2$ \citep{Ellison08, Mannucci10, Mannucci11, Cresci12, LaraLopez10, LaraLopez13, Bothwell13, AM13, Henry2013_clm, Salim14, Salim15, Sanders18, Gillman21}.    At fixed stellar mass, galaxies with higher SFRs show lower metallicities, while those with lower SFRs have more enriched ISM.  \cite{Mannucci10} dubbed this relation the ``fundamental metallicity relation'' (FMR).  Using the limited data available at the time, they argued that the FMR did not evolve out to $z\sim 2.5$; galaxies merely evolve within it.  From a theoretical perspective, the interpretation that the FMR is produced by variations in accretion, SFR, gas fractions, and feedback is supported by both semi-analytic models and numerical simulations \citep{Yates12, Dayal13, DeRossi17, Torrey18,Torrey19, Dave17, Dave19, DeLucia20}.   

In the low redshift universe ($z \la 0.3$), the MZR and FMR have been
well-characterized down to approximately $10^8$ M$_{\sun}$, using the large sample of galaxies covered by the spectroscopic Sloan Digital Sky Survey (SDSS; \citealt{Tremonti04, AM13, Brown16, Curti20}). Additionally,  extensions of the MZR to low masses have been made using nearby galaxies  \citep{Lee06,Berg12, Izotov12, Ly16}. 
Moreover, at intermediate redshifts, obtaining statistical samples is still practical.  The MZR derived from optical multi-object spectroscopic surveys reaches $M_{*} \sim 10^8-10^9$ M$_{\sun}$ for samples of around 1000 galaxies at $z\sim 0.5-1.0$ \citep{Zahid11, Maier15, Guo16a}. However, at $z>1$, observational constraints on the evolution of the MZR require near infrared spectroscopy, and have therefore been slower to develop (but see early single object spectroscopy studies; \citealt{Erb06, Maiolino08, Wuyts12, Belli13, Masters14}).   Nevertheless, in recent years, infrared grism spectroscopy with the Wide Field Camera 3 on the {\it Hubble Space Telescope} \citep{wfc3}, along with ground-based multi-object infrared spectrometers have opened up a new window on the evolution of galaxy metallicities, extending measurements of statistical samples to $z\sim 2$  (\citealt{Henry2013_wisp,  Zahid14, Sanders15, Kacprzak16, GG16, Wuyts16, Salim15, Yabe12,Yabe15, Kashino17, Sanders18}; Papovich et al., in prep).   Notably, numerous studies show that a mass-metallicity-SFR (M-Z-SFR) relation, similar to the local FMR, is also present out to $z\sim2$. 

One of the challenges faced by high redshift ($z\sim 1-2$) surveys is limited sensitivity, which leads to a survey design that favors brighter, higher mass targets.  As such, the limiting mass reached by current ground-based studies is around 10$^{9}- 10^{10}$ M$_{\sun}$ at $z>1$ \citep{Yabe15,  Salim15,  Kacprzak16, Sanders15, Sanders18, Kashino17}.   However, as we showed in \cite{Henry2013_wisp}, without pre-selection, slitless grism spectroscopy with WFC3 can measure metallicities from galaxies an order of magnitude smaller in stellar mass.   Critically, it is at the lowest masses that the effects of star formation feedback are the most prominent, as ejected ISM can more easily escape the weak gravitational potential of dwarf galaxies.  Consequently, extending measurements of the MZR and M-Z-SFR relation to lower masses can provide more powerful constraints on models of galaxy formation.     Therefore, the goal of this paper is to provide new measurements of galaxy metallicity evolution at masses ten times lower than in recent ground-based studies using statistical samples.

In this paper, we build on our previous work in \cite{Henry2013_wisp}, where we presented the first MZR from WFC3 IR grism spectroscopy.  Our earlier measurement was based on stacking spectra of 83 galaxies at $1.3 < z < 2.3$, drawn from the WFC3 Infrared Spectroscopic Parallel Survey (WISP; \citealt{Atek10, Colbert13}) and the grism coverage of the Hubble Ultra Deep Field (HUDF; \citealt{pvd_udf}).   The advantage of the WISP Survey, in particular, is the inclusion of both WFC3 IR grisms, G102 and G141.  The wavelength coverage from $0.85 \micron < \lambda < 1.7 \micron$ allows for metallicity measurements using \oii, \oiii, and \hb\ over a wide range of redshifts ($1.3 < z < 2.3$),  and consequently larger samples than G141 alone ($2.0 < z  < 2.3$ only).  

Since \cite{Henry2013_wisp}, available WFC3 grism samples have grown dramatically.    In the WISP survey, the area covered by both WFC3 IR bands (G102 and G141) has increased by more than a factor of three.  Likewise, in the well-studied fields of the Cosmic Assembly and Near-Infrared Deep Extragalactic Legacy Survey (CANDELS; \citealt{Grogin11, Koekemoer11}), the fully reduced 3D-HST spectroscopic data (G141) are now available, including extensive redshift catalogs  \citep{Momcheva16}.  Furthermore,  G102 coverage of GOODS-N (HST-GO 13420), as well as the CANDELS \lya\ Emission at Reionization Survey (CLEAR; \citealt{clear_paper1}, Simons et al., in prep) expands the sample to include the $1.3 < z < 2.0$ range over a subset of CANDELS.   Altogether, these data increase the sample size from \cite{Henry2013_wisp} by more than a factor of ten, allowing us to address evolution over $1.3 < z <  2.3$ and measure the M-Z-SFR relation at these redshifts.     Moreover, our new sample provides a meaningful constraint on theoretical models.    For the first time, we carry out a realistic comparison between observations and theory, by emulating our sample selection in the IllustrisTNG hydrodynamical simulation \citep{Marinacci18, Springel18, Nelson18, Pillepich18, Naiman18}.

This paper is organized as follows. In \S \ref{sec_data} we review the data included in this study, describing our sample selection, and measurements on individual spectra.   We describe our mass measurements, identification and removal of AGN, and provide an overview of the sample characteristics.    In \S \ref{analysis} we describe our methods for analysis, including our technique for creating composite spectra and measuring their emission lines.  In this section we also present the arguments behind our choice of metallicity calibration \citep{Curti17}, and our assessment of the current systematic uncertainties in metallicity measurements.   In \S \ref{results}  we present our resulting MZR and M-Z-SFR relations; then we compare our results to IllustrisTNG in \S \ref{theory_sec}. Finally,  \S \ref{conclusions} contains our conclusions.   Appendices include a description of our equivalent width estimates (Appendix \ref{appendix_ew}), an assessment of different methods for dust-correcting  stacked spectra (Appendix \ref{dust_stack_app}), and an outline of our Bayesian method for inferring metallicities (Appendix \ref{bayes_met}).   We use AB magnitudes and a Planck 2015 cosmology \citep{Planck15} throughout.

\section{Data and Sample} 
\label{sec_data} 
\subsection{Survey Overview} 
We select galaxies from multiple WFC3 grism spectroscopic surveys that also have multi-filter HST imaging. We require spectroscopic coverage from \oii\ $\lambda \lambda$3726, 3729 to  \oiii\ $\lambda \lambda$4959, 5007 in order to measure metallicities.  This requirement results in the selection of galaxies at $1.3 < z < 2.3$ for fields that have observations in the G102 and G141 grisms, and $2.0 < z < 2.3$ for fields with only G141 spectroscopy. Our science goals require masses of the galaxies, and therefore we select fields with multi-band photometry such that we can conduct spectral energy distribution (SED) fitting.  To this end, we select spectroscopic surveys in the CANDELS fields, all of which have significant photometric data (e.g. \citealt{Skelton14}), and fields in the WISP survey that include sufficient imaging.

In detail, the five CANDELS fields are covered by multiple imaging and spectroscopic surveys.  The photometric data are extensive, with many bands and coverage from the U-band through {\it Spitzer}/IRAC \citep{Guo13, Galametz13, Skelton14, Stefanon17, Nayyeri17,Barro19}.    For the spectroscopic data, we use G141 spectroscopy from the 3D-HST survey, which observed GOODS-S, EGS, UDS, and COSMOS to two orbit depth, and included GOODS-N G141 observations from the A Grism \ha\ SpecTroscopic (AGHAST) Survey (PI Weiner, PID 11600; two orbit depth).    Likewise, the CANDELS survey itself included deep, 21 orbit G141 spectroscopy in GOODS-S in one pointing.   In contrast to the excellent G141 coverage, no uniform G102 spectroscopy data exist.  Still, we include shallow G102 spectroscopy of GOODS-N (\citealt{Barro19}; two orbit depth) and deep G102 spectroscopy from the CLEAR survey (\citealt{clear_paper1}; 12 orbit depth) covering parts of GOODS-N and GOODS-S (which also includes multiple position angles to disambiguate contamination from overlapping spectra). If there is overlap in the G102 spectroscopy for a target, then we use the CLEAR data. Since the reduced CLEAR data are not public, we conducted our own reduction of these data (see \S \ref{phot_sec}). In total, these five fields cover an area around 680 arcmin$^{2}$. We refer to these objects as the CANDELS/3D-HST sample.

In addition to observations in the CANDELS fields, the WISP survey includes 483 individual WFC3 pointings obtained in HST's pure parallel mode.  In these fields, we limit our usage to the 151 fields that have G102 and G141 spectroscopy and include sufficient HST imaging for SED fits.   The imaging that supplements the spectroscopy for these 151 fields comprises IR imaging (typically F110W and F160W) and un-binned optical imaging with WFC3, typically in either 475X and F600LP, or F606W and F814W\footnote{Binned UVIS data cannot be corrected for charge transfer inefficiency,  nor can it be properly calibrated.  Therefore we exclude the 11 WISP fields  where UVIS optical imaging was taken in binned mode.}. Of these 151 fields, 100 also include {\it Spitzer}/IRAC channel 1 imaging \citep{Fazio04}, providing photometric coverage from 0.5-4 microns. 
Although the WISP photometry is not as extensive as in CANDELS, these data combined with the known redshifts of the sources are sufficient to obtain reliable mass estimates.  These 151 fields\footnote{The effective area of a pure parellel grism pointing is reduced from the full WFC3/IR area of 4.6 arcmin$^2 $  to around 3.6 arcmin$^2$.  Emission lines on the right hand side of the detector are excluded from the WISP catalog, because imaging outside the field-of-view would be required to disambiguate emission lines and zero order images.} cover around 540 arcmin$^{2}$.  Typically, the spectroscopy is at least two orbits in G102 and one orbit in G141, although a range of parallel visit lengths are included in the survey (\citealt{Bagley20}, Bagley et al. 2021, in prep). 

\subsection{Sample Selection} 
We select galaxies with signal-to-noise ratio,  SNR $>5$ in \oiii\ $\lambda \lambda$4959,5007 in the 3D-HST catalog from 
\cite{Momcheva16} and from the WISP emission line catalog (Bagley et al. 2021, in prep). In addition to this SNR cut, we also require detection of multiple emission lines in the WISP sample.  While \cite{Baronchelli20} show that machine learning can identify the redshifts of single-line emitters in WISP with around 80\% accuracy, we conservatively opt for a more rigorous selection to ensure the correct redshift. 
  We adopt the criterion that at least one additional line must be confirmed by visual inspection.  These lines include \ha, \hb, \oii, or an asymmetric \oiii\ line profile indicative of  blended \oiii\ $\lambda \lambda 4959, 5007$.   
  For the CANDELS/3D-HST sample, on the other hand, we allow single line emitters because the high-quality ancillary data enable photometric redshifts that are sufficient to identify the emission line and confirm the redshift (see \S \ref{inspect_fit}). In total, we select a parent sample of 1431 galaxies from the CANDELS fields and 384 galaxies from WISP, which are then reviewed (see \S \ref{inspect_fit}).  We acknowledge that an \oiii-based selection, and the requirement for multiple emission lines in some (but not all) of the sample introduce biases in metallicity.  These are addressed in \S \ref{bias_sample_sec} and \S \ref{theory_sec}.

\subsection{Spectroscopy}
The CANDELS, 3D-HST, and AGHAST spectra are publicly available from the 3D-HST collaboration on the Mikulski Archive for Space Telescopes (MAST)\footnote{\url{https://archive.stsci.edu/prepds/3d-hst/}}\footnote{\url{https://archive.stsci.edu/prepds/candels/}}   and are described in \citep{Momcheva16}.  
To these data, we add G102 observations in GOODS-N, using the fully-reduced spectra presented in \cite{Barro19}. 
Additionally, we include G102 observations from CLEAR, processing the data using the Simulation Based Extraction (SBE) pipeline created to reduce the Faint Infrared Grism Survey (FIGS) data \citep{Pirzkal17}. To summarize, a custom background subtraction was performed to remove the dispersed background from each individual observation, correcting for any variation in the background during the course of an exposure. The CLEAR imaging data were astrometrically corrected to match that of the CANDELS mosaics and these corrections were propagated to the G141 observations.  Full simulations of each of the G141 observations were then generated using the available broad band photometry and object footprints in these fields. These simulations were used to produce realistic estimates of any contamination caused by overlapping spectra, and to later calculate extraction weights to use for optimal spectral extraction \citep{Horne86}.   The spectra from multiple roll angles in the CLEAR observations were combined, excluding regions of spectral contamination. A detailed and complete description of this process is described in \cite{Pirzkal17}. 

The WISP data are also publicly available from the WISP collaboration on MAST\footnote{https://archive.stsci.edu/prepds/wisp/}, although we also made use of proprietary WISP data products (e.g.\ drizzled data at different pixel scales, as discussed in \S \ref{phot_sec}). The pipeline for spectroscopic reduction is based on the one described in \cite{Atek10}, which uses the aXe slitless grism reduction package \citep{aXe}, but is updated to include improved calibrations and methodologies (Baronchelli et al.\ 2021, in prep).  In short, the new pipeline uses Astrodrizzle, custom dark calibrations and charge transfer efficiency corrections for WFC3/UVIS as described in \cite{Rafelski15}, as well as corrections for scattered light in the WFC3/IR data, and improved source detection using all available WFC3/IR direct images. We also made use of the WISP team's upgraded line identifications and flux measurements to select galaxies covering the full set of fields as described in \cite{Bagley20} and Bagley et al. (2021, in prep). 

In all of the above surveys, the resolution of the spectra are low.   The G102 and G141 grisms have R $\sim 210$ and $130$ for point sources.  In practice, the galaxies under consideration in this paper have even lower resolution, with a line spread function that is broadened by the morphology of the source.  This results in spectra where \nii\  $\lambda \lambda 6548, 6583$ is completely unresolved from \ha, and the \oiii\ $\lambda \lambda 4959, 5007$ doublet is at best marginally resolved.

\subsection{Photometry} 
\label{phot_sec} 
We require a photometric catalog of all galaxies in our sample to enable measurements of masses in a uniform and consistent fashion using the latest SED fitting methodologies (e.g. \citealt{Pacifici16}). We incorporate the photometry catalogs from CANDELS  \citep{Guo13, Galametz13, Stefanon17, Nayyeri17,Barro19},  3D-HST \citep{Skelton14}, and WISP to create a combined photometry catalog for our selected CANDELS/3D-HST and WISP sources.    

For the CANDELS fields we utilize the CANDELS photometric catalogs. 
We give preference to these measurements, rather than those from 3D-HST,  as we found the CANDELS catalogs to be more reliable when including ground-based and {\it Spitzer}/IRAC data.  This difference is likely due to the CANDELS team's use of isophotal apertures, combined with TPHOT \citep{Merlin15, Merlin16, Nayyeri17}, rather than the fixed aperture photometry used for the 3D-HST measurements \citep{Skelton14}. However, our  parent sample does include 42 sources without matching CANDELS photometry;  for these, we instead use the 3D-HST photometry. 

The WISP photometry is obtained from the catalog in Bagley et al.\ (in prep), and is described in more detail therein. Here we provide a brief overview. All HST images are drizzled onto the same pixel scale optimized for the WFC3/UVIS images (0.04\arcsec/pixel), and cleaned of bad pixels, chip gaps, and noisy edges. The images are then 
convolved with a kernel to match the point-spread-function (PSF) in the F160W filter, using IRAF's {\tt PSFMATCH}. 
As part of the WISP reduction pipeline, SExtractor \citep{SE} is used to generate a segmentation map from the F110W and 
F160W detections.   This segmentation map is  resampled onto the same pixel scale and used as the source definitions. The photometry is performed using {\tt photutils} in Astropy to derive isophotal fluxes in all HST bands \citep{Bradley21, astropy13, astropy18}, with local sky subtraction performed in 10\arcsec\ rectangular apertures. All source flux is masked out of the sky apertures. SExtractor photometry is also performed on WFC3/IR 0.08\arcsec/pixel images (prior to PSF-matching),  in order to define the spectroscopic isophotal extraction region, and also to obtain the total magnitudes via SExtractor's MAG\_AUTO.   Aperture corrections are derived from the difference between MAG\_AUTO and the isophotal magnitude in the reddest HST band (generally F160W). 
Finally, the IRAC photometry is obtained with TPHOT matched to the F160W image. The resultant catalog contains PSF matched photometry in all filters in various apertures. 

We make modifications to the WISP photometric catalog to standardize and clean the measurements of our WISP sources before adding them to our combined catalog as follows. First, we omit IRAC entries for around 10\% of sources, where the TPHOT flag values indicates a blended source or a source near the image edge.  Second, for the HST photometry, we start with the isophotal PSF-matched photometry and apply a magnitude correction to calculate their corresponding MAG\_AUTO magnitudes (not included in the catalog for WFC3/UVIS photometry, only for the WFC3/IR 0.08\arcsec/pixel images). These MAG\_AUTO magnitudes are the total magnitudes that are needed for consistency with the CANDELS photometry, but are only provided for the near infrared photometry in Bagley et al. (in prep). 

Third, for a dozen sources with negative fluxes in the WISP catalog, where no individual  aperture correction is possible, we apply a magnitude correction to the flux value equal to the median magnitude correction of sources with SNR between 0 and 1 in the given filter. This strategy enables us to apply an aperture correction to negative flux entries, and derive upper limits instead of discarding them.  Additionally, we omit any entries for sources with a high F160W magnitude error (MAGERR\_AUTO\_F160W $>$ 15) or those without both isophotal and AUTO F160W coverage (near edges) as these result in poor magnitude corrections.  We do not apply this same aperture correction strategy to negative IRAC flux entries because the WISP catalog does not contain any flux measurements for the IRAC channels (just magnitudes), and these negative entries are therefore omitted. 

The result is a photometric catalog for both the CANDELS fields and the WISP fields with consistent aperture and PSF matched photometry. This catalog is the input for the SED fitting described in  \S \ref{mass_sec}.

\subsection{Inspection and measurement of individual grism spectra}  
\label{inspect_fit} 
We require accurate line and continuum measurements for every object, in order to measure metallicities from individual objects, and also to facilitate spectral stacking (see \S \ref{stack_sec}).  Therefore, we  (AH and MR) carried out interactive line and continuum fitting for all of the galaxies in the sample, using custom software\footnote{ \url{https://github.com/HSTWISP/wisp_analysis}} (Bagley et al., in prep).       
 Our software interactively fits cubic splines to the continuum, with a separate spline function for G141 and G102 when both are present.  This approach is ideal for handling data where contamination from overlapping spectra can produce unusual spectral shapes, even when attempts are made to model and subtract it.  In these measurements, the contamination model was not cleaned from the continuum spectra, but was instead fit and subtracted with the continuum.   Likewise, interactive inspection of each galaxy in the sample allowed us to mask out regions of contamination from zero order images, or extremely bright overlapping spectra that could not be fit well by our automated method.   It also allowed us to adjust the dividing wavelength between G102 and G141 spectra (which overlap slightly), as the optimal transition wavelength varied from object to object.   Finally, emission line fluxes were measured at the same time, by fitting Gaussian profiles where the FWHM (in pixels) were held constant among the lines\footnote{In slitless grism spectroscopy (and slit spectroscopy where the source does not fill the aperture), the line spread function is set by the morphology of the source.  Hence, the FWHM of the lines are expected to be the same {\it in pixels} in G102 and G141.  Since the dispersion in G102 that is two times higher than in G141, the same FWHM in pixels translates to two times higher spectral resolution in G102.}.   For this measurement, we fit both lines of the \oiii\ $\lambda \lambda 4959, 5007$ doublet, with a fixed doublet ratio of 2.9:1. However, for \oii\ $\lambda \lambda 3726,3729$ and \ha\ + \nii\ $\lambda \lambda 6548, 6583$ we fit single Gaussians.  In the former case, the doublet is at best marginally resolved for the resolution of our grism spectra; in the latter case, the \nii\ lines are generally weak, and the SNR of the emission line blend is not high enough to detect any non-Gaussianity in the line profile.  Importantly, this analysis combined the CLEAR data with those from 3DHST/AGHAST, ultimately providing a uniform set of line measurements and continuum+contamination subtracted spectra that we can stack.

As part of the interactive fitting,  visual inspection of emission lines also ensured a clean sample.  While the WISP catalog only includes sources that were verified by at least two people in a previous visual inspection, the CANDELS/3D-HST sample was not previously examined by members of our team.     In this step, 35 out of 384 sources were rejected from WISP and 383 out of 1431 sources were removed from the CANDELS/3D-HST sample.  The reasons were varied, but included: severe contamination due to spectral overlap with a very bright source, the absence of any visible emission lines, spectra near the edges of detectors, and cases where emission lines from multiple objects were possibly present in the 1D spectrum.   This re-analysis provided a clean sample with 349 galaxies from the WISP survey and 1048 from the five CANDELS/3D-HST fields.    

In addition to line fluxes, measurements of emission line equivalent widths are required, so that we can correct the Balmer line flux ratios for stellar absorption when we measure metallicities.    We estimate equivalent widths of the lines using broad-band photometry to estimate the continuum, as we describe in Appendix \ref{appendix_ew}.    In this appendix, we show that these measurements agree with spectroscopic measurements of equivalent width with a 1$\sigma$ scatter of 40\%.  
However, the broad-band continuum estimates are more complete for faint objects, so we adopt this method for the sake of consistency within our sample.

We verified our measured  \oiii\ fluxes and flux errors by comparing to the 3D-HST catalog  \citep{Momcheva16}.   The line fluxes that we measured from Gaussian fits were on average 15\% lower than those from  \cite{Momcheva16},  where the spectral line profile is a convolution of the source morphology and the point source line spread function.\footnote{We compared several of our line profile fits with those from 3D-HST and find that the latter are often slightly too broad.  Since the 3D-HST spectral line profiles are fixed by the broad-band source morphology, this discrepancy could be an indication that the emission line regions are more compact than the stellar continuum.  Nonetheless, on an object by object basis, the line fluxes measured by the two methods agree within the uncertainties.}  Likewise, the measurement errors from our fits were, on average, 8\% larger than those in the 3D-HST catalogs.  Combined, these systematics imply that our measurements yield emission line signal-to-noise ratios (SNR) that are, on average, 80\% of what is given in the 3D-HST catalogs.   We conclude that this level of agreement is reasonable.

Similarly, we compared our redshift identification to those from the 3D-HST catalog. Here, we followed \cite{Momcheva16}, calculating  the normalized absolute median 
deviation ($\sigma_{NMAD}$)  of the difference between our measured redshift and that from 3D-HST: $\Delta z = z_{measured} - z_{3DHST}$.  We take: 
\begin{equation} 
\sigma_ {NMAD} = 1.48 \times median \left ( \left |  {\Delta z - {\rm median}(\Delta z)} \over { 1 + z_{measured}}   \right | \right ),  
\end{equation}    
as given by \cite{Brammer08}. 
 We find $\sigma_{NMAD} = 0.0055$ for sources with $H_{160} < 24$.   If we restrict this redshift comparison to sources where our method finds \oiii\ SNR $> 5$, we find $\sigma_{NMAD} = 0.0042$.  Alternatively,  for a subset of 85 of our sources with \oiii\ SNR $>5$ and robust spectroscopic redshifts from the MOSFIRE Deep Evolution Field (MOSDEF) Survey (category 6 or 7; \citealt{Kriek15}), we calculate $\sigma_{NMAD} = 0.0014$.  Much of this uncertainty is  attributable to the low resolution of WFC3/IR grism spectroscopy, where a single pixel corresponds to $\Delta z$ = 0.009 (for \oiii\ $\lambda$5008).  
For comparison, \cite{Momcheva16} report a  higher $\sigma_{NMAD}$ of  0.0015 - 0.0045 when the 3D-HST redshift accuracy is assessed against MOSDEF and other ground-based near-infrared spectroscopic followup (e.g. \citealt{Wisnioski15}). 
We speculate that the 3D-HST redshift uncertainties may be higher due to inaccuracies that could be removed with more human supervision of the emission line fitting process (see also, \citealt{Rutkowski16}). 

Our comparison with the MOSDEF spectroscopic redshift catalog verifies the accuracy of the redshifts of the single emission line objects from the CANDELS/3D-HST sample. Of the 85 sources in common between the two samples, we note only 3 objects with $| \Delta z|  > 0.15$.  These objects are all in GOODS-N, with IDs 8537, 13286, and 21398 in the MOSDEF and 3D-HST v4.1 catalogs.   Of these, two of the MOSDEF spectra are included in the January 2021 public data release\footnote{http://mosdef.astro.berkeley.edu/for-scientists/data-releases/}.  We reviewed these two spectra, and found that the emission line detections and measured spectroscopic redshifts were not particularly convincing.  Therefore, we conclude that we cannot determine at this time whether the MOSDEF  or WFC3/IR grism spectroscopic redshifts are correct.  In either case, 82/85 objects show excellent agreement, suggesting at least $\sim$96\% accuracy on the grism+photometric redshifts of our {\it vetted} CANDELS/3D-HST sample.

Finally, as we noted above, the \oiii\ signal-to-noise that we measure in our re-analysis is typically lower than the threshold  (SNR $>5$) that we used to select the sample.      Therefore, we removed four galaxies from the WISP sample, and  265 galaxies from the CANDELS/3D-HST sample to preserve our SNR $>5$ threshold.   In total, our final sample comprises 345 galaxies from WISP and 783 galaxies from CANDELS/3D-HST, for a total of 1128 objects.      Since we have thoroughly vetted both the WISP and CANDELS/3D-HST spectroscopy, our sample is of much higher quality than the unsupervised 3D-HST spectroscopic catalog.

\begin{figure*}[!t] 
\begin{center} 
\includegraphics[scale=0.7, trim = 0 30 0 0]{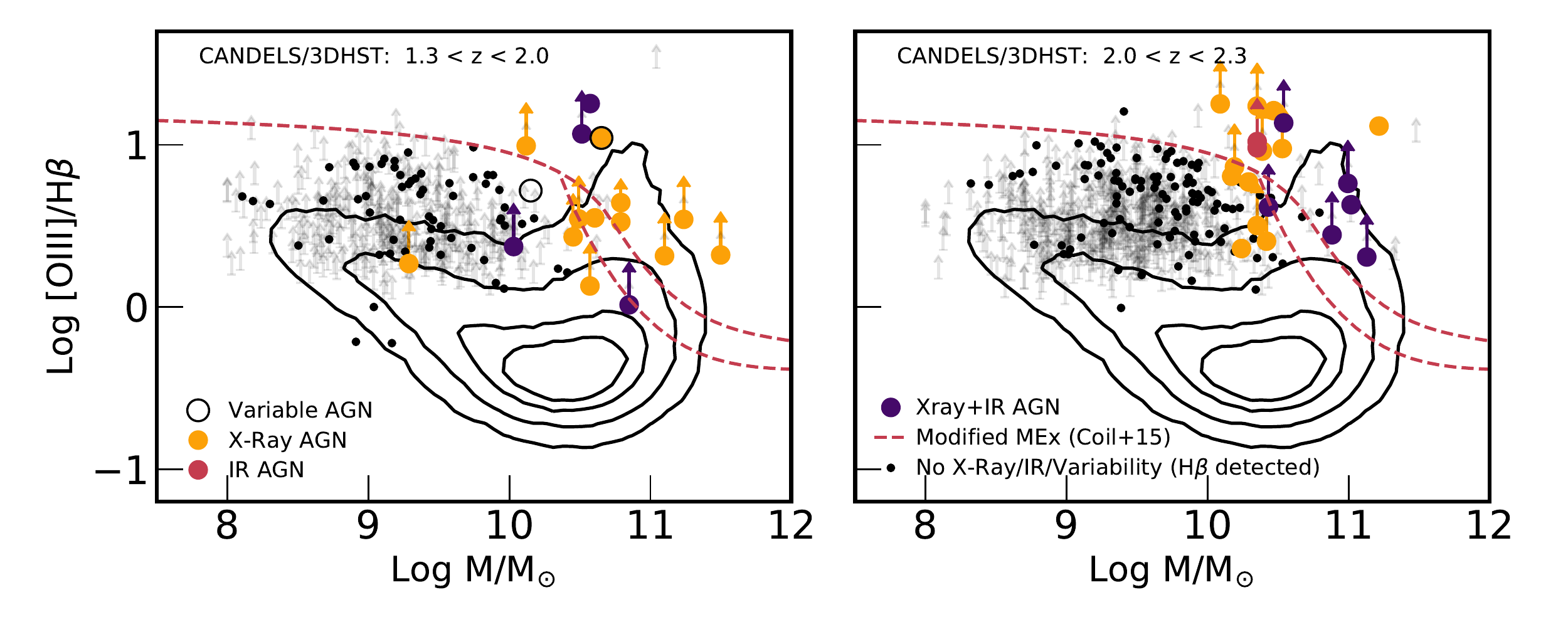} 
\end{center}
\caption{The Mass-Excitation  (MEx) diagram can be used to discriminate between star-forming galaxies and AGN.  In this diagnostic, AGN fall above the red dashed line, with the lower and upper curves representing lower and higher probabilities for hosting AGN (see \citealt{Juneau14}).   Here, we show galaxies from the CANDELS/3D-HST fields only; this approach facilitates a comparison with the known AGN in these fields.  
These AGN are identified as X-ray detected objects  (\citealt{Kocevski18}; Kocevski 2019, private communication), objects with non-stellar {\it Spitzer}/IRAC colors \citep{Donley12}, and AGN identified from their variable $z-$band photometry (GOODS-N and GOODS-S only; \citealt{Villforth10}).   Several objects meet multiple criteria.   The small black points (\hb\ detections, $> 3 \sigma$)  and grey arrows (\oiii/\hb\ 3$\sigma$ lower limits) show objects that are not classified as AGN by the X-Ray, IR or variability methods, although they may still be classified as AGN from the MEx method.    The contours show the density of low redshift galaxies, and are derived from the JHU/MPA SDSS DR7 catalog.  The dashed-line for distinguishing galaxies from AGN is from \cite{Coil15}, which follows the shape of the $z \sim 0$ relation from \cite{Juneau14}, but is shifted by their recommendation of 0.75 dex to higher mass (to the right) to account for redshift evolution.  } 
\label{agn_mex}
\end{figure*}

\subsection{SED Fitting: Stellar mass and SFR} 
\label{mass_sec} 
Stellar masses are derived by SED fitting to the photometry described in \S \ref{phot_sec}.  We opt to derive stellar masses for our sample instead of using published mass catalogs from the CANDELS or 3D-HST teams \citep{Mobasher15, Skelton14}.  This approach has the clear advantage, in that we can apply a uniform approach to the WISP and CANDELS/3D-HST data.  Likewise, prior estimates of stellar mass do not account for emission line contamination to broad-band photometry, which can cause masses to be overestimated 
\citep{Atek11}.  Re-deriving stellar masses gives us the opportunity to use the most up-to-date SED fitting methodology, making use of non-parametric star formation histories.  Indeed, \cite{Lower20} show that non-parametric star formation histories produce more accurate stellar masses than parametric models.  

We adopt the approach described in \cite{Pacifici12, Pacifici16}.  In brief, we generate model SEDs assuming star formation and chemical enrichment histories from a semi-analytical model, which allows us to span a wide range of star formation histories. Stellar population models are taken from the 2011 version of \cite{BC03},  and nebular emission lines are modeled consistently with the stars (see \citealt{Pacifici12, Pacifici16}). The attenuation by dust is computed using a two-component dust model to allow for different dust geometries and galaxy orientations \citep{CharlotFall00}. Fixing the redshifts to those derived from the grism spectroscopy, each model SED is compared to the observed photometry.  From this analysis, we derive estimates and confidence ranges for the stellar mass and the SFR using a Bayesian approach.  A \cite{Chabrier2003} IMF is assumed.  The average 68\% confidence range for stellar mass is  0.12 dex for the CANDELS/3D-HST subsample, and 0.37 dex for WISP.  Likewise, the average 68\% confidence interval for the SED-derived SFRs are  0.40 and 0.63 dex for the two subsamples, respectively.  The differences between the SED fit uncertainties in WISP versus CANDELS/3D-HST reflect the significantly larger number of filters used in the latter.

\subsection{AGN} 
\label{agn_sec} 
We aim to remove  Active Galactic Nuclei (AGN) from our sample, so that our metallicity measurements are not contaminated by non-stellar ionizing sources.  With the low resolution of our spectra, the low SNR,  and wavelength coverage that does not always reach \ha\ and \nii, discriminating between star-forming galaxies and AGN can be challenging.   
The traditional ``BPT'' diagram (\nii/\ha\ vs. \oiii/\hb; \citealt{BPT}) cannot be used to distinguish star-forming galaxies and AGN.     Fortunately, there are alternatives that are applicable to our data, each of which we consider here. First, all of the spectra in the CANDELS/3D-HST fields have X-ray observations, as well as Spitzer/IRAC photometry, both of which can identify AGN.  Second, we include AGN identified by their $z-$band photometric variability in GOODS-N and GOODS-S \citep{Villforth10}.    Third,  we can use the ``Mass-Excitation'' (MEx) diagram, which replaces the \nii/\ha\ ratio in the BPT diagram with stellar mass \citep{Juneau11,  Juneau14}.  Fourth, and finally, at $z<1.5$, we can consider a modified BPT diagram, using \sii / (\nii + \ha)  vs. \oiii/\hb.

Figure \ref{agn_mex} shows the MEx diagram for the sources in the CANDELS/3D-HST portion of our sample, divided into two panels for $1.3 <  z <  2$ and $2 < z< 2.3$.   Here, we exclude the WISP data, as we aim to validate the MEx diagram using the known AGN in the CANDELS fields.  The contours in this figure show the density of galaxies in the low-redshift relation, taken from the MPA/JHU SDSS DR7 catalog\footnote{\url{https://home.strw.leidenuniv.nl/~jarle/SDSS/}}.  We  have overplotted X-ray-identified AGN taken from the CANDELS  survey (\citealt{Kocevski18}, Kocevski et al. private communication), as well as two AGN identified through their photometric variability in \cite{Villforth10}.   We also applied the Spitzer/IRAC color selection from \cite{Donley12} to identify additional galaxies that were not in the X-ray or variable source catalogs.  For the IRAC photometry, we took measurements from the 3D-HST photometric catalog \citep{Skelton14}, requiring 3$\sigma$ detections in all four IRAC bands, and less than 50\%  contamination (from blending) to the IRAC photometry.   We considered relaxing the infrared photometry selection to include lower SNR, more contamination, or objects falling less than 1$\sigma$ outside the \cite{Donley12} color cut.    However, many of these additional objects fell well within the star-forming locus in Figure \ref{agn_mex}, suggesting that a relaxed selection returned mostly false positives.  

A comparison of AGN selection methods, including X-ray, IRAC colors, BPT, and MEx has previously been carried out for $z\sim 2.3$ galaxies by \cite{Coil15}.  Figure 1 is consistent with their conclusion that the $z\sim0$ version of the  MEx diagram cannot be used at high-redshift.  Since stellar mass serves as a proxy for \nii/\ha, and metallicity evolves with redshift, the MEx AGN selection should also evolve.  \cite{Coil15} propose, based on their X-ray and IR identified AGN, that the $z\sim0$ AGN threshold should be shifted by around 0.75 dex to higher stellar mass.  This shift is similar to the 1 dex shift that we proposed in \cite{Henry2013_wisp}; however,  since \cite{Coil15} calibrated the shift based on known AGN, we judge this estimate to be more accurate.   We show this modified selection in Figure  \ref{agn_mex}.    All of the known IR and X-ray AGN are near or above this MEx threshold, or have lower limits on \oiii/\hb\ that are consistent with a MEx-AGN  classification.   Of the photometrically variable AGN, one is an X-ray AGN and also falls above the MEx threshold, while the other appears in the star-forming region.      

Figure \ref{agn_mex} also breaks the sample into two redshift bins-- $1.3 <  z <  2$ and $2 < z< 2.3$--  in case of evolution within our sample.  
However, we see no need for different MEx thresholds to be applied in the low and high redshift bins.
 Since there is only 800 Myrs between the mean redshift in the two panels of Figure  \ref{agn_mex} ($z=1.68$ vs. $z=2.15$),  and we also do not detect significant metallicity evolution in our sample (\S \ref{evo_sec}), we conclude that there is no strong evidence for evolution of the MEx AGN selection within the redshift range that we consider.   In summary, the known AGN confirm the result seen by \cite{Coil15}:  the MEx diagnostic with a 0.75 dex shift works reasonably well at $z\sim 2$. 
 
To complement the MEx classification, we also consider a more conventional emission-line-only AGN diagnostic.  This test is only possible for the portion of our sample at $1.3 < z < 1.5$, where we have coverage of \ha\ and \sii.  While the resolution of the grism spectra blends \ha\ and the \nii\ lines,  the \nii\ is likely weak for most galaxies in our sample \citep{Erb06}, implying that \sii/(\ha + \nii) $\sim$ \sii/\ha.   This AGN diagnostic diagram is plotted in Figure \ref{agn_sii}.    As in Figure \ref{agn_mex}, we compare to known AGN: X-ray identified objects and one photometrically variable AGN at this redshift are plotted, along with those that meet the MEx criteria (for both WISP and CANDELS/3DHST). There are no IR-selected AGN in this redshift range.  Lastly, we plot a theoretical maximum for star-forming galaxies, accounting for blended \nii.   We calculated this threshold from the same Cloudy (v17; \citealt{Ferland17}) models used in \cite{Henry18}; the maximum is set by the hardest plausible ionizing BPASS v2.0 \citep{Eldridge16} stellar spectrum, having $Z  = 0.001$, an IMF extending to 300 $M_{\sun}$ and a constant star-formation rate.    We estimate that star-forming galaxies fall below a curve following the typical functional form: 

\begin{figure} 
\begin{center}
\includegraphics[scale=0.4, trim = 0 20 0 0 ]{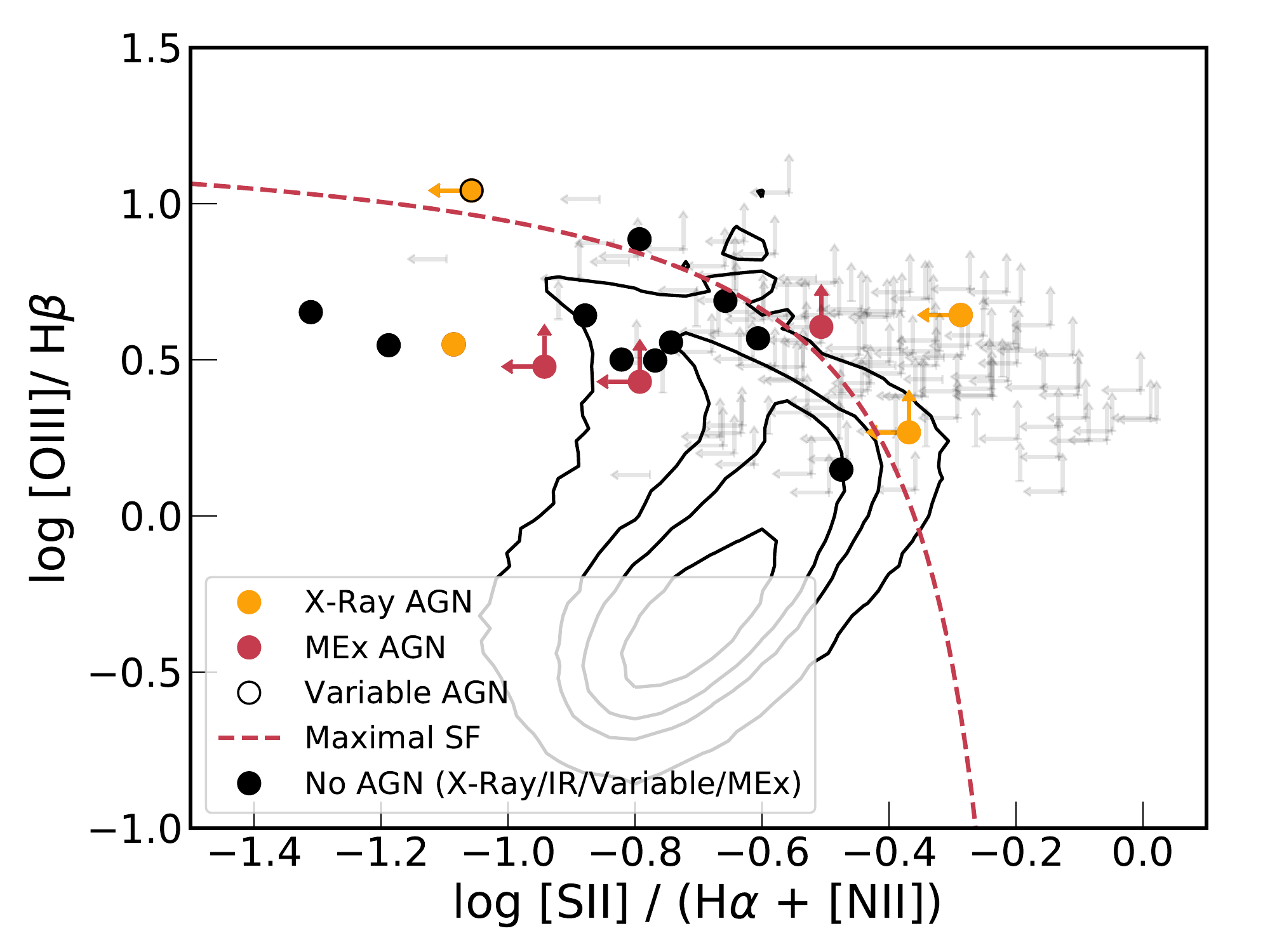} 
\end{center}
\caption{The modified BPT diagram that is measurable from WFC3/IR grism spectra, for both WISP and CANDELS/3D-HST sources at $1.3 < z < 1.5$.  Both lines of the \oiii, \sii, and \nii\ doublets are included in the line ratios plotted here.   X-ray AGN from the CANDELS/3DHST fields are shown (one of which is variable in its $z-$band photometry; \citealt{Villforth10}), along with MEx AGN from the full survey, selected via the \cite{Coil15} modified MEx threshold.  Black points and grey arrows (3$\sigma$ upper/lower limits) show galaxies that are not classified as AGN by any method.     The dashed curve shows a maximal star-forming threshold, calculated from Cloudy \citep{Ferland17} models, as given in Equation  \ref{bpt_thresh}. The contours are derived from the JHU/MPA SDSS DR7 catalog. 
 } 
\label{agn_sii} 
\end{figure} 

\begin{equation} 
{\rm log ( [OIII] 4959, 5007 /H}\beta) = {{0.28} \over { x + 0.14} } + 1.27
\label{bpt_thresh} 
\end{equation} 
where
\begin{equation}
{x} = {\rm log} { {\rm [S II] ~6716, 6731}  \over {({\rm H\alpha +  [NII]}~ 6548, 6583)  }}.
\end{equation}

\begin{figure*} 
\plotone{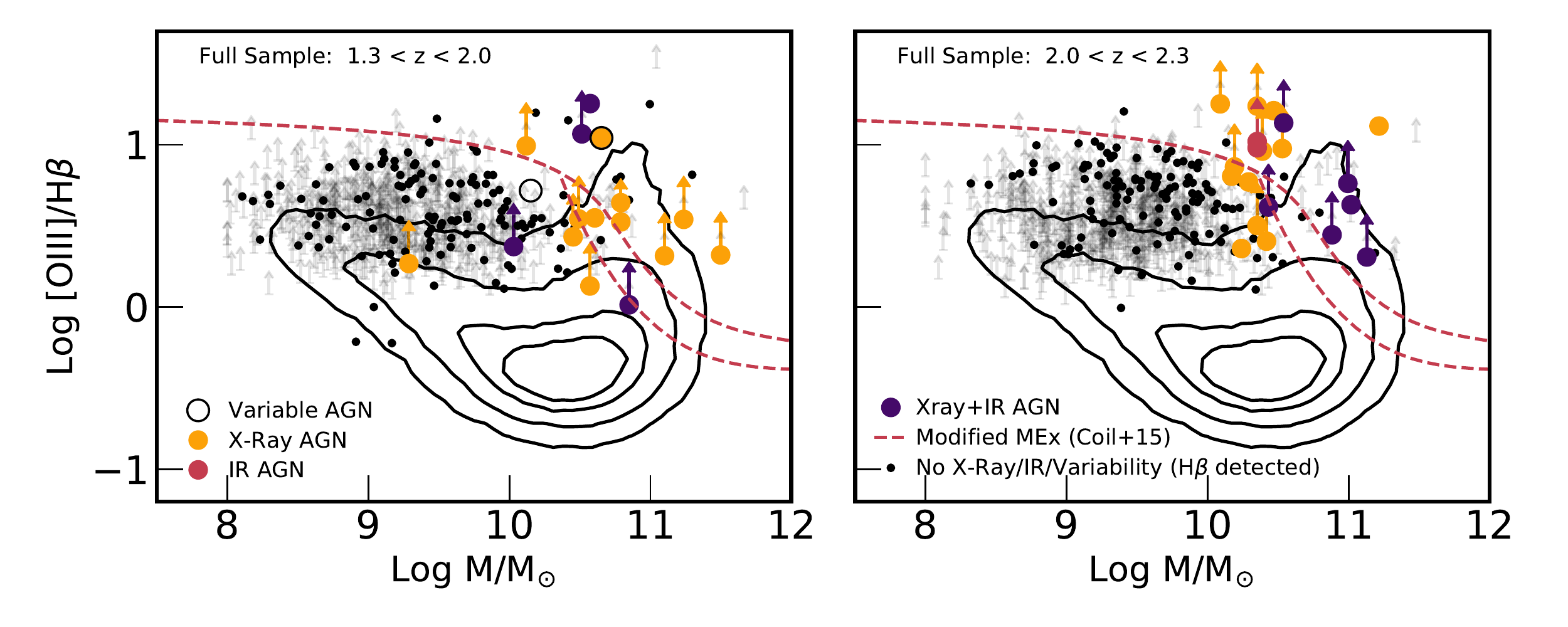} 
\caption{ The MEx diagram for the combined WISP and 3D-HST sample.  All markers have the same meaning as in Figure \ref{agn_mex}, but in this case we also include the galaxies from WISP.  For these objects, we do not have additional AGN diagnostics from X-Ray or infrared observations.   Comparing to Figure \ref{agn_mex}, we see a clear presence of objects with no prior AGN identification, but lying in the same region as the X-Ray and IR selected AGN.   The MEx diagram identifies these AGN from the WISP survey, where we have no other indicators.} 
\label{mex_all} 
\end{figure*}

As previously observed by \cite{Coil15}, we find one clear X-ray AGN showing weak \sii\ emission and low \oiii/\hb, placing its BPT measurements squarely in the star-forming region.  All of the  remaining AGN have limits on their weaker lines, such that they could fall within the star-forming region.   Therefore, we concur with the conclusions in \cite{Coil15}: the \sii\  BPT diagnostic seems to be a poor method for discriminating between AGN and star-forming galaxies at high-redshifts.  While the reason for this breakdown is unclear, it should be noted that weak \sii/\ha\ in  galaxies with otherwise normal \nii/\ha\ ratios has been proposed as a means of selecting galaxies with optically thin, Lyman Continuum (LyC) leaking ISM \citep{Alexandroff15, Wang19}.   When the ISM is density bounded, the outer regions of low-ionization state gas (\sii, \oii) can be small or absent. This scenario can also apply to AGN; indeed, \cite{Wang19} found that two of five LyC emitter candidates at $z\sim 0.3$, when selected to have weak \sii, were actually AGN.     Hence, it is plausible that the evolving conditions in the ISM of high-redshift galaxies and AGN may make \sii\ an unreliable AGN diagnostic.     
  We therefore use the MEx diagram for both WISP and CANDELS/3D-HST, along with the known X-ray, IR, and photometrically-variable  AGN in the CANDELS/3D-HST fields.     Figure \ref{mex_all} shows this diagnostic diagram for our full sample.  We now see many objects in the AGN part of the diagram from the WISP survey, where the MEx diagram is the only available indicator of AGN activity.

Finally, we note that Figures \ref{agn_mex}, \ref{agn_sii}, and \ref{mex_all} reveal that a large number of sources are ambiguous, due to \hb\ (and \sii) non-detections.   In particular, the lower limits on \oiii /\hb\  and \sii/\ha\ are consistent with being both above and below the AGN thresholds.   However, at low masses, this problem is not severe.  A number of authors have now shown that, when the high-redshift BPT diagram is considered, very few sources at low \nii/\ha\ have \oiii /\hb\ consistent with AGN \citep{Steidel14, Sanders16, Strom17, Kashino19}.     Since our analysis is based on stacking, a small minority of contaminating AGN will have a negligible impact.  However, above $M\sim 10^{10}$ M$_{\sun}$, where the MEx curves fall to lower  \oiii/\hb, the nature of the sources with \oiii /\hb\ lower limits is less clear.  Only 32 galaxies at $M > 10^{10}$ M$_{\sun}$ have 3$\sigma$ \hb\ detections and \oiii/\hb\ ratios consistent with star-formation, while 47 are ambiguous. Since X-Ray and IR AGN samples are likely incomplete at all but the highest masses, it is not clear how significant the AGN contamination is for $M > 10^{10}$ M$_{\sun}$ in our sample.  Therefore, as part of our stacking analysis in \S \ref{stack_sec}, we tested the effect of excluding these 47 galaxies with unknown AGN.   In our stack of the 32 galaxies that are clearly star-forming, the ratio of \oiii/\hb\ is decreased from 3.60 to 2.61, which corresponds to an increase in metallicity of 0.08 dex.   Much of this increase is likely a bias towards galaxies with stronger \hb\ emission and consequently lower \oiii/\hb\ ratios and higher metallicities.  Nonetheless, this test quantifies the possible impact of AGN contamination in our highest mass stack.  At most, an unknown contribution from AGN increases the \oiii/\hb\ ratio by 40\% and lowers the metallicity by 0.08 dex.  Ultimately, while we will assume that galaxies with ambiguous lower limits on \oiii/\hb\ are star-forming, we urge caution when interpreting the highest mass bin in our sample.   

Altogether, we remove  72 AGN, of which 63 meet the MEx classification, 12 are IR AGN, 35 X-ray AGN, and two have measured variability in their $z-$band photometry.  The remaining sample comprises 1056 galaxies. 

\begin{figure*} 
\plotone{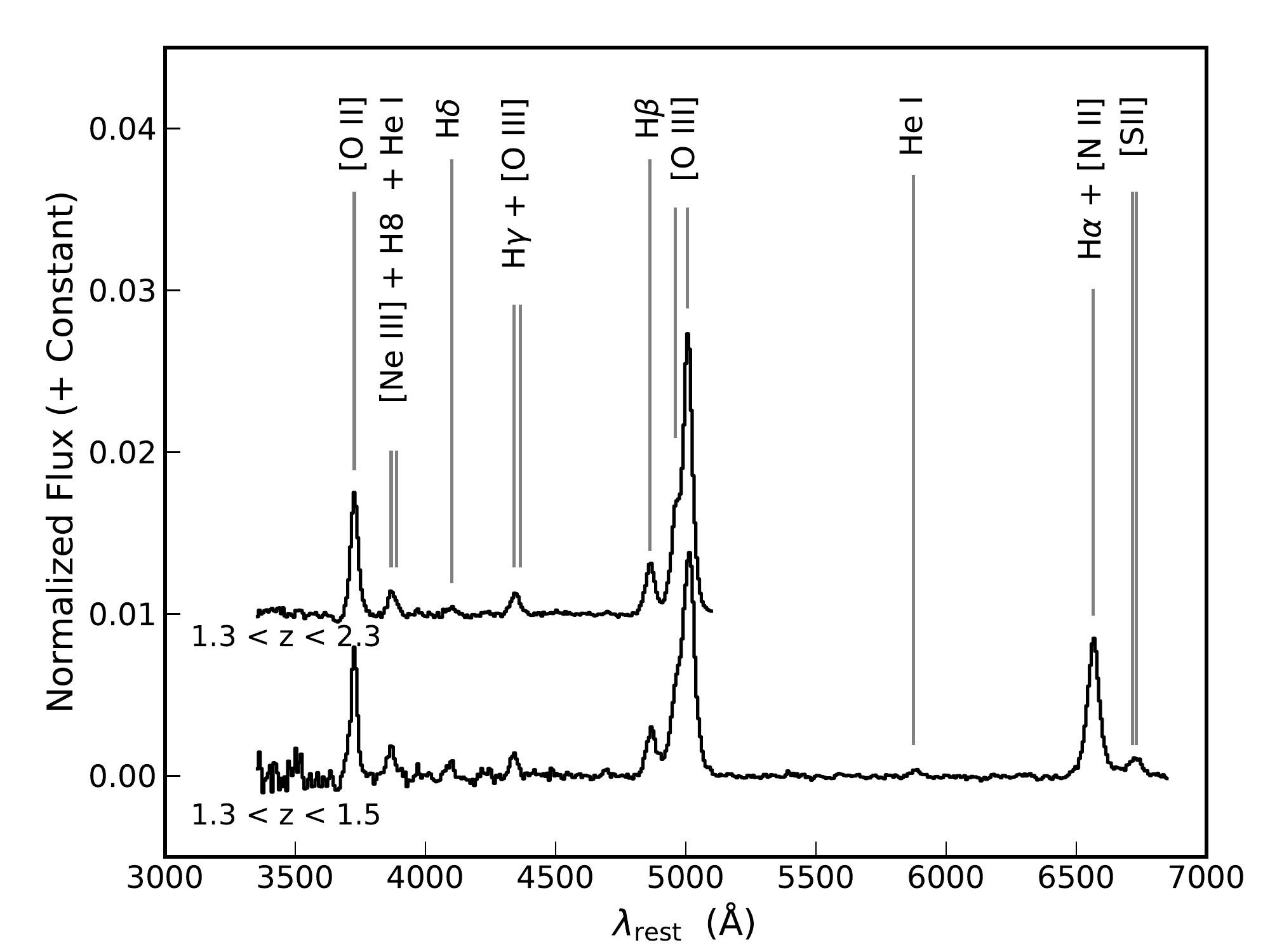} 
\caption{Stacked spectra for the full sample (top; 1056 galaxies), and the subset of the sample with coverage of \ha\ and \sii\ (bottom; 151 galaxies).  Numerous weak lines (and blends of lines) are detected.  } 
\label{stack_all_fig} 
\end{figure*}

\section{Analysis} 
\label{analysis}

\subsection{Composite Spectra}  
\label{stack_sec}  
In order to measure metallicity, we require robust measurements of multiple emission lines in our spectra.  Therefore, we create composite spectra by stacking to achieve higher SNR.  Our procedure is similar to that of \cite{Henry2013_wisp}.  First, we subtract the continuum that was determined in our interactive fitting (\S \ref{inspect_fit}).  Then, in order to avoid weighting the stacks towards galaxies with stronger emission lines, we normalized each spectrum by the measured flux of the \oiii\ $\lambda \lambda$4959, 5007 emission.  
 While this method of normalization does not remove biases owing to a range of dust extinction being present in each stack, we show in Appendix \ref{dust_stack_app} that the impact on metallicity from this effect is negligible.
  Next, we de-redshift the spectra, using a linear interpolation to shift them onto a common set of rest wavelengths.   
 Finally, we take the median of the normalized fluxes at each wavelength.  As a rule, we only consider the \ha\ + \sii\ wavelength region if the stack is restricted to galaxies at $z  \le 1.5$, where these lines are covered in the G141 spectra.    Figure \ref{stack_all_fig} shows a stack of the entire sample, alongside a stack for the subset at $z\le 1.5$.  In addition to providing  robust measurements of \oii\ and \hb\ for metallicity inferences, we detect a handful of weak emission lines:  a blend of \hg\ and  \oiii\ $\lambda$4363; \hd; blended [\ion{Ne}{3}] $\lambda$3869, He I $\lambda$3867, and H8 ($\lambda$3889); He I $\lambda$5875; and blended \sii\ $\lambda \lambda$6716, 6731.

The sample was divided into subsets for creating stacked spectra.  First, we considered stellar mass, using 5 bins of 0.5 dex:  log M/M$_{\sun} < 8.5$,  $8.5 < $ log M/M$_{\sun} < 9.0$,  $9.0 < $ log M/M$_{\sun} < 9.5$, $9.5 < $ log M/M$_{\sun} < 10.0$,    and log M/M$_{\sun} > 10.0$.       
Figure \ref{stack_massbins} shows the composite spectra for these five mass bins. 
In addition, the large numbers of galaxies in these bins allows us to test subdivision by other properties, keeping the mass bins fixed.  Therefore, we made stacks in several subsamples:
 
 \begin{figure*}[!t]
\plotone{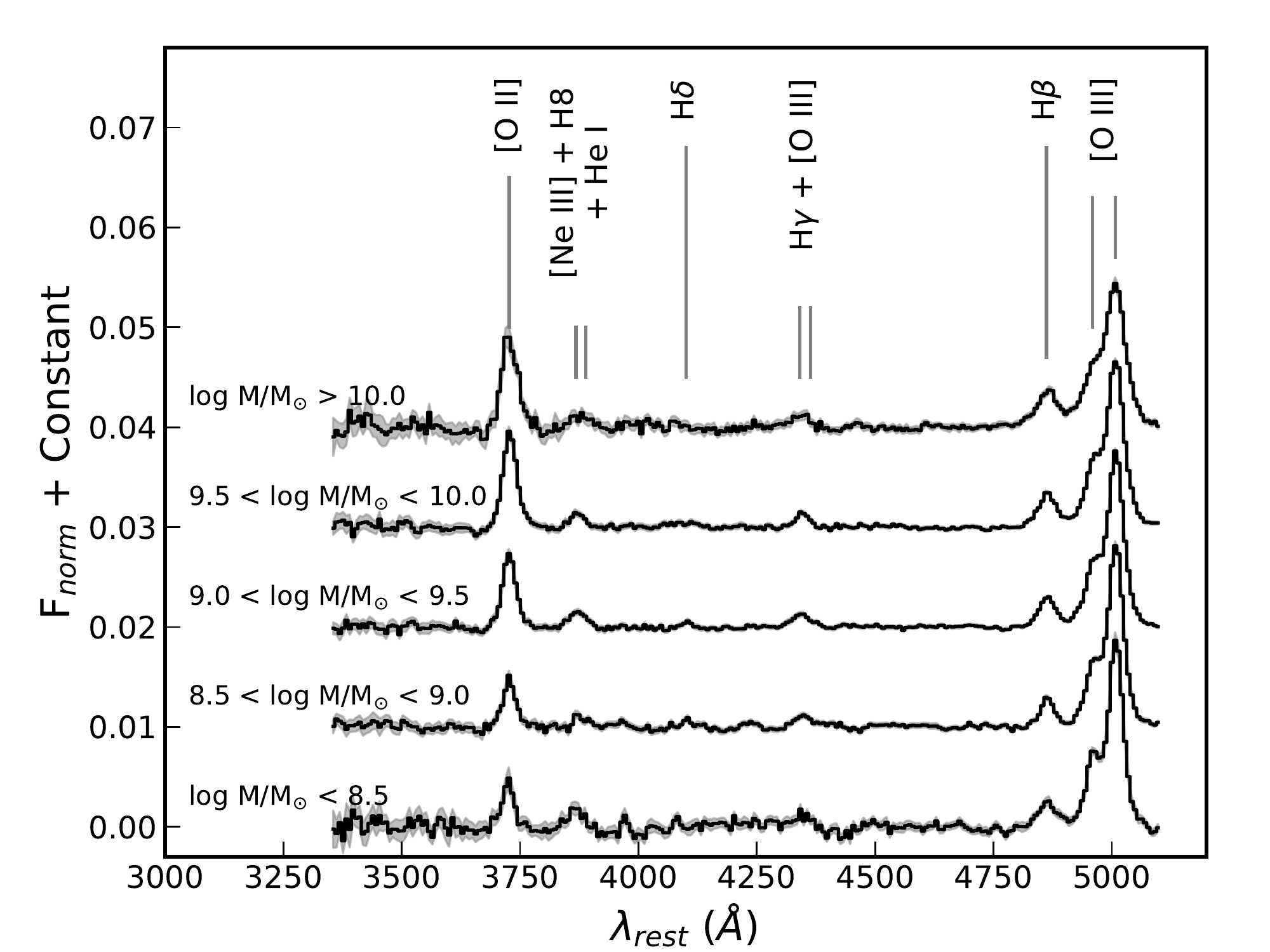} 
\caption{Stacked spectra are shown for five bins of stellar mass.  All galaxies between $1.3 < z < 2.3$ with \oiii\ SNR $>5 \sigma$ are shown.  These composite spectra are truncated just beyond \oiii, as only the lower redshift portion of the sample contributes to the longer wavelengths.  The grey shaded areas show the 1$\sigma$ uncertainty (the error on the mean), which is sometimes very small due to the large number of galaxies in the stack. } 
\label{stack_massbins} 
\end{figure*} 

\begin{deluxetable}{lc}
\tablecolumns{2} 
\tablecaption{Number of galaxies in subsample stacks} 
\tablehead{
\colhead{Sample} & \colhead{N} 
}
\startdata 
All galaxies &  68, 223, 379, 307, 79 \\ 
$1.3 < z < 1.7$   &  24, 58, 103, 69, 24  \\
$1.7 < z  < 2.3$ & 44, 165, 276, 238, 55 \\
High SFR\tablenotemark{a} &   34,  111, 189, 153, 39   \\     
Low SFR\tablenotemark{a} &   34, 112,  189,  153,  40  \\ 
High \oiii\ equivalent width\tablenotemark{b} &  34, 111,   189,  153, 39  \\
Low \oiii\ equivalent width\tablenotemark{b}  &   34,  112,  189,  153,  40   \\ 
WISP                              &  29, 77, 114, 75,  30   \\
CANDELS/3D-HST        &  39, 146,  265, 232,  49   \\ 
WISP, $1.3 < z < 1.7$  &  13, 30, 49,  29, 17   \\ 
CANDELS/3D-HST, $1.3 < z < 1.7$  &  11,  28,  54, 40,  7  \\ 
WISP, $1.7 < z < 2.3$ & 16,  47,  65, 46,   13   \\
CANDELS/3D-HST, $1.7 < z < 2.3$ & 28, 118, 211, 192,  42  \\ 
\enddata 
\label{ngals_stack} 
\tablecomments{Subsamples which we used to create stacks are given, alongside the number of galaxies  in five mass bins for each.  The mass bins are the same 
throughout: log M/M$_{\sun} < 8.5$,  $8.5 < $ log M/M$_{\sun} < 9.0$,  $9.0 < $ log M/M$_{\sun} < 9.5$, $9.5 < $ log M/M$_{\sun} < 10.0$,    and log M/M$_{\sun} > 10.0$.   } 
\tablenotetext{a}{The high and low SFR subsamples are determined by dividing at the median SED-derived SFR in each mass bin, as given in Table \ref{results_table}.}
\tablenotetext{b}{The high and low \oiii\ equivalent width subsamples are determined by dividing at the median \oiii\ equivalent width in each mass bin, as given in Table \ref{meas_table}.} 
\end{deluxetable}

 \begin{itemize} 
 \item We tested for evolution, dividing into two bins at $1.3 < z < 1.7$ and $1.7 < z < 2.3$ (discussed in \S \ref{evo_sec}). 
 
  \item We tested whether galaxies with higher SED-derived SFRs had lower metallicities than galaxies with lower SFRs. We divided the sample into sources above and below the median SFR in each mass bin (\S \ref{fmr_stack_sec}).    
   
 \item  We created stacks for galaxies in each mass bin with \oiii\ equivalent widths both above and below the median in that mass bin (\S \ref{fmr_stack_sec}). 
 
 \item  We also explored whether the different sample selections for WISP and CANDELS/3D-HST would result in different metallicities, creating separate sets of stacks for each survey.    Since the WISP and CANDELS/3D-HST samples have different mean redshifts, we also created a set of stacks  (in the same five mass bins) for each survey, at redshifts above and below 1.7 (\S \ref{bias_sample_sec}). 
 
 \end{itemize} 
 
The number of galaxies in each of the five mass bins for these subsamples are given in Table \ref{ngals_stack}.

  \begin{figure*} 
\plotone{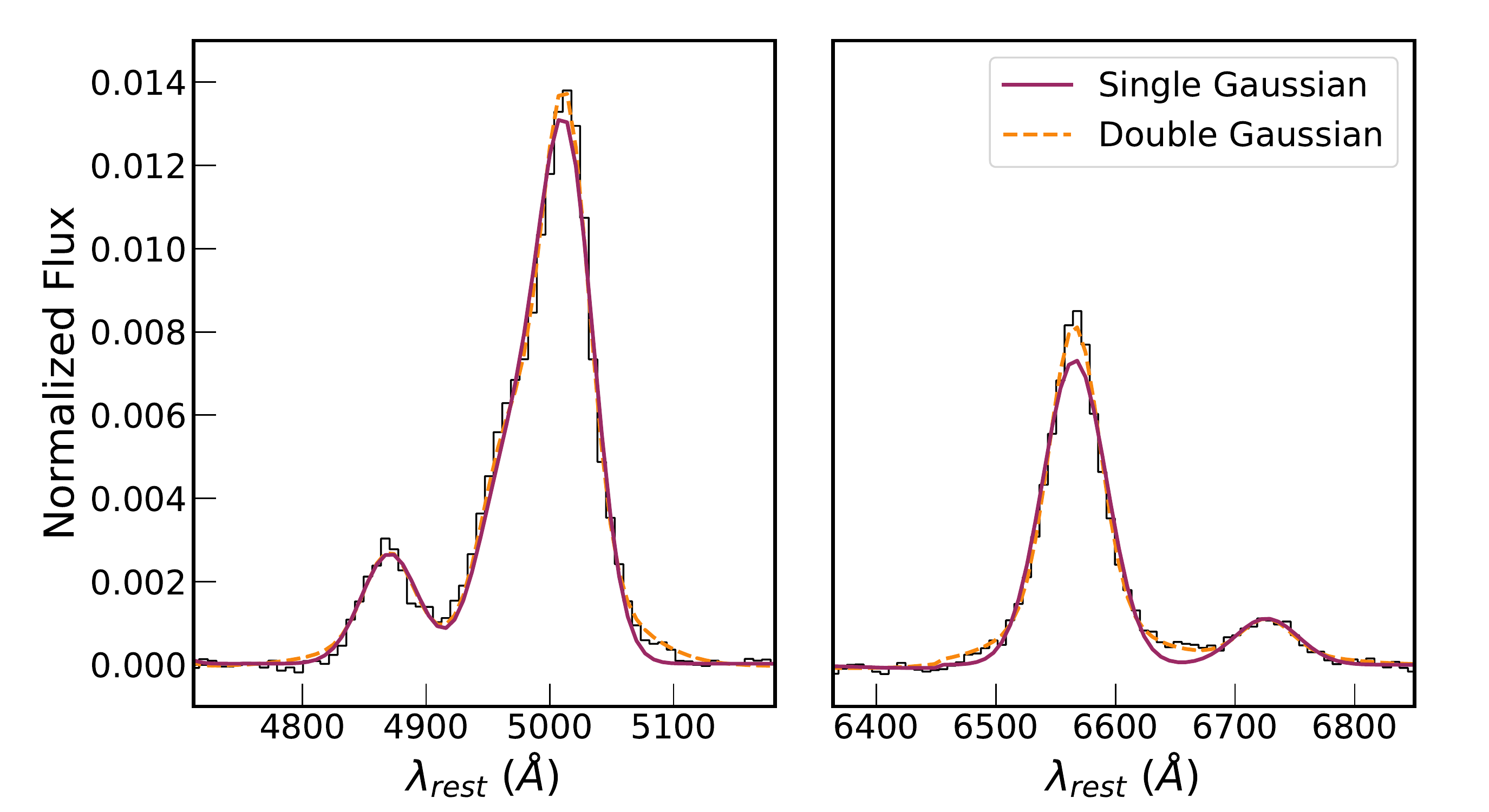} 
\caption{The stacked spectra require two Gaussians for each individual emission line, in order to fit the line profiles and provide reliable flux ratios.  Here, we show the \oiii\  and \hb\ lines (left) and \ha\ +\nii\ + \sii\ lines (right) for all the galaxies in our sample with $z< 1.5$.  The single Gaussian fails to reproduce the peaks and wings of the lines, while two Gaussians show a better fit.    }
\label{lineprof_2comp} 
\end{figure*}

In all stacks, the emission line fluxes were measured by fitting a set of Gaussian profiles to the lines in the stacked spectra.  We simultaneously fit \oii, \oiii $\lambda \lambda$ 4959, 5007, \ha\ (blended with \nii), \hb, \hg\ (blended with \oiii\ $\lambda$4363), \hd, \sii,  \ion{He}{1} 5876, and a blend of lines around \ion{Ne}{3} $\lambda$3869.  Furthermore,  we fixed the doublet ratios of \oiii\ $\lambda$5007/\oiii\ $\lambda$4959 to 2.9:1, but did not include separate lines for the closely spaced blends of \oii\ $\lambda \lambda$3726, 29 , \ha\ + \nii\ $\lambda \lambda 6548, 6583$,  \sii\ $\lambda \lambda$6716, 6731, and \hg\ + \oiii $\lambda$4363.    We also considered whether \ion{He}{1} $\lambda$6678 might be contributing to excess flux that is sometimes visible between the \ha\ + \nii\ blend and \sii.   However, this line is intrinsically 3.6 times fainter than \ion{He}{1} $\lambda$5876 \citep{Porter12}, which is already very weak in our stacked spectra.  Therefore, we conclude that contributions from this line are negligible.  Additionally, since the spectral resolution of the stacked spectra is higher at blue wavelengths\footnote{In \S \ref{inspect_fit}, we noted that the emission lines in the individual spectra were fit with Gaussians, where the FWHM is the same, {\it in pixels}, for all the lines.   The G102 grism has a dispersion and spectral resolution two times higher than G141, so this constraint implies that the bluest lines are fit with FWHM which are two times smaller in \AA.   In the stacked spectra, the wavelengths shortward of \hb\ have varying contributions from G102 and G141, resulting in a spectral resolution that increases at blue wavelengths.},  we do not require the lines to have the same FWHM,  except for closely spaced pairs (\oiii\ and \hb, \ha\ and \sii).  We did, however, require the FWHM of the individual lines to be within a factor of two of the \oiii\  line width.  We also allowed a small shift of the emission line centroids, 
within $\pm 10$ \AA\ in the rest frame, in order to accommodate systematic uncertainties in the grism wavelength solution. 
Finally, since we subtracted the continuum in the individual spectra, we generally did not need to account for it when measuring the lines in the stacked spectra.   However, occasionally we see a  small residual continuum that might be present around \hg\ + \oiii\ $\lambda$4363.  Therefore, we modeled a flat residual continuum, spanning several hundred \AA, under these particular lines.

\begin{deluxetable*} {rccccccccccc}[!t]
\tablecolumns{10}
\tablecaption{Measurements from Stacked Spectra} 
\tablehead{
\colhead{log (M/M$_{\sun}$)}    & N & \colhead{\oiii/\hb}  & \colhead{\oii/\hb} &  \colhead{\oiii/\oii}  &  \colhead{\hg/\hb}   & \colhead{\hd/\hb}  &
\colhead{F(\oiii) }  &  \colhead{$W$(\hb) }   & \colhead{$W$(\oiii) }  \\ 
(1)  & (2)  & (3) &  (4) & (5)  &   (6)      &   (7) & (8) & (9)  & (10)
} 
\startdata
$< 8.5$    & 68    & $6.86 \pm 1.18$   &  $0.88 \pm 0.22$ &    $7.78 \pm 1.43$  &  $0.83 \pm 0.22$ &  $1.16 \pm 0.69$ & 7.4  & $170 \pm 54$  & $1210 \pm 320$    \\ 
8.5  - 9.0  &  223  & $8.95 \pm 0.97$   &   $1.39 \pm 0.21$ &    $6.43 \pm 0.65$  &  $0.61 \pm 0.16$ &  $0.23 \pm 0.10$ & 8.3 & $80 \pm 14$  & $740 \pm 100$  \\ 
9.0  - 9.5 &  379  & $5.98  \pm 0.41$  &  $1.50 \pm 0.12$  &    $3.97 \pm 0.19$  &  $0.48 \pm 0.10$  & $0.09 \pm 0.03$ & 8.8  & $57 \pm 7$  & $350 \pm 40$  \\   
9.5 - 10.0 & 307   & $5.20 \pm 0.26$   &  $1.74 \pm 0.11$  &    $2.99 \pm 0.13$  & $0.26 \pm 0.05$   & $0.16 \pm 0.09$ & 9.9   & $36 \pm 15$   &  $200 \pm 80$     \\
$>10.0$ & 79    & $3.60 \pm 0.39$   &  $1.31 \pm 0.17$  &    $2.75 \pm 0.22$  & $0.21 \pm 0.08$  &  $0.03 \pm 0.05$ & 14 & $37 \pm 5$   &  $140 \pm 10$
\enddata
\label{sample}
\tablecomments{{\bf (1)} The stellar mass range for each bin. {\bf (2)} The number of galaxies in each bin.  {\bf (3-7)}  Flux ratios measured from the stacked spectra shown in Figure \ref{stack_massbins}.  Ratios represent observed quantities only, and are not corrected for dust or stellar absorption.   Ratios involving \oiii\ and \oii\ include both lines of the doublets. The \hg\ measurement includes a contribution from \oiii\ $\lambda$4363.  {\bf  (8)} The median \oiii\ flux for the galaxies in the stack, in units of $10^{-17}$ erg s$^{-1}$ cm$^{-2}$;  both lines of the doublet are included.    {\bf (9) }  Inferred rest frame equivalent width of \hb\ emission in \AA, estimated as described in Appendix \ref{appendix_ew}.  No correction for stellar absorption is applied.   {\bf (10)} Median rest frame equivalent width of \oiii\ for galaxies in the stack, in units of \AA.  The uncertainty represents the error on the mean.   } 
\label{meas_table} 
\end{deluxetable*}

Figure \ref{lineprof_2comp} shows that, for any given line, a single Gaussian is a poor representation of the line profile in our stacked spectra.   The single Gaussian does not reach the peak of the line profile, and it is too narrow in the line wings.     This mis-match between the data and the single Gaussian model is characteristic of all of the stacks that we present in this paper.  In retrospect, it is not surprising that the individual slitless spectra-- whose line profiles are a convolution of the source morphology and the point source line spread function-  do not make a perfect Gaussian profile when they are stacked to reach high signal-to-noise. Therefore, we added a second broader Gaussian component to the model fitting, in order to get an accurate sum of the line fluxes.    The FWHM of the broad component, relative to the narrow component, is required to be the same for all of the lines, while the amplitudes of the broad components are allowed to vary (among positive values).    Visual inspection of all of the stacked spectra used in this paper show excellent fits when this second component is included.   Figure \ref{lineprof_2comp} shows a representative example of the improvements gained by adding a secondary component.   

\begin{figure*}[!t]
\begin{center} 
\includegraphics[scale=0.85]{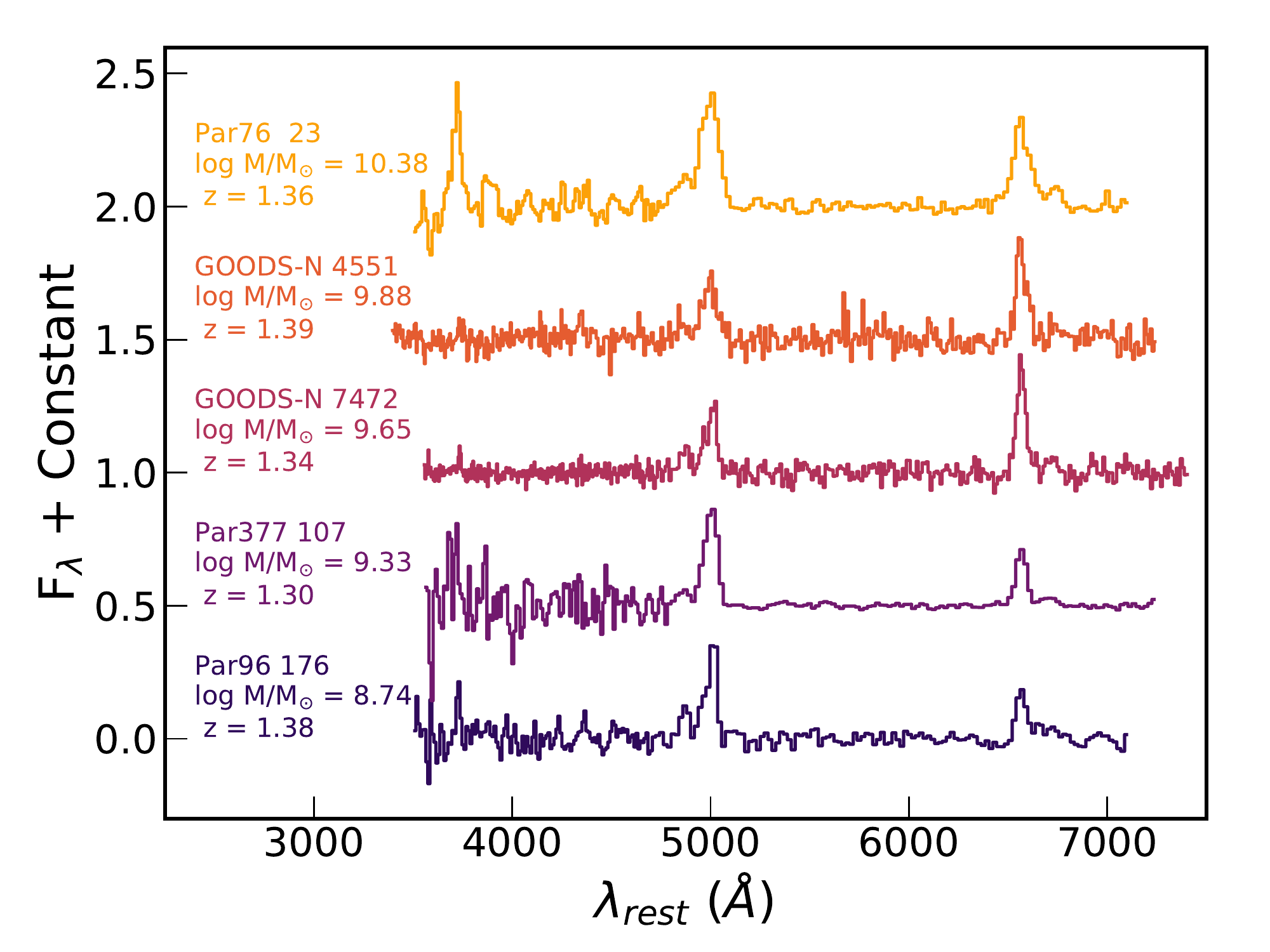} 
\caption{Continuum subtracted spectra of five objects with \ha\ SNR $>10$.     These objects are representative of the 49 individual spectra at $1.3 < z < 1.5$ that we consider in this paper.   Three objects-- Par96 176, Par 377 107, and Par76 23-- are from WISP, while GOODS-N 7472 and GOODS-N 4551 have G141 spectra from 3D-HST and G102 spectra from CLEAR.    The spectra are in units of 10$^{-17}$ erg s$^{-1}$ cm$^{-2}$ \AA$^{-1}$, and are continuum subtracted (with an arbitrary constant added to display them).  The spectrum of Par96 176 is multiplied by a factor of four for visualization purposes.     \label{specplot_ind} } 
\end{center} 
\end{figure*}

Measurement uncertainties on the line fluxes in the stacked spectra are obtained by bootstrapping with replacement. 
In brief, for each sample of $N$ galaxies that are stacked,  we draw $N$ random galaxies from that sample, allowing individual objects to be selected more than once.   Then we create a new stack from these objects, and measure the lines.  We repeat this procedure 1000 times, and calculate the standard deviation on the line fluxes that are measured from each stack.     
Measurements from the composite spectra in five mass bins are given in Table \ref{meas_table}.    

Finally, we note that equivalent widths  in Table \ref{meas_table} are estimated for the emission lines in the stacked spectra using broad-band photometry and the sample average line fluxes from the stacks.  Our method is described in more detail in Appendix \ref{appendix_ew}.

\begin{deluxetable*} {cccccccccc}[!t]
\tablecolumns{10}
\tabletypesize{\footnotesize}
\tablecaption{Measurements from Individual Spectra with \ha\ SNR $>10$} 
\tablehead{
\colhead{ID}    & \colhead{RA} & \colhead{Dec}  & \colhead{z} &  \colhead{\oii}  &  \colhead{\hb}   & \colhead{\oiii}  &  \colhead{\ha +\nii\ }    & \colhead{$W$(\oiii) }   & \colhead{$W$(\ha)}  \\ 
(1)  & (2)  & (3) &  (4) & (5)  &   (6)      &   (7) & (8) & (9)  & (10) 
} 
\startdata
Par76-23               & 201.83432          & 44.519665       &  1.3595      & 48.4 $\pm$ 4.8         &  22.1  $\pm$ 6.2     &  108.0 $\pm$  5.2      &   74.1 $\pm$  3.4     &   322   &  221 \\ 
GOODS-N 4551   & 189.14907792    &  62.16037039   &  1.3873    & 8.3 $\pm$ 2.3           &   4.1 $\pm$  2.0       &  19.6 $\pm$  1.7  &   16.5 $\pm$  1.1    &    196 & 216  \\
GOODS-S 7472  &  189.32295809   &    62.1790002     & 1.3360     & 8.0 $\pm$   1.6      &   5.7 $\pm$  1.1         &   18.0 $\pm$ 1.2    &   13.9 $\pm$  0.6     &    245    & 276  \\
Par377-107            &  255.382965    &  64.136452       & 1.3022     &  18.4 $\pm$  4.0       & 9.1 $\pm$   2.8        &  70.10 $\pm$   2.7  &  32.4 $\pm$   1.2    &   1630  & 755 \\
Par96-176           &  32.362873  &     -4.718067          &    1.3798    & 3.1 $\pm 0.8$           &  3.9 $\pm$ 0.9       &    14.6 $\pm 1.0$      & 6.3 $\pm$ 0.5          & 841  &  361 
\enddata
\tablecomments{Measurements for the 49 spectra with \ha\ SNR $>10$, as described in \S \ref{ind_spec}.  Columns are defined as follows: {\bf (1)} The field name and object ID.  {\bf (2-3)} RA and Dec, J2000, given in decimal degrees.  {\bf (4)}  Redshifts, as measured from the grism spectra.  {\bf  (5-8)} Line fluxes, in units of $10^{-17}$ erg s$^{-1}$ cm$^{-2}$;  both lines of the \oiii, \oii\, and \nii\ doublets are included.  No corrections for dust extinction or stellar absorption are applied. {\bf (9) }  Rest frame equivalent width of the \ha\ and \oiii\ emission in \AA, calculated as described in Appendix \ref{appendix_ew}.  No correction for stellar absorption is applied.  Uncertainties on the emission line equivalent widths of individual objects are around 40\%.  The full table is available online. } 
\label{meas_table_ind} 
\end{deluxetable*}

\subsection{Individual Spectra}
\label{ind_spec} 
While the majority of the sample have low SNRs,  a modest-sized subset have sufficient quality for analysis without stacking.  
In particular, we find 49 objects with \ha\ SNR $> 10$ and $1.3 < z < 1.5$, where we have full spectral coverage from \oii\ to \ha.    The high SNR of these objects ensures meaningful constraints on \ha/\hb\ ratios, dust corrected \ha\ luminosities, and SFRs, while minimizing Eddington bias from low SNR sources scattering into the sample \citep{Eddington13}.  While the median SNR on the \hb\ emission line flux is low ($\sim 2.5$), our analysis method marginalizes over this uncertainty, and the \ha\ SFRs that we obtain are still more precise than the SED-derived SFRs.   Five representative examples of these objects are shown in Figure \ref{specplot_ind}.    

The emission line fluxes for these galaxies are taken from the fitting procedure that we described in \S \ref{inspect_fit}, and equivalent widths are measured from broad-band photometry (see Appendix \ref{appendix_ew}).
These measurements are given in Table \ref{meas_table_ind} for a subset of our sample, and provided for all of the 49 high SNR objects in a  machine readable format online. 
To derive nebular gas properties, we use the Bayesian method described in Appendix \ref{bayes_met}.  This technique simultaneously constrains dust, metallicity, and contamination of \ha\ emission by \nii, while marginalizing over uncertainties due to Balmer line stellar absorption.  In particular, metallicity is measured using the \cite{Curti17} calibration, and dust extinction is inferred from the \ha/\hb\ ratio, using a \cite{Calzetti00} extinction curve.  Then we use the \cite{Kennicutt98} calibration to obtain SFRs from \ha\ luminosities, dividing by 1.8 to convert to a \cite{Chabrier2003} IMF.

\begin{figure}
\begin{center} 
\includegraphics[scale=0.43]{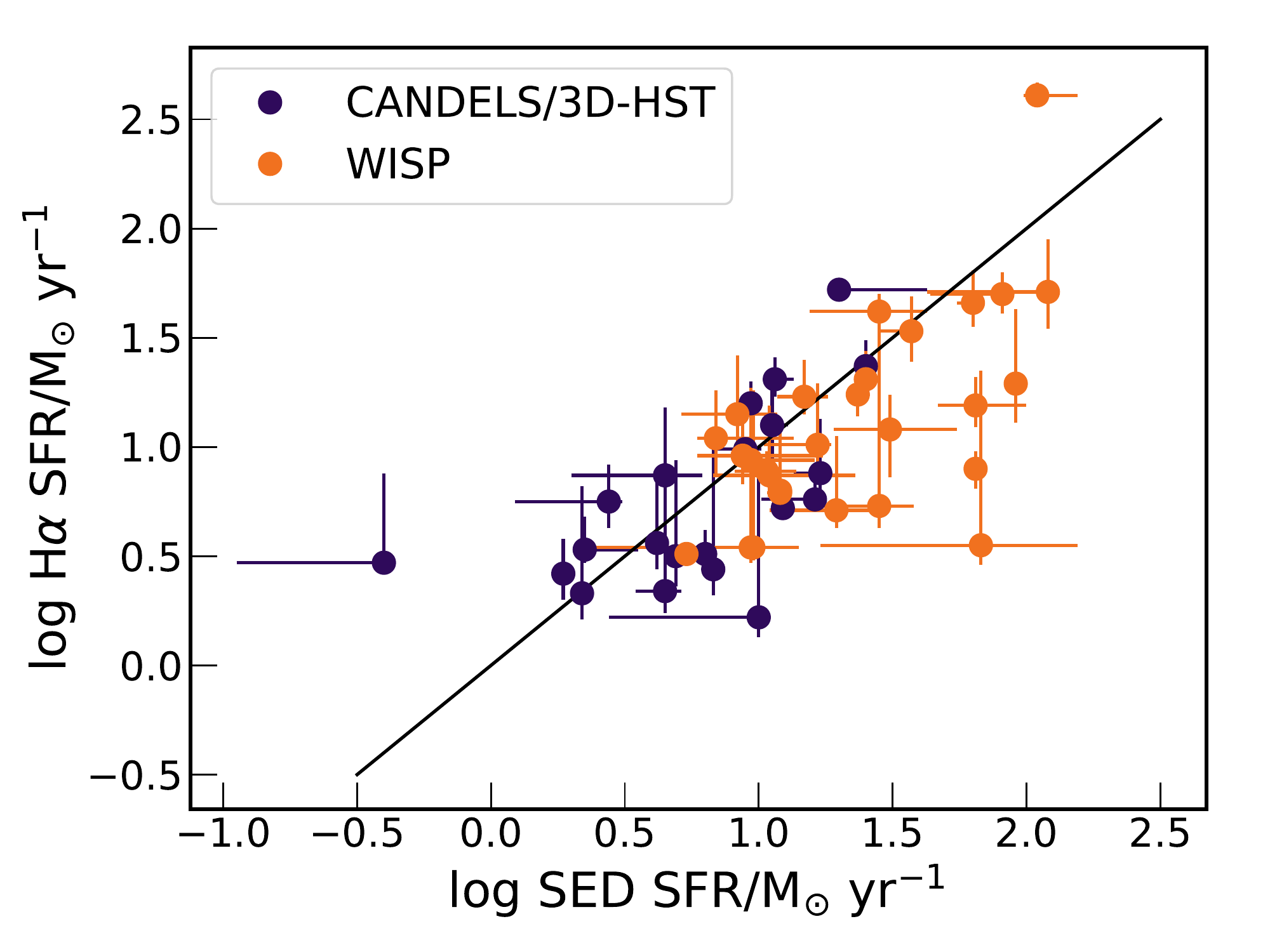} 
\caption{The SED-derived SFRs are compared to \ha\ derived SFRs for the 49 objects at $1.3 < z < 1.5$ with \ha\  S/N $>$ 10.  Both SFRs are corrected for dust extinction.   The black line shows the 1:1 relation. Compared to the CANDELS/3D-HST sample, the WISP measurements are shifted up and to the right:  their SED-derived SFRs are systematically 0.25 dex higher than their \ha\ SFRs, and their \ha\ SFRs are, on average, 0.35 dex higher than those of the CANDELS/3D-HST objects.   Note that the error bars are asymmetric in this plot, and in some cases the most likely solution is near the 1$\sigma$ upper or lower bound on the SFR. \label{hasfrs}  } 
\end{center} 
\end{figure}

Since most of our sample has only SED-derived SFRs, we can use the \ha\ derived SFRs to assess the accuracy of the SED-based measurements.   In Figure \ref{hasfrs}, this comparison shows good agreement between the two methods for the CANDELS/3D-HST sample.  There is no systematic offset between the two methods; the SED-derived SFRs are larger than \ha\ SFRs, on average, by only 0.02 dex.    The scatter between the two methods is 0.36 dex, which is somewhat larger than the uncertainties on the SED-derived SFRs (half of the 68\% confidence interval on the SED-derived SFRs for CANDELS/3D-HST is 0.2 dex).   
However, additional scatter can easily be explained by different time scales sampled by \ha\ emission and continuum emission from young stars (e.g. \citealt{Lee2011, Guo16b,  Mehta17, Emami19}).    For the WISP sources, on the other hand, the SED-derived SFRs are systematically higher than the \ha\ SFRs by 0.25 dex;  this difference is not surprising, as the WISP fields have only five bands of imaging, compared to the extensively-sampled SEDs in the CANDELS/3D-HST fields.  For the 49 objects under consideration here,  the SED fits tend towards higher extinction in WISP compared to 3D-HST (mean $A_V = 0.40$ vs. $A_V = 0.26$).   This difference is likely systematic error in the SED fitting rather a physical difference in the two samples; propagated to ultraviolet wavelengths, it explains the higher SFRs in the WISP objects.   We take the systematic uncertainties on SED-derived SFRs into account when we consider the M-Z-SFR relation from stacked spectra in \S \ref{fmr_stack_sec}.

\subsection{Sample Characteristics}  
\label{sample_char_sec} 
\begin{figure*}
\begin{center} 
\includegraphics[scale=0.6]{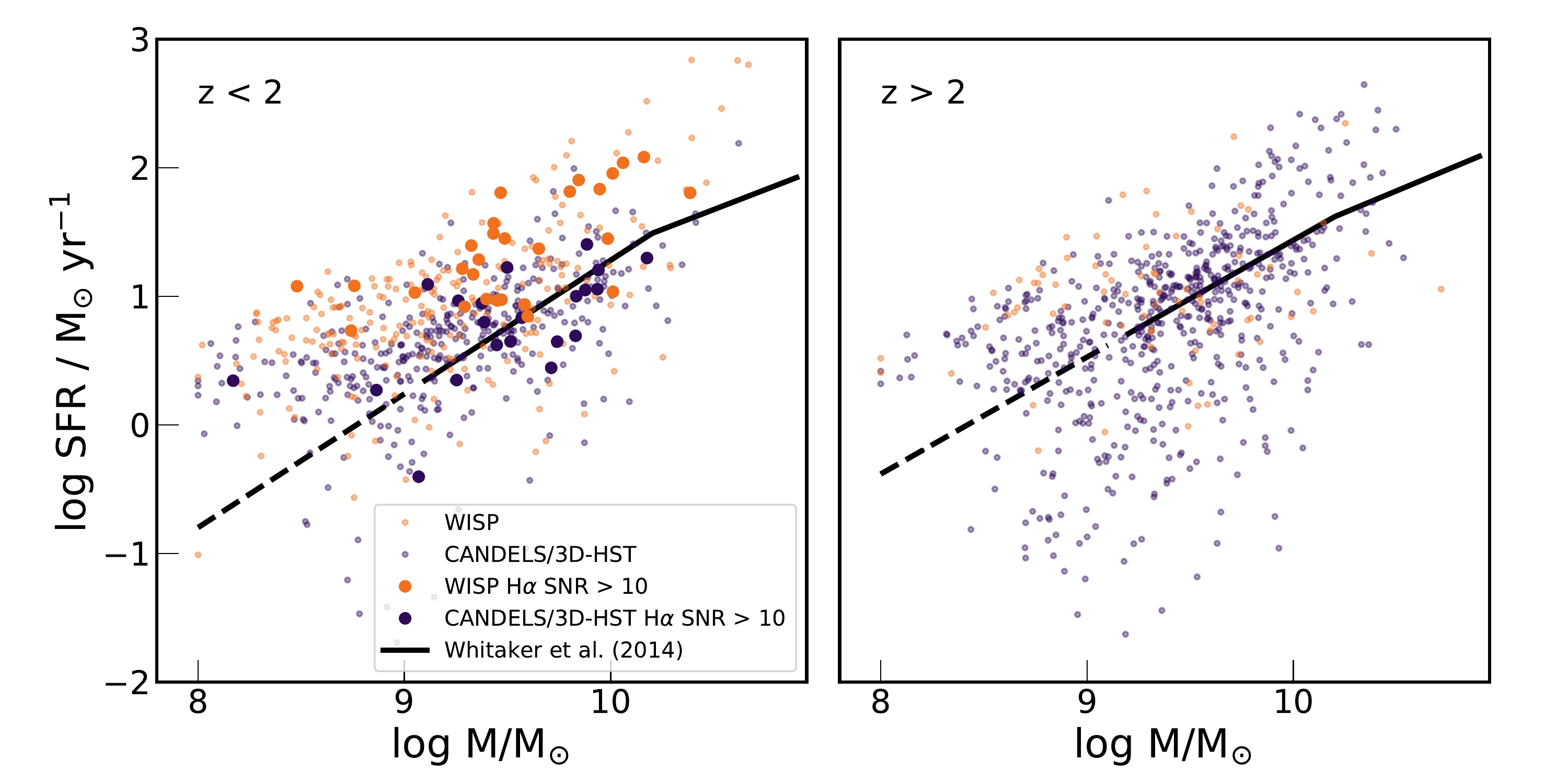} 
\caption{ The relationship between SFR and $M_{*}$ is shown, for star-forming objects in our sample at $1.3 < z< 2$ (left), and $2 < z < 2.3 $ (right).  Both stellar masses and SFRs are derived from SED fits.  The dark purple points indicate objects from the CANDELS/3D-HST sample, while the orange points show objects from the WISP survey.   The larger points in the left panel highlight the 49 objects with \ha\ SNR $> 10$, still showing the SED-derived SFRs.  The star-forming main sequence from \cite{Whitaker14} is shown by the black curves, for $1.5 < z < 2.0$ (left) and $2.0 < z  < 2.5$ (right).   The dashed line indicates an extrapolation of this measurement.  \label{msfig}     }
\end{center} 
\end{figure*}

\begin{figure*} 
\begin{center} 
\includegraphics[scale=0.55]{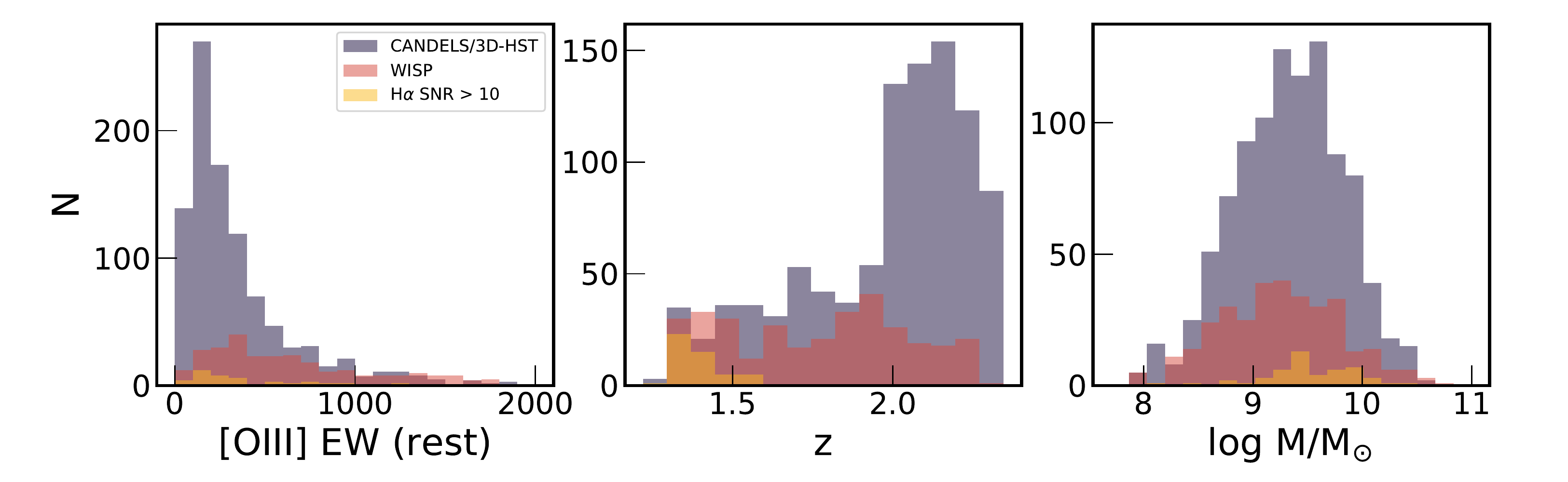} 
\end{center} 
\caption{The distribution of rest-frame \oiii\ equivalent width ($\lambda \lambda $4959,5007 summed; left),  redshifts (center), and masses (right) are shown. The WISP sample has somewhat different properties than the CANDELS/3D-HST objects, due to observational design and sample selection.  The gold histograms show the same properties, but for the 49 objects with \ha\ SNR $> 10$, where we obtain constraints on dust and metallicity without stacking.  These objects have similar properties to the parent sample, but are skewed towards slightly higher masses.   } 
\label{redshift_EW_dist} 
\end{figure*}

Figure \ref{msfig} shows the star-forming main sequence for our sample.  Here, the SFRs and masses are derived using SED fitting, as described in \S \ref{mass_sec}. We show $z<2$ and $z>2$ in the left and right panels, respectively, and show the WISP and CANDELS/3D-HST samples as gold and dark purple points, respectively.  We also highlight the 49 objects at $1.3 < z < 1.5$ with high SNR spectra discussed in \S \ref{ind_spec} with larger points (left panel only).  These objects do not show any clear difference from the parent sample in the stellar mass -- SFR space.    We compare our results to the main sequence from \cite{Whitaker14}, which we tentatively extrapolate below log M/M$_{\sun} \sim 9.2$.   This comparison shows that our full sample is somewhat biased towards higher SFRs, especially at the lowest masses.  This  bias also appears  strongly for WISP galaxies, especially at $z<1.5$; however, as we showed in \S \ref{ind_spec}, some of this effect may be due to a systematic overestimation of the SED-derived SFRs in the WISP data.  

Figures \ref{hasfrs} and \ref{msfig} compare objects from the WISP survey with objects from CANDELS/3D-HST.   This distinction shows that the \oiii\ emission line selection is different for these two surveys:  the objects from the WISP survey have higher SFRs than the 3D-HST galaxies, by an average of 0.35 dex (from the \ha\ SFRs in Figure \ref{hasfrs}).   This result is due to different selection techniques. For WISP, the objects are required to have clear redshift identification, on the basis of a second line.  For CANDELS/3D-HST, on the other hand, single lines are included when their photometric redshift indicates that the line is \oiii. As noted in \S \ref{inspect_fit}, this strategy is possible in CANDELS/3D-HST (but not WISP), because the extensive photometry in the CANDELS fields provides robust photometric redshifts.  Consequently, relative to CANDELS/3D-HST, the WISP survey is less complete to objects with overall weaker emission lines and lower SFRs.  

Figure \ref{redshift_EW_dist} shows the distributions of \oiii\ equivalent width, redshift, and stellar mass for the star-forming sample.    The \oiii\ equivalent widths, shown in the left panel, are given for the sum of the $\lambda$4959 and $\lambda$5007 lines.   Histograms are shown for the CANDELS/3D-HST objects, WISP objects, and the 49 objects with \ha\ SNR $> 10$.
As noted above, the inclusion of single-line emitters with robust photometric redshifts from CANDELS/3D-HST adds objects with lower equivalent widths, whereas the higher equivalent width tail is similar for the two surveys.   While the WISP sample appears more biased towards highly star-forming objects, the redshift distribution in the center panel of Figure \ref{redshift_EW_dist} highlights its importance.     Due to inclusion of the G102 grism in all of the WISP fields, the broader wavelength coverage nearly doubles the available sample at $1.3 < z < 2.0$ (when combined with the GOODS-N and CLEAR G102 data).   In comparison, the high SNR objects are similar to the parent sample in \oiii\ equivalent width and redshift, but are somewhat higher in stellar mass.

Lastly, the distribution of masses in the right panel of Figure \ref{redshift_EW_dist} gives a sense of the mass completeness of the samples.  Previously, \cite{Whitaker14} reported that the star-forming main sequence from CANDELS imaging was complete to log M/M$ _{\sun}$ $\sim$ 9.2-9.3 at these redshifts.  As evident by the turn-over in the mass distribution, our sample is roughly consistent with this completeness, for both WISP and CANDELS/3D-HST.   Hence, at the lowest masses in our sample, we are sensitive to the sources with only the highest SFRs.  This quality is also apparent in Figure \ref{msfig}, where the sample lies primarily above (albeit, an extrapolation of) the star-forming main sequence at $M \la 10^9$ M$_{\sun}$.    We take this incompleteness into account by modeling our sample selection in the IllustrisTNG simulation in \S \ref{theory_sec}.

\subsection{Considerations on Strong Line Metallicity Calibrations} 
\label{met_cal} 
The measurement of gas-phase metallicities from the spectra of galaxies has been a subject of debate.  Primarily, two types of calibrations have been used to infer metallicities from strong emission lines:  theoretical calibrations, based on photoionization models \citep{KD02, KK04, Strom18}, and empirical calibrations tied to direct-method metallicities derived from electron temperature ($T_e$) sensitive auroral lines \citep{PP04, Pil05, Pil12, Pil16, Curti17}.      Critically, large systematic errors are apparent between the different calibrations, even among the theoretical/empirical classes.   \cite{KE08} showed that the MZR of SDSS galaxies differs in shape and normalization when different calibrations are used, with a systematic offset as high as 0.7 dex.     While it is plausible that the photoionization models represent an oversimplification, some authors have also argued that metallicities based on the direct method are biased, as emission can be dominated by regions with higher $T_e$ (e.g. \citealt{Stasinska02, Bresolin07, Peimbert07, GarciaRojas07}).   
Other authors have argued that photoionization models can be brought in line with direct-method metallicities if the electron energies follow a $\kappa$-distribution, rather than a Maxwell-Boltzmann distribution \citep{Binette12, Nicholls12, Dopita13}.

Despite these challenges, recent studies have begun to converge on a range of reasonable calibrations in the local universe.   In \ion{H}{2} regions and nearby galaxies, comparisons between direct metallicities and supergiant metallicities show similar values \citep{Kudritzki16, Bresolin16, Davies17}.      
These results imply that empirical $T_e$-based metallicity calibrations (e.g. \citealt{PP04, Curti17}) should be preferable to photoionization models.

For high redshift galaxies, the possibility for evolution of the local metallicity calibrations is a cause for concern.  It is suspected that the physical conditions in \ion{H}{2} regions are different at high  redshifts, as high redshift galaxies are offset from the low redshift locus in the \nii/\ha\ vs.\ \oiii/\hb\ line diagnostic diagram (the BPT diagram; \citealt{BPT}).   This offset implies that, in high redshift galaxies, metallicities derived from \nii/\ha\ will differ from metallicities derived using oxygen-based indicators--- even if a self-consistent empirical calibration is used for the two diagnostics (e.g. \citealt{Maiolino08, Curti17}).   Several explanations for this evolution 
have been suggested, including:  contamination by AGN \citep{Wright09, Trump11, Trump13}, higher electron density \citep{Shirazi14}, higher ionization potential \citep{Kewley13, Kewley15}, harder ionizing spectra coupled with non-solar O/Fe ratios \citep{Steidel14, Steidel16, Strom17, Strom18}, and elevated N/O ratios \citep{Masters14, Masters16, Shapley15, Strom17}.   These effects could impact metallicity measurements, if low redshift strong-line calibrations are applied blindly to high-redshift galaxies. 

Taking these systematics into account, \cite{Strom18} derived a new strong-line calibration for high redshift galaxies.  They used photoionization modeling of around 200  galaxies at $z\sim2-3$ in the Keck Baryonic Structre Survey \citep{Steidel14}.  The key differences between their approach and earlier photoionization modeling (e.g.\ \citealt{KD02}), were the inclusion of harder ionizing spectra that include binary stars (e.g. BPASS; \citealt{Eldridge16, BPASS}), a decoupling of the nebular metallicity and ionizing stellar metallicity (to emulate variations in O/Fe ratios), and allowing the N/O ratio to vary independently of O/H.  \cite{Strom18} derived the metallicity for each galaxy in their sample, and then provided relations between their derived metallicity and commonly observed line ratios.  While their \nii/\ha\ calibration shows some evidence for evolution, their $R_{23}$ calibration\footnote{$R_{23}$ = (\oiii\ $\lambda \lambda$4959, 5007+ \oii\ $\lambda \lambda$3726, 3729)/\hb}  is similar to the one derived from local \ion{H}{2} regions reported by \cite{Pil12},  after the latter is corrected upwards by 0.24 dex.    

 \begin{figure*}[!t]
\plotone{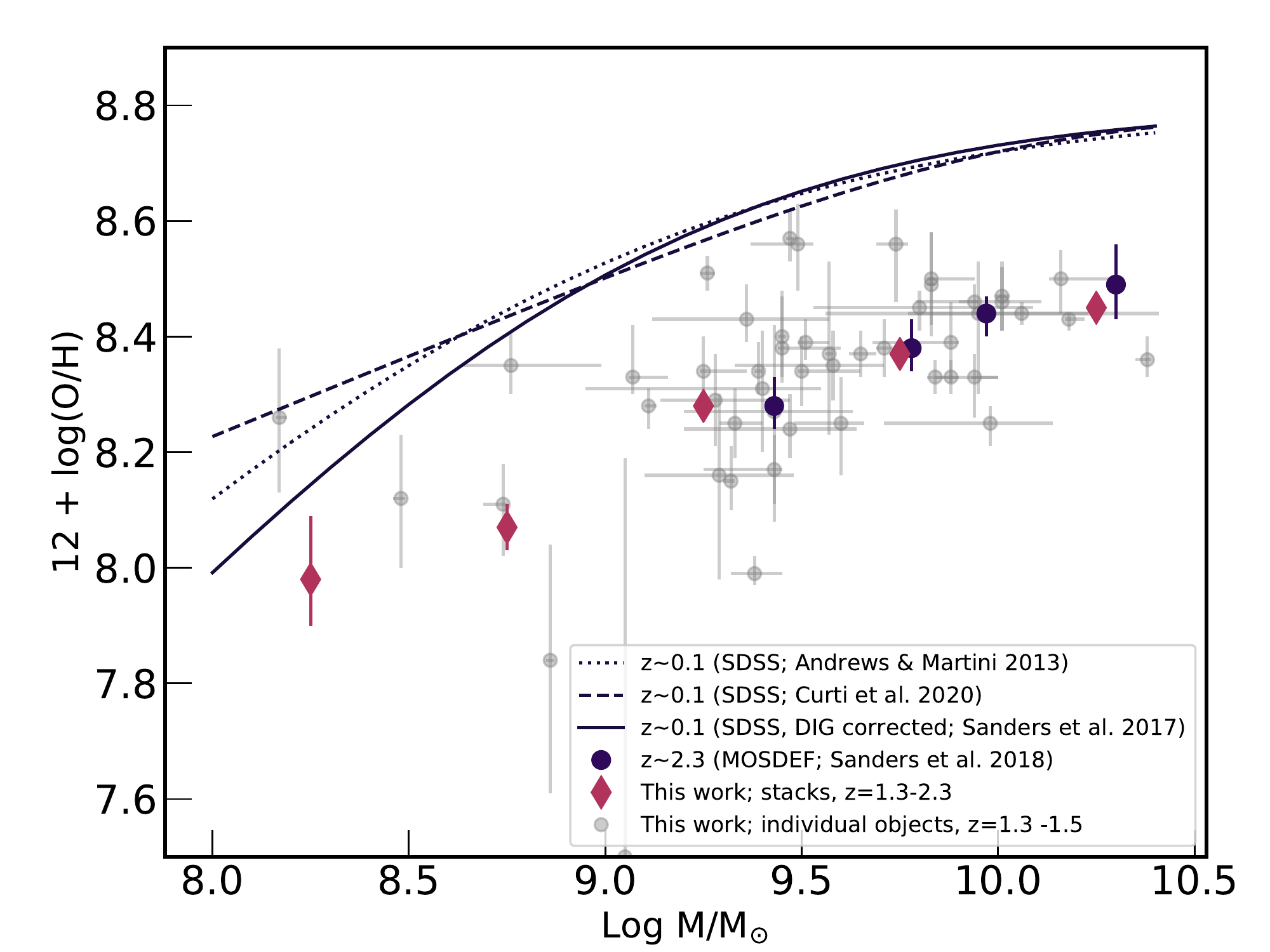} 
\caption{The MZR is shown for our stacked spectra, as well as the objects at $1.3 < z < 1.5$ with \ha\ SNR $>$ 10.   The mean redshift of the sample in the stacks is $z = 1.9$.  Our results show excellent agreement with the MOSDEF survey, as reported by \cite{Sanders18}; we have recalculated the metallicities from their stacked measurements, using the \cite{Curti17} calibration.  Other samples at $z\sim2$ are not shown, as they require the use of \nii-based metallicity calibrations, which may evolve with redshift  \citep{Masters14, Masters16, Shapley15, Strom17}.} 
\label{mzr_highz} 
\end{figure*}

 Overall, the agreement between the latest photoionization models and nearby  \ion{H}{2} regions suggests a convergence of oxygen abundance indicators,  with systematic uncertainties greatly reduced from the 0.7 dex reported by \cite{KE08}.   Further supporting this conclusion, recent detections of \oiii\ $\lambda$4363 in high redshift galaxies confirm that low redshift empirical strong line calibrations agree with direct-method based measurements, within the uncertainties \citep{Jones15a, Gburek19, Sanders19}.   These findings suggests that empirically derived strong-line calibrations, tied to direct-method metallicities are applicable at high redshfifts.  Given this assessment, we adopt the empirical calibration from \cite{Curti17}, as it is well suited to the Bayesian methodology that we describe in the next section.  The $R_{23}$ calibration from  \cite{Curti17} gives metallicities that are around 0.2 dex lower than those from \cite{Strom18}, more in line with the \ion{H}{2} regions from \cite{Pil12}.  Some of this offset may be attributable to different handling of dust depletion in photoionization models compared to direct-method measurements.

\subsection{Bayesian Inference of Metallicity and Dust Extinction} 
We use a Bayesian methodology to derive nebular gas properties from our measurements.  A detailed description of our calculation is presented in Appendix \ref{bayes_met}.
 In brief, we model metallicity, dust extinction, Balmer line stellar absorption, and contamination of \hg\ by emission from \oiii\ $\lambda$4363.    
 As noted above, metallicities are derived using the \cite{Curti17} calibration.  We do not apply a correction for diffuse ionized gas (DIG), as the contribution is expected to be minimal for highly star-forming, compact galaxies the redshifts of our sample \citep{Sanders17}.   The dust extinction is calculated from the Balmer decrement, assuming a \cite{Calzetti00} extinction curve.  For sources at $1.3 < z < 1.5$, we use \ha/\hb, as well as measurements or upper limits on H$\gamma$/\hb\ and H$\delta$/\hb.  For the higher redshift stacks, we do not have coverage of \ha, so the dust constraints from the Balmer lines alone are poor.  In these cases, we adopt a prior based on the lower redshift measurements of dust extinction in stacked spectra for $1.3 < z < 1.5$ (see Appendix \ref{bayes_met}).  We note that \hb\ stellar absorption and the relative strength of  \oiii\ $\lambda$4363 are poorly constrained by this method.  Therefore, we marginalize over these nuisance parameters to provide realistic uncertainties on metallicity and dust extinction.    We do not consider these poorly constrained quantities further. 

We applied this methodology to the stacked spectra discussed in \S \ref{stack_sec} and shown in Figure \ref{stack_massbins}, as well as the 49 individual high SNR spectra described in \S \ref{ind_spec}.   Results for stacks of the full sample, divided into five mass bins, are given in Table \ref{results_table}.   Likewise, results for the individual high SNR spectra are highlighted in Table \ref{results_table_ind} and presented  in machine readable format online.

\section{Results} 
In this section, we present the results of our MZR and M-Z-SFR measurements, for both stacks and individual galaxies.    In \S \ref{mzr_z2}, we compare to previous results at similar redshifts, while in \S \ref{evo_sec} we discuss the evolution of the MZR.    Then, we present the M-Z-SFR relation from stacked spectra in \S \ref{fmr_stack_sec}, and the 49 high SNR individual objects in \S \ref{fmr_ind_sec}.   Finally, we address biases in our sample selection in \S \ref{bias_sample_sec}.  

\label{results} 
\subsection{The MZR at $z \sim 1-2$} 
\label{mzr_z2}

\begin{deluxetable*} {rccccccc}[!t]
\tablecolumns{8}
\tablecaption{Derived Quantities from Stacked Spectra} 
\tablehead{
\colhead{ log(M/M$_{\sun}$)}   & \colhead{N}  & \colhead{$\langle$log(M/M$_{\sun})\rangle$}    &\colhead{$\langle$log(SFR/M$_{\sun}$ yr$^{-1} ) \rangle$}   &   \colhead{E(B-V)}    & \colhead{$W$(\hb)$^{*}$} & \colhead{\oiii $\lambda4363$/\hg} & \colhead{12 + log(O/H)}  \\ 
(1) & (2)  & (3) & (4) & (5) &   (6)  & (7)  & (8)
} 
\startdata
$< 8.5$      & 68  &  8.27   &0.51   &   $0.0^{+0.21}_{-0.00} $   &  $0.0^{+3.75}_{-0.0} $   &    $0.46^{+0.00}_{-0.24}$ &  $7.98^{+0.11}_{-0.08}$    \\ 
8.5  - 9.0   &  223  & 8.77  &  0.61     &  $0.0^{+0.13}_{-0.00}  $   & $5.85_{-3.45}^{+0.0} $     &    $0.40^{+0.24}_{-0.22}$ &  $8.07_{-0.04}^{+0.04}$   \\ 
9.0  - 9.5    &  379  &   9.26 &0.80    &  $0.33_{-0.06}^{+0.06}$ & $3.0_{-1.2}^{+1.5} $     &    $0.28^{+0.18}_{-0.20}$  & $8.28^{+0.02}_{-0.02}$     \\   
9.5 - 10.0   &  307 & 9.73   &  1.11   &  $ 0.43_{-0.07}^{+0.06}$  & $5.85_{-2.1}^{+0.0} $ &      $0.00_{-0.00}^{+0.20}$  & $8.37^{+0.01}_{-0.02}$  \\
$>10.0$     &   79 & 10.20   & 1.45   & $0.66_{-0.06}^{+0.06} $  & $3.15_{-1.35}^{+1.80}$ &     $0.00_{-0.00}^{+0.24}$  & $8.45_{-0.02}^{+0.01}$
\enddata
\tablecomments{ Derived quantities for stacked spectra in five mass bins. {\bf (1)} The stellar mass bin for each stack. {\bf (2)}  The number of galaxies in each bin.   
{\bf (3)}  The mean stellar mass of the galaxies in each bin. 
{\bf (4)} The mean SED-derived SFR for the galaxies contributing to the stack.   {\bf (5-8)}   Properties from our Bayesian inference of nebular dust extinction, stellar \hb\ absorption equivalent width, the \oiii\ $\lambda$4363/\hg\ ratio, and metallicity.  Errors on each parameter denote the 68\% confidence interval, and are marginalized over the three parameters that are not under consideration.   A measurement uncertainty of zero (in one direction) indicates that the most likely solution was found at the edge of the physically allowed parameter space (see Appendix \ref{bayes_met} for a definition of the allowed parameter space).   } 
\label{results_table}
\end{deluxetable*}

\begin{deluxetable*} {cccccc}[!t]
\tablecolumns{6}
\tabletypesize{\footnotesize}
\tablecaption{Derived Quantities from individual spectra with \ha\ SNR $>10$} 
\tablehead{
\colhead{ID}    & \colhead{log M/M$_{\sun}$} & \colhead{log SFR/M$_{\sun}$ yr$^{-1}$ (SED)}  &  \colhead{log SFR/M$_{\sun}$ yr$^{-1}$ (\ha)} &  \colhead{E(B-V)$_{gas}$} &  \colhead{12 + log(O/H)}    \\ 
(1)  & (2)  & (3) &  (4) & (5)  &   (6)    
} 
\startdata
Par76-23 & $10.38_{-0.03}^{+0.01}$ & $1.80_{-0.06}^{+0.02}$ & $1.66_{-0.11}^{+0.15}$  & $0.17_{-0.12}^{+0.12}$ &  $8.36_{-0.03}^{+0.04}$  \\ 
GOODS-N 4551 & $9.88_{-0.02}^{+0.02}$ & $1.05_{-0.02}^{+0.06}$ & $1.10_{-0.17}^{+0.23}$  & $0.26_{-0.22}^{+0.19}$ &  $8.39_{-0.07}^{+0.07}$  \\ 
GOODS-N 7472&  $9.51_{-0.02}^{+0.06}$ & $0.65_{-0.35}^{+0.14}$ & $0.87_{-0.07}^{+0.08}$  & $0.14_{-0.07}^{+0.07}$ &  $8.39_{-0.03}^{+0.04}$  \\ 
Par377-107  & $9.33_{-0.04}^{+0.07}$ & $1.17_{-0.10}^{+0.09}$ & $0.23_{-0.08}^{+0.17}$  & $0.10_{-0.10}^{+0.16}$ &  $8.25_{-0.06}^{+0.06}$   \\ 
Par96-176    & $8.74_{-0.05}^{+0.05}$ & $0.73_{-0.02}^{+0.09}$ & $0.51_{-0.04}^{+0.05}$  & $0.00_{-0.00}^{+0.04}$ &  $8.11_{-0.09}^{+0.07}$  
\enddata
\tablecomments{Dervied quantities for the 49 spectra with \ha\ SNR $>10$.    {\bf (1)} Field Name and Object ID;   {\bf (2-3)} Stellar mass and SFR from SED fits, as described in \S \ref{mass_sec};   {\bf (4) }SFR derived from \ha, as described in \S \ref{ind_spec};  {\bf (5-6)} Nebular dust extinction and metallicity, derived simultaneously, as described in Appendix \ref{bayes_met}.  A measurement uncertainty of zero (in one direction) indicates that the most likely solution was found at the edge of the physically allowed parameter space.  The full version of this table is available online.     } 
\label{results_table_ind} 
\end{deluxetable*}

Figure \ref{mzr_highz} shows the MZR that we derive for our stacked spectra at  $1.3 < z < 2.3$ and individual galaxies at $1.3 < z < 1.5$.       The mean redshift of the full sample is $z=1.9$.   While we aim to compare our measurements with others in the literature at similar redshifts ($z \sim 1-2$), much of this work relies on \nii\ (e.g.\ \citealt{Erb06, Yabe15, Kashino17, Wuyts12, Gillman21}).    Given the uncertainties surrounding nitrogen abundances in high-redshift galaxies \citep{Masters14,Shapley15, Strom18}, we restrict our comparison to oxygen-based indicators.   This limits the comparison considerably, to data from \cite{Sanders18}, \cite{Henry2013_wisp}, and \cite{GG16}.     For these data,  we use the published emission line fluxes (or flux ratios) to recalculate metallicities using the \cite{Curti17} calibration, in order to maintain consistency with our measurements.    We use a simple maximum likelihood estimator to derive metallicities and their uncertainties from reported dust and stellar absorption corrected R$_{23}$ and $O_{32}$ ratios\footnote{Frequently, $O_{32}$ is defined as $\lambda 5007$/\oii\ $\lambda \lambda 3726, 29$, excluding \oiii\ $\lambda 4959$.  The choice of definition does not matter, as long as one is self-consistent.  In this paper, we use  $O_{32}  \equiv$ \oiii\ $\lambda \lambda 4959, 5007$/\oii\ $\lambda \lambda 3726, 39$ since we don't resolve the doublet.  The  $O_{32}$ metallicity calibration from \cite{Curti17} is adjusted accordingly. } 
and measurement errors.    Curiously, the \cite{Henry2013_wisp} and \cite{GG16} samples show metallicities which are higher than the present MZR.   We believe this to be a sample bias resulting from the requirement for the detection of multiple emission lines, which we discuss further in \S \ref{bias_sample_sec}.  Therefore, in Figure \ref{mzr_highz}, we focus on a comparison with the results from \cite{Sanders18}.

Figure \ref{mzr_highz} shows that our results are in excellent agreement with the stack-based measurements from \cite{Sanders18} at masses where the samples overlap,  even though they have slightly different mean redshifts ($z = 1.9$ for the present data and $z = 2.3$ for \citealt{Sanders18}).   Critically, we extend the measurement an order of magnitude lower in stellar mass.   Additionally, the 49 objects with high SNR spectra show good agreement with the stacked spectra, even though they are at a lower redshift ($1.3 < z< 1.5$).

\begin{figure*} 
\begin{center} 
\includegraphics[scale=0.6]{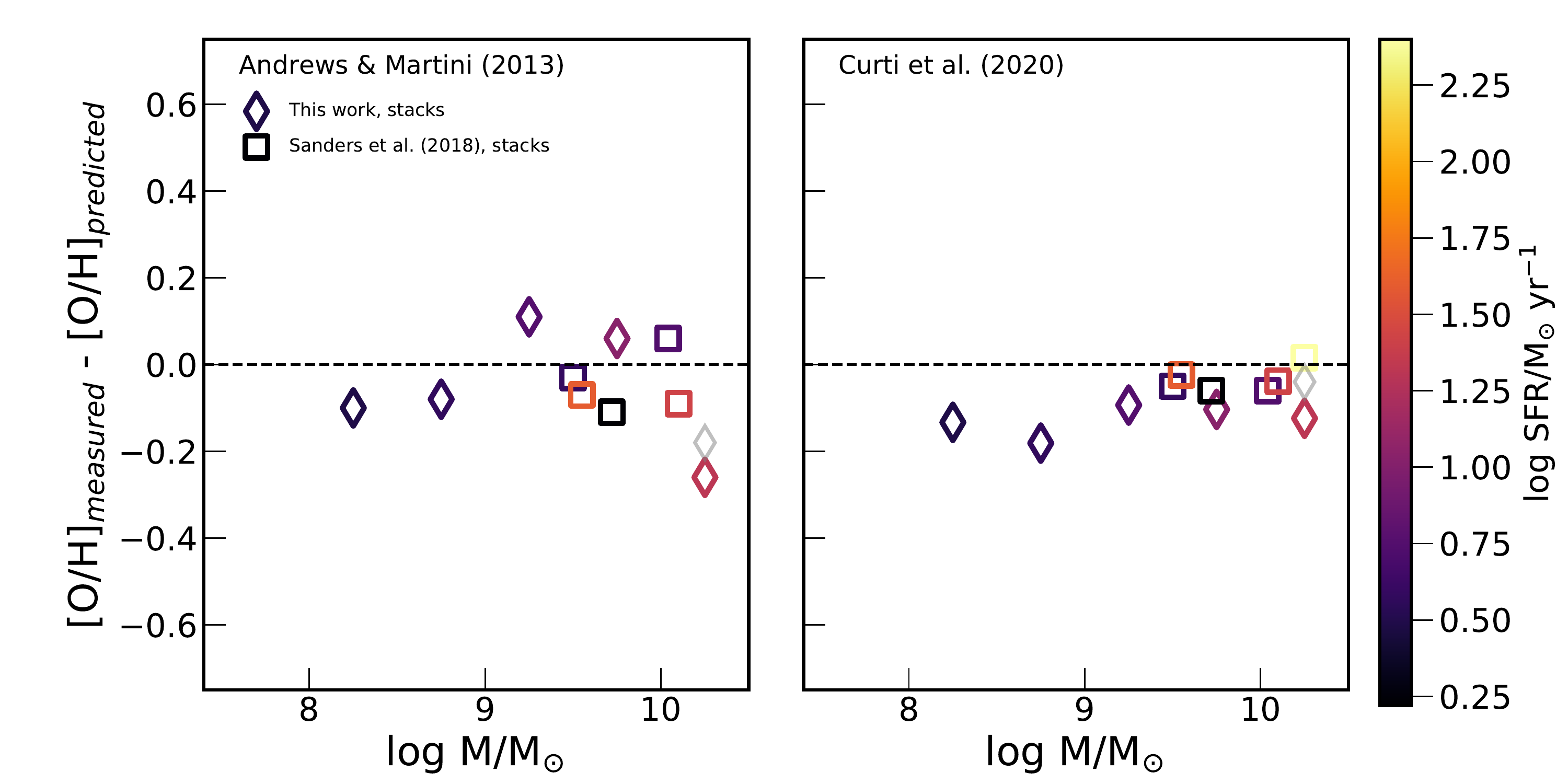}  
\end{center}
\caption{Residuals from the local M-Z-SFR of \cite{AM13} (left) and \cite{Curti20} (right).  Points show the results for stacked spectra in this paper  (diamonds)  and Sanders et al.\ (2018; squares).  The \cite{Sanders18} results are shown for the stacks divided into three specific SFR (SFR/$M_{*}$)  bins and two mass bins, with metallicities recalculated on the \cite{Curti20} calibration.  
Colors indicate SFR.  For the stacks in this paper, the mean SED-derived SFR in each bin is used, whereas for \cite{Sanders18} the SFRs are derived from \ha.  They grey diamond shows the 0.08 dex shift to higher metallicity that we derive when we exclude galaxies with ambiguous AGN contribution.  } 
\label{fmr_resid} 
\end{figure*}

\subsection{Redshift Evolution of the MZR}
\label{evo_sec}
Figure \ref{mzr_highz} also shows the evolution of our MZR from $z\sim 2$ to  $z\sim0.1$, by comparison with results from the SDSS.  We highlight two measurements,  both of which are empirically tied to direct-method $T_e$ metallicities:   \cite{AM13}, which measured the metallicity in stacked spectra where auroral lines are detected and \cite{Curti20}, which applies the \cite{Curti17} strong-line calibration to individual SDSS galaxies.    These two SDSS MZR measurements are very similar.  We also show an updated MZR reported by \cite{Sanders17}, which corrects the \cite{AM13} MZR for the DIG (and also to use more recent atomic data; see references in \citealt{Sanders17}).  This local relation is similar to the \cite{AM13} and \cite{Curti20} measurements over most of the mass range in Figure \ref{mzr_highz}, although it has a steeper slope at low masses.         Above log M/M$_{\sun} \sim 8.5$, we find a metallicity evolution of around 0.3 dex.   The shape of the $z\sim 2$ MZR appears similar to the low-redshift relation from \cite{AM13} and \cite{Curti20}, but may be flatter than the DIG corrected MZR from \cite{Sanders17}.  The larger metallicity error in the lowest mass bin make it difficult to ascertain whether the $z\sim2$ MZR takes a different shape than the local relation.

Given our large sample size {\it and} large redshift range, we also have the ability to measure metallicity evolution within our sample.   Therefore, we divided the sample in two redshift bins: $1.3 < z < 1.7$ and $1.7 < z < 2.3$.  Each redshift bin is divided into the same mass bins that we used for the whole sample, as indicated in Tables \ref{meas_table} and \ref{results_table}.      
We see no evidence for metallicity evolution in the redshift range that we probe.   The higher and lower redshift MZRs are consistent with one another, as well as the MZR for the full sample (Table \ref{results_table}), within the measurement uncertainties.  (We do not show these results in tabular form or a figure, as they are indistinguishable from Table \ref{results_table} and Figure \ref{mzr_highz}.)    We conclude that any evolution over the redshifts spanned by our sample must be smaller than (or comparable to) our measurement uncertainties.  This result is sensible if metallicity evolution is (to first order) linear with time.  We measure only 0.3 dex (a factor of two) of metallicity evolution over the 9 Gyrs between $z=1.9$ and $z=0.1$, while the time spanned between the mean redshifts of our high and low redshift bins ($z=1.5$ and $z=2.0$) is only 1 Gyr. Hence, we might expect a 0.05 dex increase in metallicity between our high and low redshift bins, which is indeed comparable to our uncertainties.

\subsection{The M-Z-SFR relation from stacked spectra} 
\label{fmr_stack_sec} 
At lower redshifts, the MZR shows a secondary dependence on SFRs (or gas fractions), such that, at fixed mass,  galaxies with higher SFRs (more gas-rich objects) have lower metallicities  \citep{Ellison08, Mannucci10,  Mannucci11, LaraLopez10, LaraLopez13, Bothwell13, AM13, Henry2013_clm, Cresci12, Salim14, Hirschauer18, Curti20}.   In this section, we aim to quantify this relation using stacked spectra with our full sample at $1.3 < z < 2.3$. 

We begin by asking whether our MZR relation is consistent with a non-evolving M-Z-SFR relation.   Figure \ref{fmr_resid} shows the metallicity residuals between our stack measurements and the local relations from \cite{AM13} and \cite{Curti20}.  Here, we adopt the median of the SED-derived SFRs for the galaxies in each mass bin\footnote{ Formally, the local M-Z-SFR relation is derived from \ha\ SFRs rather than  SED-based SFRs.  Nonetheless, evidence of the relation has previously been seen when SED-based SFRs are used \citep{Henry2013_clm}. As we showed in Figure \ref{hasfrs}, the SED-derived SFRs for the CANDELS/3D-HST subset of our sample (the majority) show no systematic offset from their \ha-derived SFRs, so the median SED-derived SFR in each mass bin should be adequate for comparing to the M-Z-SFR relation.}.  For comparison to \cite{AM13}, we identify the corresponding mass and SFR bin in their tabulated relation, whereas for \cite{Curti20}, we evaluated their relation for the masses and SFRs of our stacks. We choose the version of the \cite{Curti20} relation for total (aperture corrected) SFRs, in order to obtain a measure of the global properties of galaxies.  This choice also ensures consistency with \cite{AM13}.      We do not apply a correction for contribution from the DIG in local galaxies, as the galaxies in the portion of the local M-Z-SFR relation that match our high-redshift sample are expected to have a minimal DIG fraction and negligible shift in metallicities \citep{Sanders17}. 

In comparison to the local M-Z-SFR relation, Figure  \ref{fmr_resid} shows that our metallicities from stacked spectra (open diamonds) agree with \cite{AM13} within  $\pm$ 0.1 dex in the four lower mass bins, but are 0.26 dex lower than the local relation in the highest mass bin.   As we noted in \S \ref{agn_sec}, the contribution from AGN in this bin is uncertain.   Excluding galaxies where the AGN contribution is ambiguous increases our measured metallicity in this stack 0.08 dex.  This reduction in the residual from the local relation is shown by the grey diamonds in Figure \ref{fmr_resid}.   However, even in this case, this bin shows the largest residual compared to the \cite{AM13} M-Z-SFR relation.  Moreover, the increase in metallicity is at least partly due to a bias from including only the strongest \hb\ lines in the stacked spectrum.   Hence, we conclude that the difference between our sample and the local M-Z-SFR relation from \cite{AM13} at log $M/M_{\sun} > 10$  is not a result of AGN contamination.    Curiously, the large residual at high masses is not mirrored in the right panel of Figure \ref{fmr_resid}, where we compare our measurements to the local parameterization given by \cite{Curti20}.  In this case, our metallicities from stacked spectra are systematically lower than the local relation by an amount between 0.10 and 0.17 dex.

\cite{Sanders18} also compare stacked spectra at $z\sim 2.3$ to the local M-Z-SFR relation given by \cite{AM13}.  They report that their observations are offset to metallicities 0.1 dex lower than the local relation.   However, \cite{Sanders18} make this comparison using nitrogen-based metallicity indicators.  Therefore, for consistency with this work, we re-evaluate the metallicities for their M-Z-SFR stacks using the \cite{Curti17} calibration for $R_{23}$ and $O_{32}$.   These results are shown as open squares in Figure \ref{fmr_resid}.   The bin with the highest mass and SFR is excluded from the left panel, as it does not have a corresponding measurement in \cite{AM13}.   Similar to our stacked spectra, the \cite{Sanders18} metallicities fall within 0.1 dex of the local M-Z-SFR relation from \cite{AM13}, and do not show a systematic difference.   On the other hand, their data fall very close to the local M-Z-SFR relation from \cite{Curti20}, showing better agreement than our stacked spectra. 

In short, both our stacked spectra for $M/M_{\sun} < 10$,  as well as those from \cite{Sanders18} agree with the local M-Z-SFR relation from \cite{AM13} within $\pm 0.1$ dex, but when we use the \cite{Curti20} measurement of the local M-Z-SFR relation, our stacks show a systematic offset.  Hence,  it is impossible to determine whether the M-Z-SFR relation evolves, because we find different results when comparing to different measurements of the local relation (both from the SDSS).    Additionally, the metallicity calibrations used to compare high-redshift samples to the local M-Z-SFR seem to matter, as we do not find the same 0.1 dex metallicity offset reported by \cite{Sanders18}.

We next aim to determine whether we see evidence of an M-Z-SFR relation {\it within our sample}.  For this test, we divide each of our five mass bins into bins above and below the median SED-derived SFR in that bin. We follow the procedures described in \S \ref{stack_sec} and Appendix \ref{bayes_met}, creating stacks,  measuring the emission lines, and inferring metallicities.  Since the SED-derived SFRs are more precise for the CANDELS/3D-HST objects than the WISP objects, we tried this stacking exercise both with and without the WISP objects.      Regardless, we find no significant difference between the high-SFR and low-SFR stacks:  the high SFR MZR agrees with the low SFR MZR at 1-2$\sigma$.  We also tried stacking galaxies with high and low \oiii\ equivalent widths, dividing each mass bin into galaxies above and below the median \oiii\ equivalent width in that bin.   The result was the same:  the metallicities were consistent between the high and low equivalent widths.   

The lack of an M-Z-SFR relation in our stacking analysis is somewhat surprising, since a relation has been reported at $z\sim2$ by  \cite{Salim15} and \cite{Sanders18}.   One possibility is that our SED-derived SFRs may not be accurate enough to distinguish galaxies with high SFRs from those with low SFRs.   We used a simulation to test whether this assessment is correct.  In brief, we created a mock sample of galaxies with our stellar mass range, and SFRs defined by the \cite{Whitaker14} star-forming main sequence at $2.0 < z < 2.5$.  Then, we added 0.3 dex of {\it intrinsic} scatter to the SFRs, and calculated the metallicities using these SFRs with the \cite{Curti20} M-Z-SFR relation.  Then, to model our measurement errors, we added 0.2 dex of additional SFR scatter-- half of the typical 68\% confidence interval for SED-derived SFRs of the CANDELS/3D-HST objects.  We used these SFRs to divide the sample into galaxies which we select to be above and below the median SFRs.   In this way, some galaxies which should be above/below the median SFR are scattered into the opposite SFR bin.  We find that the mean metallicities of galaxies that we select to be in the high SFR bins are only around 0.05-0.06 dex lower than the galaxies in the low SFR bins.  This difference is only somewhat larger than the 1$\sigma$ uncertainty on the metallicities that we measure in our stacks (Table \ref{results_table}).   Hence, we conclude that more accurate measurements of SFR  are needed to quantify the M-Z-SFR relation from stacked spectra.

\begin{figure*} 
\begin{center} 
\includegraphics[scale=0.55]{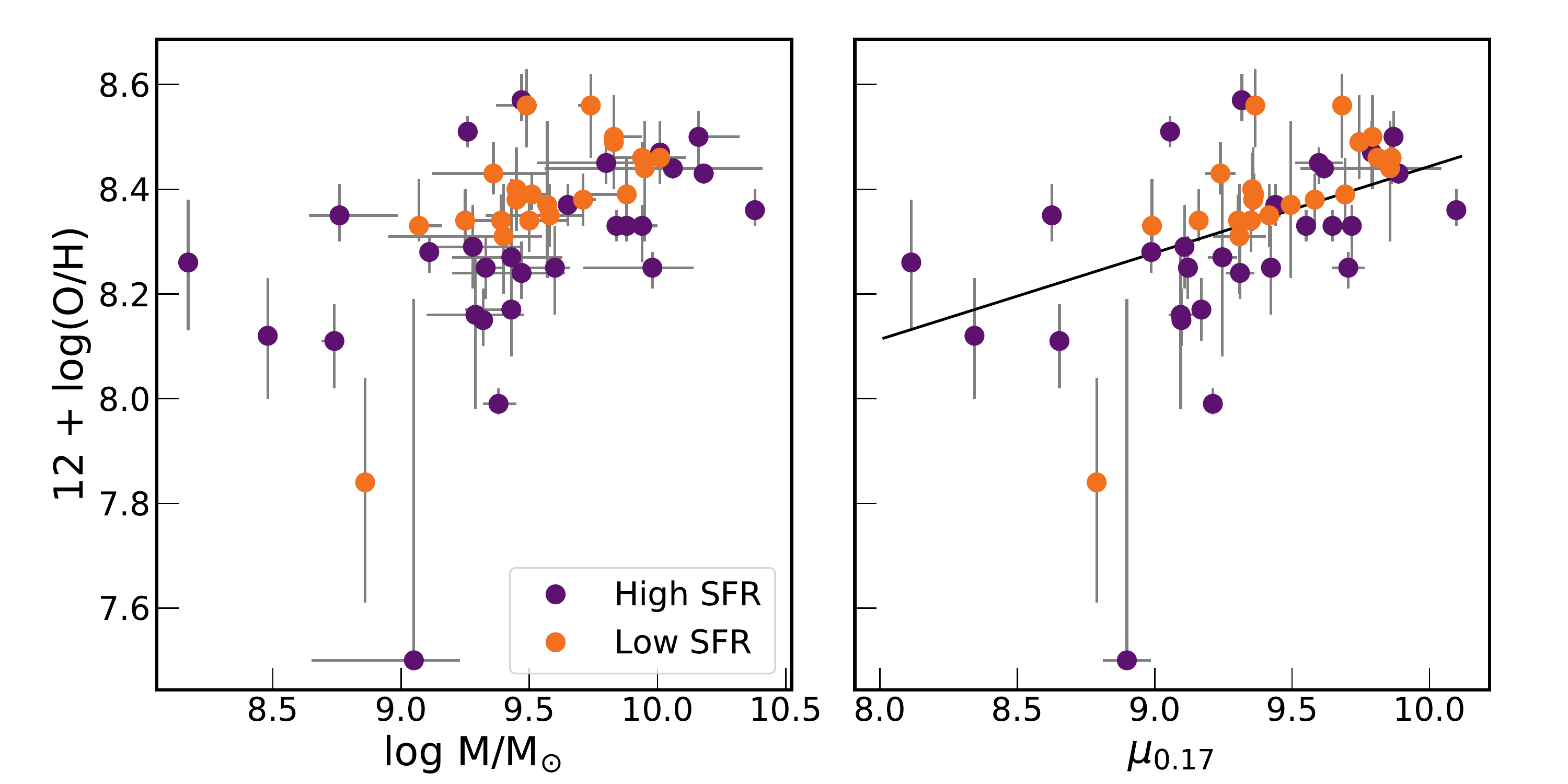} 
\end{center} 
\caption{{\it Left--} An M-Z-SFR relation is detected in our data for individual galaxies with SFRs derived from \ha. Points show 49 galaxies with \ha\ SNR $>10$. We divide the sample into high and low SFR bins, containing objects that are above/below ${\rm log(SFR/M_{\sun} yr^{-1})}$ = 0.68 ${\rm log(M/ M}_{\sun})$ - 5.56.  This line is determined from a fit to the masses and \ha\ SFRs in the subset of these galaxies at log(M/ M$_{\sun}) > 9.2$. {\it Right--}  The projection of least scatter for the same galaxies. Following \cite{Mannucci10}, we define $\mu_{\alpha} = {\rm log(M/ M}_{\sun}) - \alpha {\rm log(SFR/M_{\sun} yr^{-1})}$; $\alpha = 0.17$ minimizes the scatter.   The solid line shows a linear fit to metallicity as a function of $\mu_{0.17}$, and is given in Equation \ref{proj_eq}.       }
\label{fmr_ind} 
\end{figure*}

\subsection{The M-Z-SFR relation from individual objects at $1.3 < z < 1.5$}
\label{fmr_ind_sec} 
\begin{figure*}
\plotone{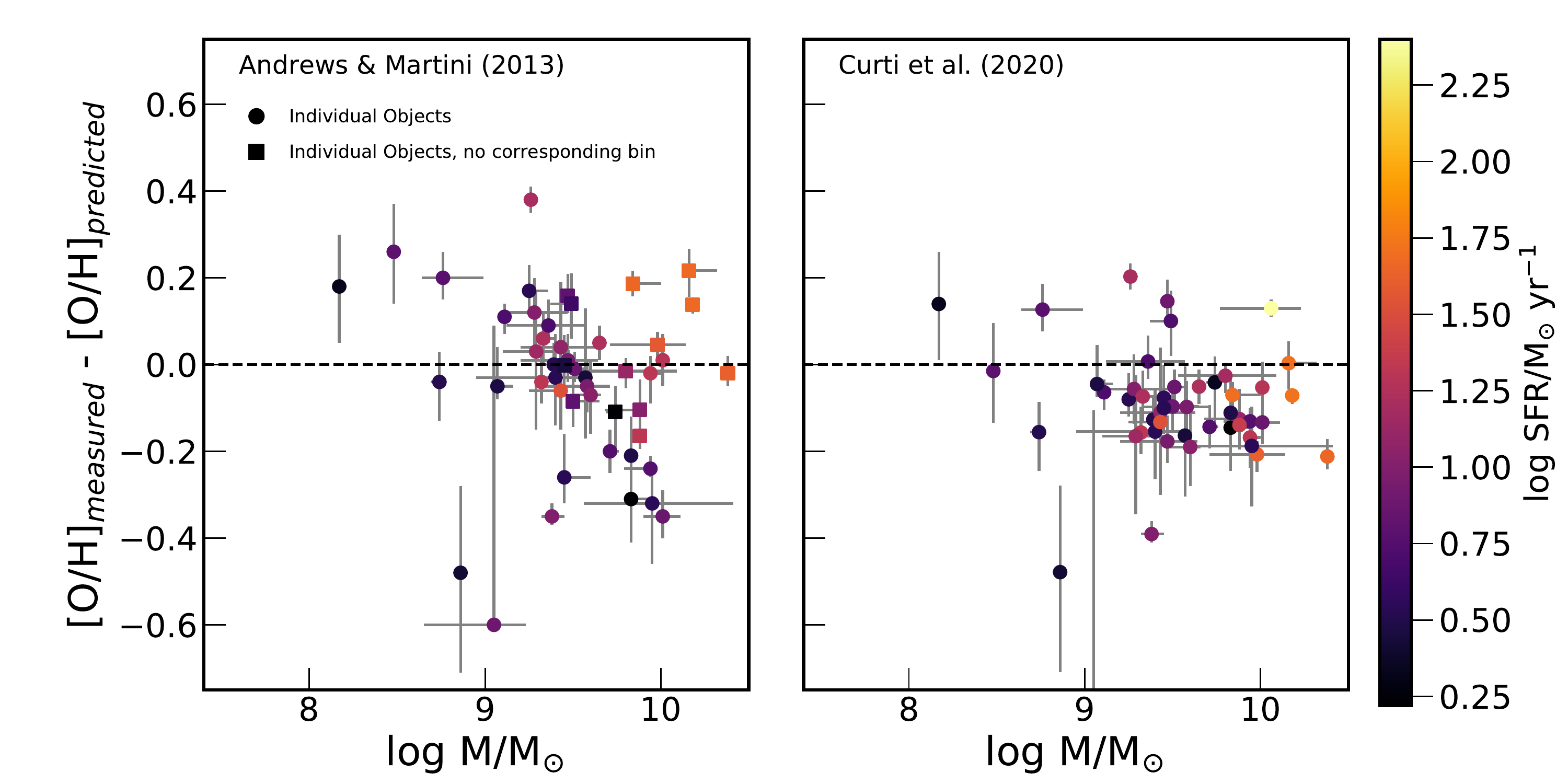} 
\caption{The residuals from the local M-Z-SFR relation are shown for individual objects at $1.3 < z < 1.5$ with \ha\ SNR $>10$ (points).  The left panel shows residuals from the \cite{AM13} relation, where galaxies are matched to the nearest mass and SFR bin in the local sample.   The right panels shows residuals from the parametric relation using total (aperture corrected SFRs) reported by \cite{Curti20}.    Squares in the left panel denote the 13 objects that do not have a corresponding bin in \cite{AM13}; the predicted metallicities for these sources are calculated by extrapolating the reported MZR in appropriate SFR bins  (see Table 4 in \citealt{AM13}). All SFRs are derived from dust-corrected \ha\ emission. Lastly, we note that error bars are asymmetric, and in some cases  the statistical uncertainties are smaller than the points.    } 
\label{fmr_resid_ind} 
\end{figure*}

As noted in \S \ref{ind_spec}, 49 galaxies at redshifts $1.3 < z < 1.5$ have \ha\ SNR $>10$, facilitating dust and metallicity measurements without stacking, as well as providing SFRs from \ha.   The median 68\% confidence interval on these \ha\ derived SFRs is 0.3 dex-- marginaly better than the SED-derived SFRs from the CANDELS/3D-HST data (0.4 dex).   Figure \ref{fmr_ind} (left) shows the M-Z-SFR relation for these galaxies, now using \ha\ SFRs.   We bisect the sample into high and low SFR objects, using a linear fit to the data at log(M/ M$_{\sun}) > 9.2$ (where the sample is larger and the metallicity errors are smaller).  This fit yields the relation ${\rm log(SFR/M_{\sun} yr^{-1})}$ = 0.68 ${\rm log(M/ M}_{\sun})$ - 5.56.     Galaxies with SFRs above (below) this line are shown as orange (purple) points in Figure  \ref{fmr_ind}.   In contrast to the stacked spectra, we do see evidence of an M-Z-SFR relation.  At a given stellar mass, galaxies with higher SFRs have lower metallicities, particularly at $M \ga 10^9$ M$_{\sun}$, where the numbers of galaxies with high signal-to-noise \ha\ detections are higher.   With only a handful of galaxies at $M < 10^9$ M$_{\sun}$ in Figure \ref{fmr_ind}, detecting an M-Z-SFR relation at these masses and redshifts will require larger samples with higher SNR spectra. Deeper WFC3/IR grism spectroscopy, as well as new observations with the {\it James Webb Space Telescope} can improve statistics in this low mass regime.

We next explore how well these individual objects agree with the M-Z-SFR relations from the literature.   As we did with the stacked spectra, we compare to 
the local relations from \cite{AM13} and \cite{Curti20}, showing residuals from these local SDSS relations in  Figure \ref{fmr_resid_ind}.  
 Here, we use the published tabular data from \cite{AM13} in the left panel, and the parametric relation for total SFRs from \cite{Curti20} in the right panel.   In the former case, 13 objects do not have matching bins in local M-Z-SFR relation from \cite{AM13}; for these, we extrapolated the parameterized MZR that is given in bins of SFR (see Table 4 of \citealt{AM13}).  In this comparison, our data show systematic differences from both local measurements.    On one hand, our metallicities are 0.1-0.2 dex lower than the local M-Z-SFR relation from \cite{Curti20}, which uses strong line metallicities.   The left-hand panel, on the other hand, reveals a linear trend in the residuals from the M-Z-SFR relation reported by \cite{AM13}, indicating that the M-Z-SFR relation from from our data has a different shape than the local one derived with direct-method metallicities.   
The substantial negative residual for the stack of all galaxies with log M/M$_{\sun} > 10$ highlighted in Figure \ref{fmr_resid} may be a reflection of this trend.

Interpreting a potential disagreement with the local M-Z-SFR relation is difficult.    While an offset could imply evolution of the relation, the masses and SFRs occupied by high-redshift galaxies are not well sampled in either the \cite{AM13} or \cite{Curti20} relations.   In the latter case, our data fall on an extrapolation of the surface that they fit to their data.  Indeed, \cite{Curti20} point out that the accuracy of their parameterization at low masses and high SFRs -- where high-redshift galaxies lie -- is only 0.3 dex. 
It is worth reiterating that the MZR evolves by a similar amount from $z\sim2$ to $z\sim0$,  so these systematic errors in the local M-Z-SFR are significant.  Hence, we arrive at a clear need:  while the high-redshift M-Z-SFR relation will be an obvious focus of JWST, measuring its evolution will be challenging unless we can find larger samples of analogous galaxies in the local universe.    

It is also instructive to consider scatter in metallicities, which is only possible with the individual high SNR spectra.  Both the scatter around the MZR, and the scatter relative to the M-Z-SFR relation constrain models for galaxy formation.  A key question, which we can address, is how much the MZR scatter can be reduced by considering an M-Z-SFR relation.  For this analysis, \cite{Mannucci10} introduced the projection of least scatter, such that metallicity is a function of $\mu_{\alpha} = {\rm log(M/ M}_{\sun}) - \alpha {\rm log(SFR/M_{\sun} yr^{-1})}$.    Using our data for the 49 galaxies with \ha\ SFRs, we find that scatter is minimized for $\alpha=0.17 \pm 0.07$.  This value is smaller than the value of $\alpha = 0.32$ that was originally reported for the local M-Z-SFR relation by \cite{Mannucci10}.   Other studies that use strong-line metallicity measurements have reported $\alpha = 0.19$ \citep{Yates12}, and $\alpha=0.18$ \citep{Hirschauer18}.    Alternatively, \cite{AM13} find $\alpha = 0.66$ minimizes scatter in {\it stack-based} measurements that use direct metallicities\footnote{This measurement of $\alpha$ is not reduced significantly by accounting for the DIG contribution in local galaxies. \cite{Sanders17} report that a DIG correction decreases $\alpha$ from the \cite{AM13} measurements only marginally, to $\alpha = 0.63$. }.  They argue that larger $\alpha$-- implying that metallicities depend more strongly on SFRs---is a characteristic of direct metallicity measurements of the M-Z-SFR relation. They showed that it is not a consequence of stacking, as the M-Z-SFR relation derived from strong-line calibrations with stacked spectra are characterized by $\alpha = 0.1-0.3$.

The projection of least scatter for our sample is shown in the right panel of Figure \ref{fmr_ind}.  
  Curiously, the value of $\alpha$ that minimizes the scatter  in the MZR still leaves the low SFR galaxies with higher metallicities than the high SFR galaxies in the projection of least scatter.  We verified that this result does not change when $\alpha$ is determined only from the galaxies at $M > 10^{9}$ M$_{\odot}$ and 12 + log(O/H) $> 8.0$.  This finding may indicate that the projection of least scatter is a poor functional representation of the data.  
  Nonetheless, we provide a linear fit to the data, in order to characterize the metallicity as a function of $\mu_{\alpha}$ and facilitate comparisons with other studies:   
 \begin{equation} 
12 + {\rm log(O/H)}  = 6.74 + 0.17 \mu_{0.17}.
\label{proj_eq} 
\end{equation}
The scatter about this relation has an RMS of  0.16 dex, reduced from 0.17 dex in the MZR (relative to an MZR with the shape from \citealt{AM13}). 
This small reduction in scatter is characteristic of results that have been derived using strong-line metallicity diagnostics applied to individual galaxies. 
 For example, \cite{Curti20} find that scatter is reduced from 0.07 dex in the MZR to 0.054 dex relative when SFRs are accounted for and \cite{Hirschauer18} find a reduction from an RMS of 0.182 to 0.178 for a sample of nearby emission line selected galaxies.
 On the other hand,  \cite{AM13} find that when direct-method metallicities are used, the scatter among stacked measurements is reduced more significantly, from 0.22 dex about the MZR to 0.13 dex in the projection of least scatter.  Nonetheless, we acknowledge that scatter among stacks is not the same thing as scatter among galaxies, so these differences are difficult to interpret.   Still, a  greater reduction in scatter, along with the larger $\alpha$, suggest a stronger relation between SFR and metallicity (at fixed mass) when direct-method metallicities are used.   As \cite{AM13} point out, this behavior may indicate that systematic errors on strong-line calibrations are correlated with SFR.  Our work is not immune to these systematic errors.   This points to an increased need for direct metallicities, both at low and high redshifts.  The latter will soon be possible with JWST.

\subsection{Metallicity Biases from Selection Effects} 
\label{bias_sample_sec} 
The existence of an M-Z-SFR relation implies that sample selection effects may impact metallicities, especially for samples that are biased towards high SFRs.    As we discussed in \S \ref{sample_char_sec}, our sample is not mass complete.  Below M/M$_{\sun} \la 9.2-9.3$, galaxies fall mostly above an extrapolation of the star-forming main sequence from  \cite{Whitaker14}.  Likewise, as we showed in Figure \ref{redshift_EW_dist}, the CANDELS/3D-HST sample contains more objects with low \oiii\ equivalent widths compared to WISP.   As we showed in Figure \ref{hasfrs}, for the 49 objects at $1.3 < z <  2.3$ with \ha\ SNR $>10$, the \ha\ SFRs from WISP objects are, on average, 0.35 dex higher than the those of CANDELS/3D-HST objects.

In order to assess the impact of selection effects in our inhomogeneous sample, we created separate stacked spectra for the WISP and CANDELS/3D-HST objects.  We used the same five mass bins that we have used throughout this analysis (see Tables \ref{meas_table}  and \ref{results_table}).  We find that the metallicities from these stacks all agree at the 1-2$\sigma$ level, and there is no systematic difference between the two surveys\footnote{Since the WISP and CANDELS/3D-HST sources have different mean redshifts, we also measure the MZR in four sets of stacked spectra: WISP at $z>2$, $z<2$, and CANDELS/3D-HST at $z>2$, $z<2$.  Still, we find no significant differences in the metallicities of these subsamples.}. 

The agreement between the metallicities in the WISP sub-sample and the CANDELS/3D-HST sub-sample can be explained by the fact that the SFR dependence in the M-Z-SFR relation that we measure is not particularly strong.    Following Equation \ref{proj_eq}, if the WISP sources have SFRs that are 0.35 dex higher,  we expect their metallicities to be lower than CANDELS/3D-HST by only 0.01 dex.   This difference is smaller than the precision of our metallicity measurements (Table \ref{results_table}).  Even SFRs that are 1 dex higher than the main sequence---as we might see towards the lowest masses in our sample (see Figure \ref{msfig})---only result in metallicities that are lower by 0.03 dex. 

The dependence of metallicity on SFR may be larger if we were able to use direct metallicities in our high-redshift sample.  
At $z\sim 0.1$, the M-Z-SFR relation shows a larger spread in metallicity with changing SFR when it is derived by stacking spectra to measure direct metallicities \citep{AM13}. These authors report that the projection of least scatter has a slope of 0.43, with $\alpha$ = 0.66.  Therefore, a 1 dex increase in SFR corresponds to a 0.3 dex decrease in metallicity, although the differences between the WISP and CANDELS/3D-HST metallicities would still be small (0.03-0.09 dex).  Nonetheless, when we apply a strong-line calibration to our sample of 49 objects with high SNR spectra, we find a weak relation between SFR and metallicity (at fixed mass); therefore, we do not expect to observe a measurable bias towards low metallicities in our sample.

\begin{figure}
\begin{center} 
\includegraphics[scale=0.4]{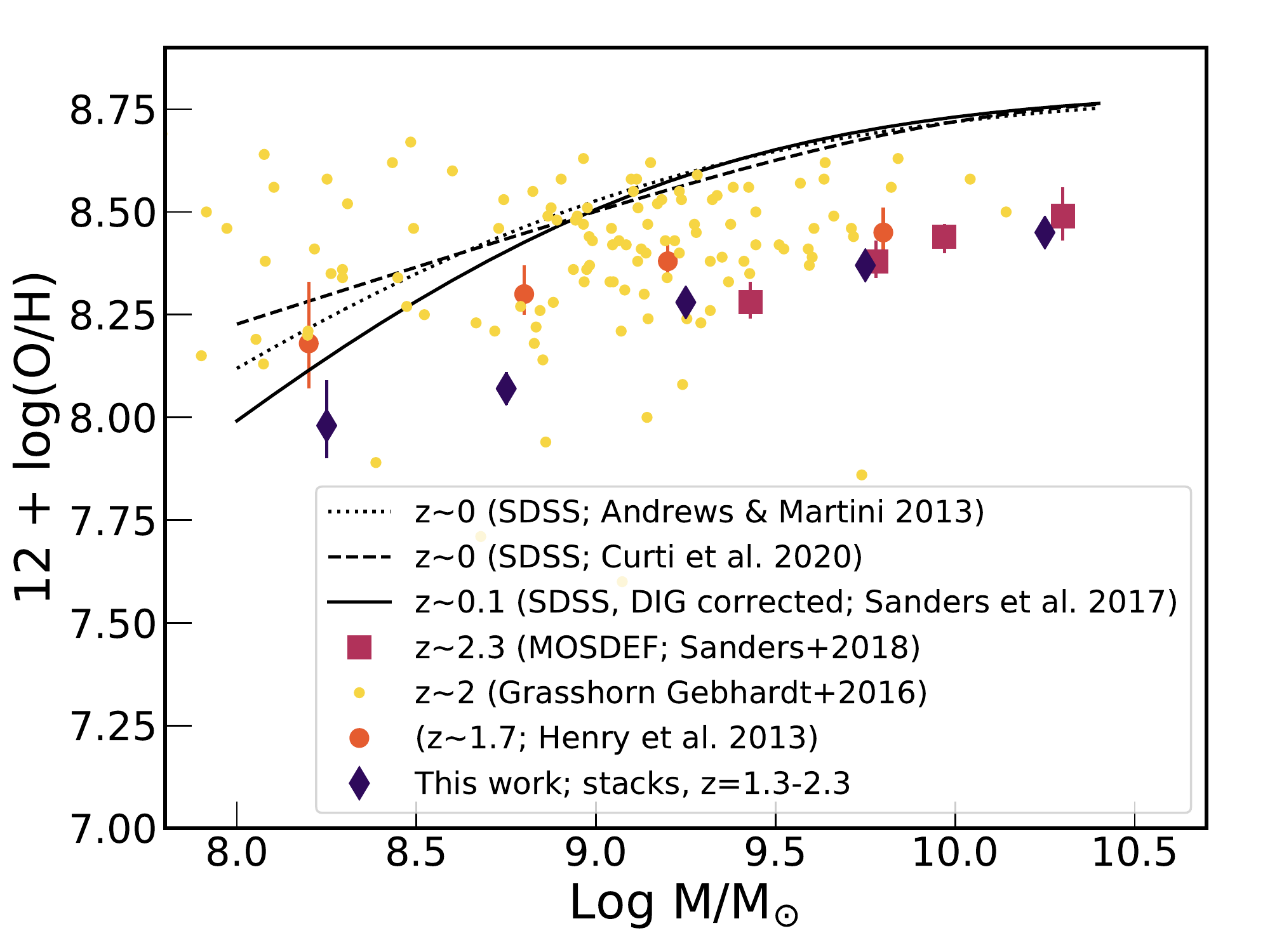} 
\caption{Sample selection effects can impact the metallicity measurements in low SNR spectra when a second line is required to confirm the redshift and measure the metallicity. Here, we compare our new measurements to two WFC3/IR grism-based studies that require confirming emission lines: \cite{Henry2013_wisp} and \cite{GG16}.  This selection effect, coupled with low SNR spectra, seems to result in an MZR which is biased towards high metallicities.     The metallicities from all of the high redshift studies shown here have been calculated on the \cite{Curti20} calibration, using published line ratios.     \label{mzr_sel} } 
\end{center}
\end{figure}

In addition to mass-completeness, metallicities can also be biased when spectroscopic surveys require the detection of multiple emission lines.  
In WFC3/IR grism spectra, where the SNR is often not particularly high, secondary emission lines used to confirm redshifts-- typically \hb\ or \oii\-- often have SNR $\sim 2-3\sigma$.  In this case, Eddington bias leads an overestimate of the average fluxes of these lines, due to statistical scatter above the detection threshold \citep{Eddington13}.  Artificially increasing the average \hb\ and \oii\  fluxes will result in decreased $R_{23}$, \oiii/\hb, and  $O_{32}$ ratios.   These biases tend to increase metallicity inferences (provided that galaxies are on the upper branch of  $R_{23}$ and \oiii/\hb, as they are in most of our sample).

To quantify the bias due to the requirement for multiple emission lines, we re-calculated the MZR using the \cite{Curti20} calibration with the published $R_{23}$ and $O_{32}$ values in two previous WFC3 IR grism studies that were subject to this bias: \cite{Henry2013_wisp} and \cite{GG16}.  The MZRs from these two works are compared to our new results in Figure \ref{mzr_sel}.   On average, the metallicities are around 0.1-0.2 dex higher.   (Although, we note that the current MZR and \cite{Henry2013_wisp} are consistent at around the 2-3$\sigma$ level.) 
Curiously, this difference contrasts with the WISP-only stacks from our present sample, which we showed to be consistent with CANDELS/3D-HST-only stacks and the MZR of the full sample.   A major difference between our prior WISP sample and and the current one is that the most recent line lists adopt a higher SNR threshold for the primary (strongest) emission line, and should include far fewer spurious emission line galaxies (Bagley et al., in prep).   We conclude that there is a great amount of subjectivity in the sample selection with low SNR spectra when a second line is required to confirm the redshift.  The present WISP sample seems to be less biased than \cite{Henry2013_wisp} and \cite{GG16}, as it shows negligible differences from CANDELS/3D-HST.

 Lastly, we acknowledge that an \oiii-based selection can impart a metallicity bias, as the strength of this line is critically sensitive to metallicity.  As such, we might expect our sample to be biased towards objects with the highest \oiii/\hb\ ratios and metallicities around 12 + log(O/H) $\sim 8.0$.  Therefore, we next consider the effects of \oiii\ selection by modeling our survey in the IllustrisTNG  hydrodynamical simulation \citep{Marinacci18, Springel18, Nelson18, Pillepich18, Naiman18}.

\section{Comparison to Theoretical Models} 
\label{theory_sec}
The MZR and M-Z-SFR relation are key observational constraints on theoretical models describing the evolution of galaxies.  
As we argued in \S \ref{met_cal}, metallicity calibrations have improved considerably over the past several years, indicating that we are now better poised to compare observations and theoretical models.  One thing that remains uncertain, however, is how selection effects, in the presence of an M-Z-SFR relation, impact our ability to compare observations and models.   While recent simulations report success in matching the slope and evolution of the MZR \citep{Ma16, DeRossi17, Dave17,Dave19, Torrey19}, they do not account for the possibility that observed metallicities may be biased by sample selection effects. 

In this section, we address this shortcoming by modeling our sample selection in the IllustrisTNG hydrodynamical simulation 
\citep{Marinacci18, Springel18, Nelson18, Pillepich18, Naiman18}.   We use the TNG100-1 simulation, which has a volume of 106.5$^3$ comoving Mpc$^3$, and select 
star-forming galaxies with  $M>10^{8}$ M$_{\sun}$.  We include only central galaxies, as satellite galaxies have metallicities heavily influenced by their environment, and are unlikely to be resolved from central galaxies at $z=2$ (Torrey 2020, private communication).   From these snapshots, we take masses, SFRs, and metallicities for 24,108 galaxies at $z\sim 0.1$, 30,485 galaxies at $z\sim 1.5$ and  29,578 galaxies at  $z\sim 2.0$.  We take the SFR-weighted metallicities from the simulation, as these most accurately reflect the luminosity-weighted measurements that are made from emission line spectra.   We refer the reader to \cite{Torrey19} for details pertaining to the IlustrisTNG MZR.   

 In order to model emission line based selection, we convert these quantities into observed emission line fluxes. 
We first use SFRs to determine intrinsic \ha\ luminosities, assuming the \cite{Kennicutt98} relation between SFR and \ha\ luminosity, and multiplying by 1.8 to convert from a \cite{Chabrier2003} IMF used by IllustrisTNG to a \cite{Salpeter} IMF used by  \cite{Kennicutt98}.  Unreddened \ha\ line fluxes are then calculated from the redshift of the sources in the simulation, assuming a Planck 2015 cosmology \citep{Planck15}. 

We next calculate the unreddened \hb\ line flux assuming an intrinsic Balmer decrement of \ha/\hb\ = 2.86. The unreddened \oiii\ line flux is inferred from the relation between metallicity and \oiii/\hb\ from \cite{Curti17},  adding 0.07 dex of scatter in log (O/H), relative to the calibration, as reported by \cite{Curti17}. 

Finally, we dust redden all three emission lines.  In practice, we should be able to estimate dust mass from simulation data, using measures of gas and metal content.  However, translating this estimate to dust extinction is highly dependent on uncertain assumptions (e.g.\ \citealt{Narayanan18}).  Therefore, we adopt an empirical approach.   For $z\sim0.1$, we use the relation between $A_{H\alpha}$ and stellar mass in SDSS galaxies \citep{Garn10}, including 0.28 magnitudes of scatter in the reddening.   At higher redshifts, however,  measurements of the Balmer decrement at $z\sim 1-2$ do not detect strong correlations between dust extinction and galaxy properties \citep{Dominguez13, Theios19}.    Therefore, for $z\sim 1.5$ and 2.0, we assign dust  extinction randomly, drawing from an exponential distribution with an e-folding length of E(B-V)$_{gas}$ = 0.4, roughly consistent with the distribution reported by \cite{Theios19}.  Reddening is then applied using the \cite{Calzetti00} extinction relation.

\begin{figure*} [!t]
\begin{center} 
\includegraphics[scale=0.6]{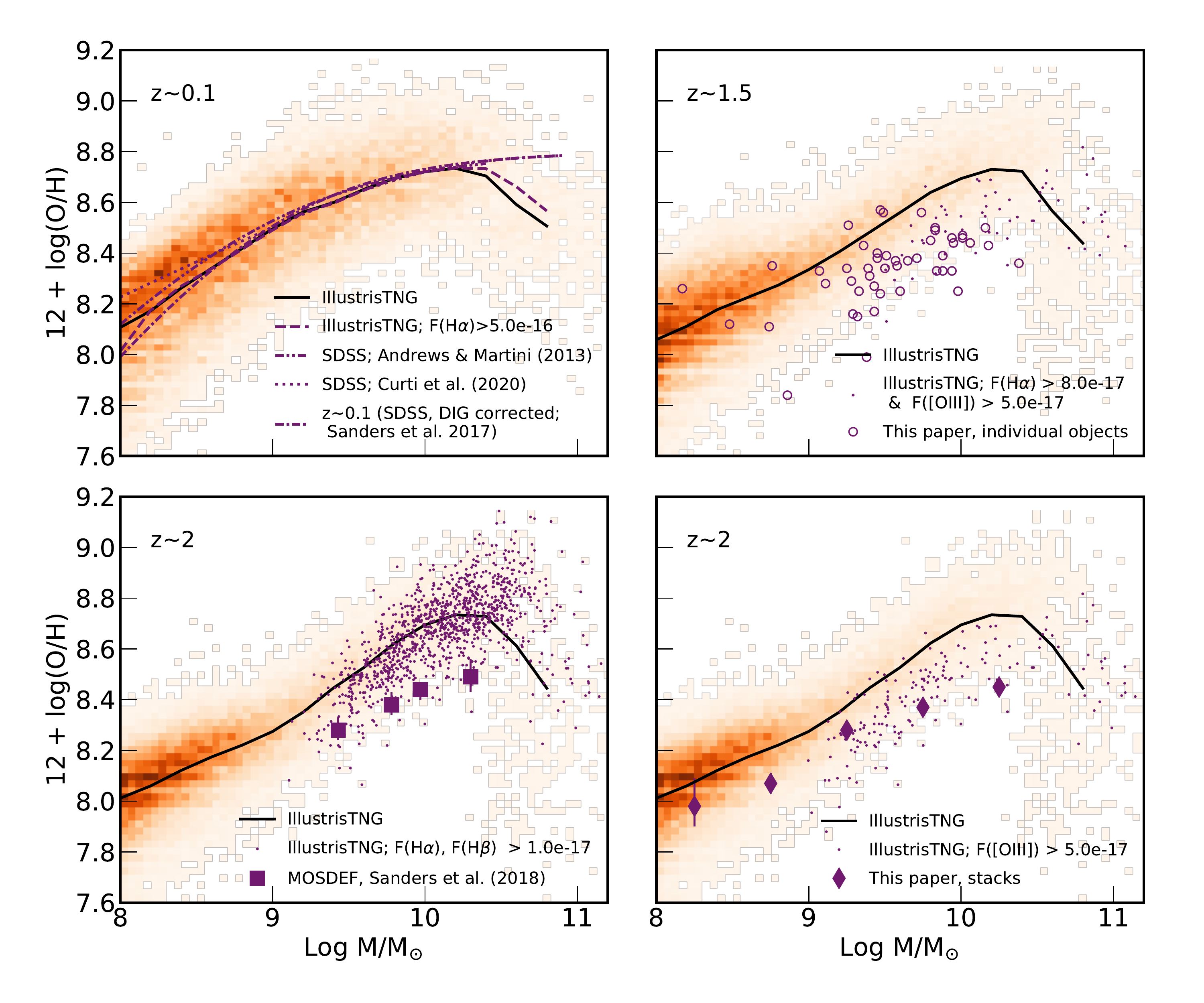} 
\caption{Line flux limited surveys can result in a sample selection that is not representative of typical galaxies in a simulation.  The MZR from IllustrisTNG \citep{Torrey19} is shown as the orange shaded region for the $z\sim0.1$ (upper left), $z\sim 1.5$ (upper right), and $z\sim 2$ (bottom).     The black solid line shows the mean of the simulation data.  All simulated metallicities have been shifted downwards by 0.2 dex, adjusting for uncertainties in the nucleosynthetic yield of oxygen and matching the $z\sim 0.1$ MZR around $M  = 10^{10} M_{\sun}$.  Small points illustrate the effect of  line-flux limited selections on the simulation.  These  represent the SDSS (F(\ha) $> 5  \times 10^{-16}$ erg s$^{-1}$, cm$^{-2}$, upper left),  MOSDEF   (F(\ha) and F(\hb) $> 1  \times 10^{-17}$ erg s$^{-1}$ cm$^{-2}$, lower left), and the present WFC3 grism surveys (F(\oiii) $> 5  \times 10^{-17}$ erg s$^{-1}$ cm$^{-2}$ for stacks, lower right, and F(\ha) $> 8  \times 10^{-17}$ erg s$^{-1}$ cm$^{-2}$ for individual objects, upper right).   For comparison, observational data are shown.  The upper left panel includes the $z\sim0.1$ MZR from \cite{AM13} and \cite{Curti20}, as well as the DIG corrected SDSS measurement from \cite{Sanders17}.  The $z\sim 2$ measurements from MOSDEF are shown the the lower left  \citep{Sanders18}, and the stacks and individual objects from this work are shown in the right panels.   \label{tng_mzr}     } 
\end{center} 
\end{figure*}

This approach provides us with modeled \ha, \hb, and \oiii\ fluxes for each simulated galaxy.   We next emulate sample selections by applying cuts to the emission line fluxes, to generate four mock samples of galaxies.  For the present WFC3 grism selected sources that we use in stacks, we adopt F(\oiii) $> 5 \times 10^{-17}$ erg s$^{-1}$ cm$^{-2}$.     This flux limit corresponds to the approximate 50\% completeness of our catalog.  Similarly, for the 49 objects with \ha\ SNR $> 10$, we take  
F(\ha) $> 8 \times 10^{-17}$ erg s$^{-1}$ cm$^{-2}$.   These limits are an approximation, as the WISP data have a range of exposure times, some of which are significantly longer than the CANDELS/3D-HST observations. 
We also compare to the flux limits used by \cite{Sanders18}  in the MOSDEF survey, where we estimate\footnote{\cite{Kriek15} report typical 5$\sigma$ sensitivity for MOSDEF of $1.5\times10^{-17}$ erg s$^{-1}$ cm$^{-2}$, and  \cite{Sanders18} select their sample by requiring 3$\sigma$ detections in \ha\ and \hb.} that \ha\ and \hb\ fluxes are greater than $1 \times 10^{-17}$ erg s$^{-1}$ cm$^{-2}$. Lastly, we consider the SDSS for a local comparison sample. We use the 50\% completeness in the \ha\ fluxes reported in the DR7 MPA/JHU catalog, which corresponds to around F(\ha) $>6\times 10^{-16}$ erg s$^{-1}$ cm$^{-2}$.     We note that more detailed modeling, accounting for broad-band magnitude limits, equivalent width cuts, non-uniform survey sensitivity (a particular concern for WISP), and slit mask/fiber plug plate designs are not addressed here.  We leave this more sophisticated analysis to future work.

Figure \ref{tng_mzr} shows the IllustrisTNG MZR at $z\sim 0.1$ (upper left), $z\sim 1.5$ (upper right), and $z\sim 2.0$ (bottom).     The shaded regions show the relation for all star-forming galaxies, with the black solid curve indicating the mean.  These are the same relations discussed in detail in \cite{Torrey19}; we have decreased the normalization of the modeled metallicities by 0.2 dex, in order to better match the $z\sim 0.1$ SDSS relation.   This re-normalization is generally required, as the nucleosynthetic yield of oxygen remains uncertain \citep{Nomoto13}. The small points show different sample selections that we apply to the simulation, in order to match comparison observations:   the SDSS $z\sim0.1$ measurements from \cite{AM13} and \cite[upper left panel]{Curti20}, and the $z\sim 2$ measurement from MOSDEF \citep[lower left]{Sanders18}, along with the individual objects at $z\sim 1.3-1.5$ (upper right)  and the stacked spectra (lower right; mean $z = 1.9$) in the present work. 

At low redshifts, the \ha\ flux selection does not modify the modeled MZR--- i.e., the curves in the upper left panel of Figure \ref{tng_mzr} fall on top of each other at all masses.  Given the relative brightness of nearby galaxies, the SDSS is sensitive to a representative sample of galaxies at $M>10^{8}$ M$_{\sun}$---56\% of simulated galaxies have modeled \ha\ fluxes 
above the selection threshold.  Likewise, IllustrisTNG does a fair job at matching the slope of the SDSS MZR, as the observations overlap closely with the model curves.

\begin{figure*} 
\begin{center} 
\includegraphics[scale=0.7]{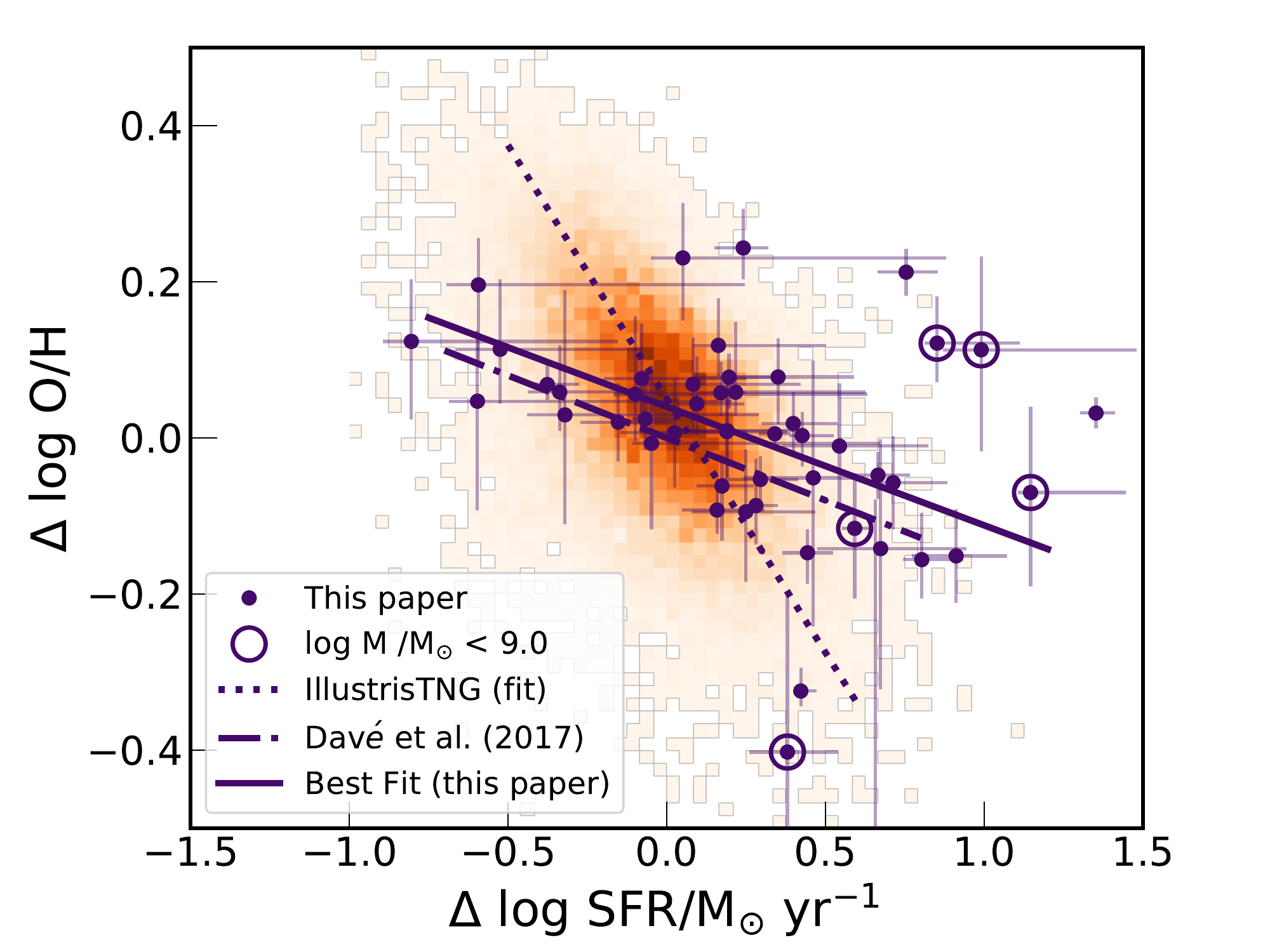}
\end{center}
\caption{The M-Z-SFR relation at $z=1.5$ in IllustrisTNG has a dependence on SFR that is too steep when compared to observations. The shaded region shows the simulated deviations from the star-forming main sequence, $\Delta$log SFR vs the deviations from the MZR, $\Delta$log O/H.   Both are calculated with respect to polynomial fits to the simulated scaling relations.  The slope of the deviations, shown by the dashed line, is around $\sim -0.65$.   Points show our observations, which are fit with a linear relation having a slope of $-0.15$ (solid line).  The observations are closer to the slope of $-0.16$ in the simulation reported by \cite{Dave17} (dot-dashed line).  \label{tng_fmr} } 
\end{figure*} 

At higher redshifts, emission line flux limits play a more significant role.   The observations in this paper, shown in the right hand panels, show an MZR that is around 0.2 dex lower than the IllustrisTNG MZR  {\it derived from the entire simulation.}  However, when we model the \oiii-based selection in the simulation,  we see better agreement between theory (small points) and observations for both the stacked spectra (filled diamonds) and the individual objects  with \ha\ SNR $>10$ (open circles).   The \oiii-based selection picks out objects with the lowest metallicities, which we expect for galaxies in the mass and metallicity regime that we consider. 
We also notice in the right had panels of Figure \ref{tng_mzr}  that modeling the sample selection removes simulation galaxies at the lowest masses probed by our data. This difference is likely a reflection of our simplified model of sample selection effects, but could also be an indication that simulation galaxies do not exist in the right densities for the masses, metallicities, and SFRs probed by this portion of sample.    We leave this question to future work. 

In contrast to the results from the present survey, the lower left panel of Figure \ref{tng_mzr} shows that the \ha\  + \hb\ selection modeled after MOSDEF selects galaxies that are mostly representative of the full simulation. However, the model data with selection effects applied have metallicities around 0.2 dex higher than the observations. {\it Hence, while accounting for selection effects brings the IllustrisTNG model in line with the MZR in the present work, it does not do so for the MOSDEF MZR. }   Curiously,  even though our MZR shows good agreement with the MOSDEF observations (see Figure \ref{mzr_highz}), when we account for selection effects, IllustrisTNG predicts that our MZR should have a lower normalization than the one from MOSDEF. 
   
The disagreement between the observed and simulated $z\sim 2$ MZRs in Figure \ref{tng_mzr} suggests a coupling between mass, metallicity, and SFR that is difficult to disentangle {\it from the MZR alone.}   While modeling the \oiii-based selection lowers the metallicities that are predicted from the simulation, this is not the only effect that is in play.  The selection effects, when applied to the simulation, predict that our \oiii-selected sample should have higher SFRs than MOSDEF,  by about 0.2 dex at  $9.75 < $ log M/M$_{\sun} < 10.0$ (average log SFR/M$_{\sun}$ yr$^{-1}$ = 1.16 vs. 0.97).   This trend would also lead to lowered metallicities, given the presence of the M-Z-SFR relation in both the observations and the simulation \citep{Torrey19}.   However, our observations show very similar SFRs to those of MOSDEF in the mass range where they overlap (cf.\ Table \ref{results_table} vs Table 1 in \citealt{Sanders18}).    Therefore, we next compare the simulated M-Z-SFR relation with observations.

 Figure \ref{tng_fmr} shows the simulated M-Z-SFR relation, plotting the deviations from the MZR and the star-forming main sequence, $\Delta$log O/H versus $\Delta$log SFR/ M$_{\sun}$ yr$^{-1}$ , as proposed by \cite{Dave17}.  In this case, we use a snapshot from IllustrisTNG at $z=1.5$, for ease of comparison with our 49 high SNR objects at $1.3 <  z < 1.5$.   The slope shows how changes in SFR translate to changes in metallicity.  Excluding simulation sources at log M/M$_{\sun} > 10.2$, where the model for quenching becomes substantial and the simulations do not match the observations, we find a slope of -1.0 in the simulation data (dotted line).     The same quantities for our observations are shown by the points, where we have calculated $\Delta$log O/H relative to a linear fit to the MZR from our stacked spectra, and $\Delta$log SFR/ M$_{\sun}$ yr$^{-1}$ relative to the star-forming main sequence of \cite{Whitaker14}\footnote{The star-forming main sequence from \cite{Whitaker14}  is reported for $1.0 < z < 1.5$ and $1.5 < z < 2.0$.  We take the average of these measurements to represent our sample at $1.3 < z < 1.5$.}.    A linear fit to our data for the 49 high SNR objects (accounting for errors in both $\Delta$log O/H and  $\Delta$log SFR/M$_{\sun}$ yr$^{-1}$) yields a slope of -0.15 $\pm 0.05$.  This measurement is close to the slope of -0.16 found by \cite{Dave17}\footnote{We note that \cite{Dave17} express this diagram in terms of $\Delta$log sSFR/ yr$^{-1}$ rather than $\Delta$log SFR/ M$_{\sun}$ yr$^{-1}$.  These quantities are equivalent, as they are calculated at a fixed mass.  }. 

Figure \ref{tng_fmr}  suggests that our data follow a shallower slope than inferred from the simulated galaxies in IllustrisTNG.   
A 1 dex increase in SFR (or sSFR) will translate to a 1 dex decrease in metallicity in the simulation, but only a 0.15 dex metallicity decrease in our observations.   This weak SFR dependence is similar to what we identified from the projection of least scatter in \S \ref{fmr_ind_sec} and Equation \ref{proj_eq}.  In fact, it follows from Equation  \ref{proj_eq}  (the projection of least scatter) that the slope in Figure \ref{tng_fmr}  should be 0.17 $\times$ $\alpha_{0.17} = 0.03$.  While the SFR dependence that we find from the projection of least scatter is even weaker than what we see in the  $\Delta$log O/H versus $\Delta$log SFR/ M$_{\sun}$ yr$^{-1}$ plot , they are both much weaker than in IllustrisTNG.  
The slight disagreement between these two observational methods is not surprising, as the calculations have different systematics (e.g. $\Delta$log SFR is calculated with respect to \citealt{Whitaker14}; Equation \ref{proj_eq} is assumed to be linear). 

While the M-Z-SFR relation at low redshift shows a stronger SFR dependence when inferred  from stacked spectra with direct metallicity measurements \citep{AM13}, the amplitude of this effect is not likely to be large enough to explain the discrepancy between our observations and IllustrisTNG.   When \cite{AM13} report the projection of least scatter, they find $\alpha$ = 0.66, and the linear relationship between $\mu_{\alpha}$ and 12+log(O/H) has a slope of 0.43; hence their data suggest that a correlation between $\Delta$log O/H and $\Delta$log SFR/ M$_{\sun}$ yr$^{-1}$ in the SDSS should have a slope of approximately $-0.3$.  Similarly, the derivative (with respect to SFR) of the M-Z-SFR relation from \cite{Curti20} is around -0.17 at M$ =  10^{9.0} -10^{10.0}$ M$_{\sun}$.  Hence, locally, the difference between strong-line calibrations applied to individual galaxies and direct metallicities measured from stacks amounts to steepening of the slope in Figure  \ref{tng_fmr} by around 0.1.    Therefore, it seems unlikely that measuring direct metallicities at $z\sim2$ would bring the high-redshift M-Z-SFR relation in line with the steeper SFR dependence in IllustrisTNG.   The slope of the simulated data in Figure \ref{tng_fmr} is too steep.

We conclude that modeling selection effects and comparing the M-Z-SFR relation in simulations and observations  provide more rigorous tests than the the MZR alone.  In addition to more careful modeling of selection effects, future comparisons to simulations beyond IllustrisTNG would be informative.

\section{Summary and Conclusions} 
\label{conclusions} 
We have measured the MZR and M-Z-SFR relation at $1.3 < z < 2.3$, using a sample of 1056 star-forming galaxies selected based on \oiii\ line emission in their WFC3/IR grism spectra.  This sample, drawn from the WISP survey and CANDELS/3D-HST, has a mean redshift of $z = 1.9$, and is four times larger than previous MZR samples at $z\sim2$ \citep{Sanders18}.  The WFC3/IR grism selection reaches stellar masses an order of magnitude lower ($10^{8}$ M$_{\sun}$) than ground-based IR spectroscopic surveys at this redshift.  To measure metallicities, we stack the spectra in five mass bins, between $ 8 < $log M/M$_{\sun} < 10.5$, but also consider 49 objects at $1.3 < z < 1.5$ and \ha\ SNR $>10$,  where our observations cover the full suite of optical emission lines from \oii\ to \ha\ + \sii.  

The MZR shows good agreement with previous oxygen-based measurements at $z\sim2$ \citep{Sanders18}, in the mass regime where they overlap.   Stacking our full sample in five mass bins, we find that metallicities evolve by 0.3 dex  from $z\sim 2$ to $z\sim0$.  In contrast, we see no evidence for redshift evolution over the two Gyrs spanned by our sample; stacked spectra at $1.3 < z < 1.7$ and $1.7 < z < 2.3$ show good agreement. 

We also measure the M-Z-SFR relation in our data.   We use SED-derived SFRs to create stacked spectra having higher and lower SFRs at each mass bin.    This exercise shows no evidence for an SFR dependence: the MZR derived from high SFR galaxies is consistent with the MZR from low-SFR galaxies.  We show that uncertainties on SED-derived SFRs make it difficult to fully separate the sample into high and low SFR bins.  However, if we instead consider the 49 galaxies with more accurate \ha-derived SFRs, we uncover an M-Z-SFR relation at $1.3 < z  <1.5$.   In comparison to low redshifts,  we find metallicities that are systematically 0.1 dex lower than the local M-Z-SFR relation derived from strong lines  in individual galaxy spectra \citep{Curti20}, but which show a linear trend in their residuals from the M-Z-SFR relation when it is measured from direct metallicities in stacked spectra \citep{AM13}.  It is difficult to determine whether there is real evolution, because small samples of low redshift galaxies at the masses and SFRs of our sample imply that the local M-Z-SFR relation is poorly constrained in this regime. 

The 49 objects with \ha\ SFRs allow us to measure the projection of least scatter of the M-Z-SFR relation (metallicity as a function of
 $\mu_{\alpha}$, where $\mu_{\alpha} = $ log M/M$_{\sun}$ - $\alpha$ log SFR / M$_{\sun}$ yr$^{-1}$).   We find an optimum fit corresponding to $\alpha  = 0.17$, although the scatter is reduced only by a small amount: from 0.17 dex about the MZR,   to 0.16 dex about the M-Z-SFR relation.    Both the low value of $\alpha$, and the very small reduction in scatter, are consistent with low-redshift studies that use strong emission line metallicity calibrations \citep{Mannucci10, Yates12, AM13, Hirschauer18}.  
The weak SFR-dependence in our M-Z-SFR relation implies that our sample, while incomplete to low-SFR objects at low-masses, is biased to low metallicities by 0.03 dex at most.    However, it is important to note that direct-method abundances could yield a different result. The low-redshift M-Z-SFR appears to have a stronger dependence on SFR when direct-method abundances are used (\citealt{AM13} find $\alpha=0.66$  using direct abundances in stacked SDSS spectra, and $\alpha = 0.1-0.3$ using strong lines in the same stacks).    These authors argue that this stronger SFR dependence from direct abundances indicates that strong-line calibrations are subject to systematic uncertainties that correlate with SFRs.   We speculate that the M-Z-SFR relation at high redshifts may have a stronger SFR dependence if it were be measured with direct-method metallicities.   As a corollary, our MZR may be more biased by incompleteness than we have inferred from the strong-line M-Z-SFR relation. 

Finally, we compare our results with theoretical models of the MZR and M-Z-SFR relation.   Because of the presence of an M-Z-SFR relation, directly comparing an observed MZR with models, as is often done, can be misleading.  Therefore,  we have applied sample selection effects to the IllustrisTNG simulation,  thereby more accurately modeling the MZR from the present data, as well as the MZR from \cite{Sanders18}.    This exercise suggests that the selection effects in our present sample are uncovering the objects with the highest SFRs and lowest metallicities, whereas the selection used by \cite{Sanders18} uncovers more typical $z\sim2$ galaxies.   Still, modeling these selection effects brings the high-redshift MZR from IllustrisTNG more in line with the observations that we present in this paper.  
Curiously, however, IllustrisTNG predicts that the galaxies observed by \cite{Sanders18} should have higher metallicities than the galaxies in our WFC3/IR selected sample.  Yet, the MZR measurements from these two studies show excellent agreement in the mass range where they overlap.

 To further explore this discrepancy between the model and the data, we compared the M-Z-SFR relation in IllustrisTNG with our 49 galaxies with \ha\ SFRs.   Following \cite{Dave17}, we show the relationship between 
  $\Delta$log O/H and $\Delta$log SFR/ M$_{\sun}$ yr$^{-1}$ for both our observations, and the IllustrisTNG simulation data.   We find that the slope in $\Delta$log O/H vs. $\Delta$log SFR/ M$_{\sun}$ yr$^{-1}$ plane is around $-1.0$ in the simulation data, but is -0.15 in the observations.     The slope of the simulated data is much too steep; i.e., metallicity depends too strongly on SFR in IllustrisTNG.    On the other hand, the shallow slope that we measure shows good agreement with the simulated M-Z-SFR relation from \cite{Dave17}.  While direct metallicity measurements at $z\sim2$ could bring observations closer to IllustrisTNG, we argued in \S \ref{theory_sec} that--based on measurements of the local M-Z-SFR relation-- the amplitude of this systematic effect is too small to explain our discrepancy with IllustrisTNG.

In conclusion, we reiterate that the metallicities of galaxies remain a key constraint on galaxy formation models.  This work has provided new measurements of the evolution of the MZR and the M-Z-SFR relation, while also uncovering discrepancies in the simulated M-Z-SFR relation from IllustrisTNG.   Nonetheless, we also identify some clear needs.  If we are going to understand the evolution of the MZR and M-Z-SFR relation, we need statistical samples of galaxies with direct-method abundances at high redshifts.  These data, which will soon be within reach of JWST, will be key to clarifying the strength of the SFR dependence in the M-Z-SFR relation at high-redshift.

\acknowledgements  
AH, MR, DE, and BE acknowledge support from HST-AR 14580.   We thank Paul Torrey for assisting with the interpretation of IllustrisTNG data.   We are grateful to Bahram Mobasher and Kit Boyett for thoughtful comments that improved this manuscript.   AJB acknowledges funding from the ``FirstGalaxies" Advanced Grant from the European Research Council (ERC) under the European Union's Horizon 2020 research and innovation programme (Grant agreement No. 789056).  Y.-S. Dai acknowledges the support from NSFC grant number 10878003.  
This work is based on data obtained from the Hubble Space Telescope, as part of the  3D-HST Treasury Program (GO 12177 and 12328 and the CANDELS Multi-Cycle Treasury Program, as well as GO 11600, 11696, 12283, 12568, 12902, 13352, 13517, 14178,  13420,  and 14227.   Support for these programs was provided by NASA through a grant from the Space Telescope Science Institute, which is operated by the Association of Universities for Research in Astronomy, Incorporated, under NASA contract NAS5-26555.

\appendix  
\section{Estimating Emission Line Equivalent Widths} 
\label{appendix_ew}
Estimates of emission line equivalent widths are necessary for both stacked spectra and individual spectra, as the correction for Balmer line stellar absorption is proportional to the emission equivalent width.  We describe how this correction is applied in Appendix \ref{bayes_met}.   Here, we detail the method used to estimate equivalent widths and their uncertainties. 

Slitless grism spectra are imperfect for measuring the equivalent width of emission lines, as inaccurate sky subtraction or contamination subtraction can result in large systematic errors.  Moreover, it is not unusual to detect lines without any continuum, in which case the spectra can only provide a lower limit on the equivalent width.   An alternative approach is to measure the continuum from the broad-band infrared images.  However, this method does not directly measure the continuum flux at the specific wavelength of the emission line, and the extraction aperture is not guaranteed to be the same for the emission lines and the continuum.   The broad-band fluxes are also likely contaminated by line emission.     

Given these caveats, we proceed cautiously, and compare two different methods for equivalent width estimation. 
 The first method is to measure equivalent width directly from the spectra.  For this measurement, we subtract the contamination model, and count spectral continuum detections as cases where the continuum SNR $> 0.5$ per pixel.  The second method is to use the broad-band photometry to estimate the continuum at the wavelength of \oiii. In this case, we use the WFC3/IR photometry: F110W and F160W for WISP, and F125W and F160W for the CANDELS/3D-HST fields.  In each galaxy, the broad bands photometry is corrected for contamination from emission lines, using our measured fluxes for \oiii, \oii, and \ha.  We also include a small correction for \hb\ and \sii, assuming the average ratios for the sample:   \oiii /\hb = 7.1, and \sii /(\ha + \nii) = 0.16 (with the latter ratio for $z< 1.5$ only).   Then, we use a linear interpolation between the corrected continuum fluxes to estimate the continuum at the wavelength of \oiii, thereby obtaining the equivalent width in the line.    Uncertainties on the continuum at \oiii\ are estimated by a Monte Carlo simulation, where the WFC3 photometry and the contribution from emission lines were perturbed according to their measured errors.  
 
\begin{figure} 
\plotone{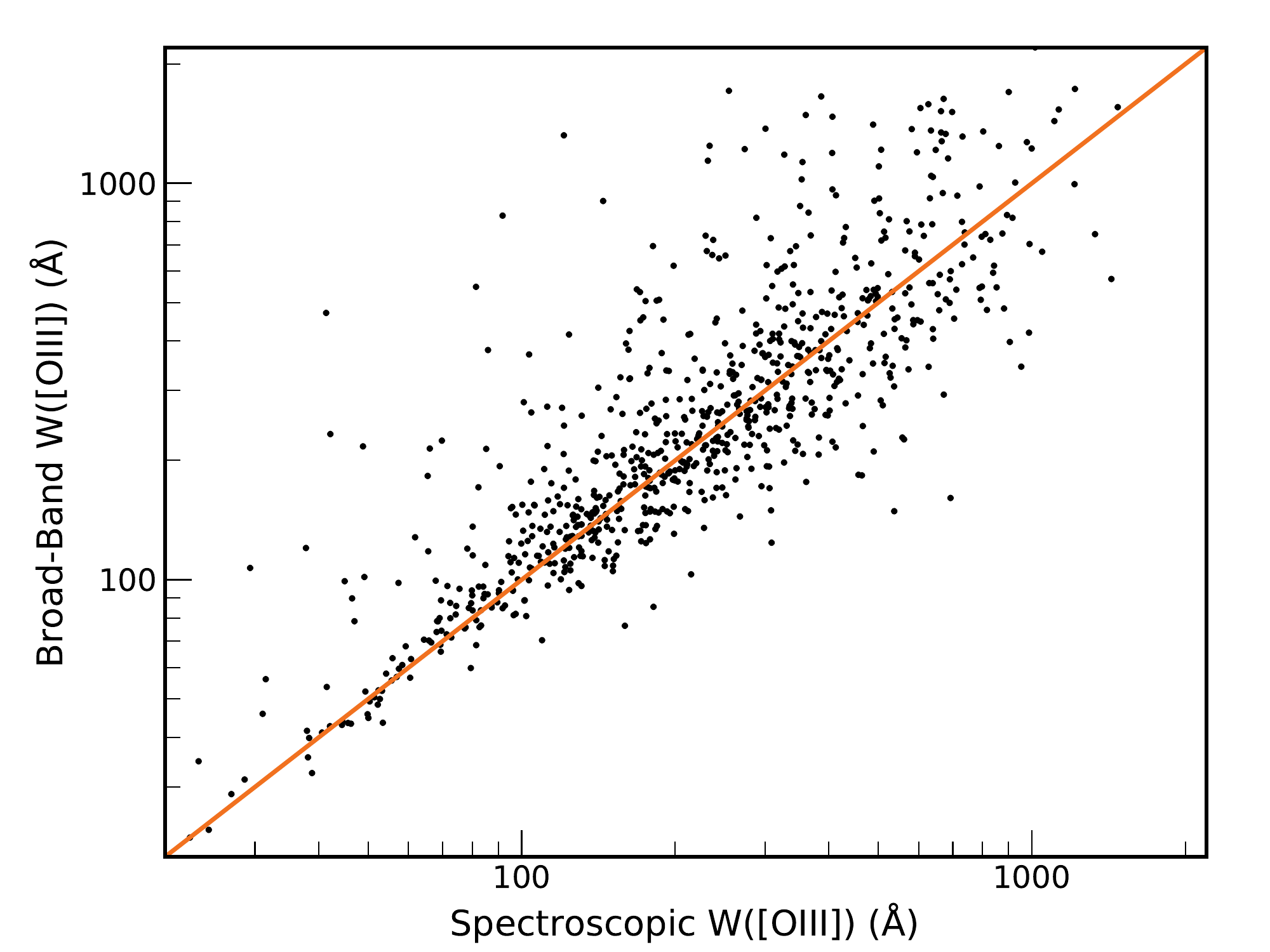} 
\caption{Rest-frame \oiii\ $\lambda \lambda$4959, 5007 emission equivalent widths are measured using continuum from spectra (abcissa), and continuum from emission line corrected broad band photometry (ordinate).   Measurements are shown for objects where the continuum is detected in the grism spectroscopy.   The solid line shows the 1:1 relation.  The broad-band continuum measurements give equivalent widths with no systematic difference from the spectra, but are more complete to sources with fainter continua. \label{ew_methods} }  
\end{figure} 
 
Figure \ref{ew_methods} compares the rest-frame \oiii\ $\lambda \lambda$4959, 5007 equivalent width measurements from the spectra with the measurements from broad-band photometry.  Only sources that have continuum detections in the grism spectra with SNR $> 0.5$ per pixel are shown. 
This analysis shows that there is no systematic difference between the two methods for estimating equivalent widths.   Likewise, the fractional error -- the RMS of ($W_{spec} - W_{broadband}) /W_{broadband}$ --  is 40\%.   Since we detect continuum from more galaxies when we use the broad-band photometry, we adopt this method for all galaxies in the sample. 

The approach described above provides an estimate of the lines that are detected in individual spectra.  However, when we rely on stacking,  \oiii\ is  often the only line that we consistently detect.  Therefore, we require a different method to estimate characteristic equivalent widths of the lines that we observe only in the stacked spectra.    If the average spectral slope was flat through the rest-frame optical, we could simply scale the median \oiii\ EW of the stacked galaxies by the measured flux ratios in the stack.   However, this is not necessarily a good approximation, except for perhaps \hb, since it is close in wavelength to \oiii.  Therefore, we estimate the continuum at the wavelengths for \ha, \hb, \hg, and \hd, using the same linear interpolation that we described above for estimating the continuum at \oiii\ from F110W or F125W and F160W.   Then, we take (showing \hg\ for an example): 
\begin{eqnarray} 
{W_{\rm stack}(H\gamma) } = W_{\rm stack}([\rm{O~III]})  \left ( { {F(H \gamma)}  \over {F([\rm{O~III}]) } } \right )_{stack}   \nonumber  \\ 
\times   \left ({{F_{\lambda}(\rm{5007})}  \over {F_{\lambda}(4341) } } \right )_{\rm stack}. ~~~~~~~~~~~~~~~~
\label{ew_stack_eq} 
\end{eqnarray} 
Here, $W_{\rm stack}([\rm{O~III]})$ is the median of the \oiii\ equivalent widths that are included in a given spectral stack, $(F(H \gamma) / F([\rm{O~III}]))_{\rm stack} $  is the relative line flux ratio measured from that stack, and  $(F_{\lambda}(5007)) /F_{\lambda}(4341) )_{\rm stack} $ is the median of the continuum flux ratios for the galaxies in that stack. In calculating the medians of stacked samples, we exclude the small minority of 25 galaxies where the broad-band flux estimates at \oiii\ are measured at less than 3$\sigma$ significance.    

The uncertainties on the equivalent width estimates for lines other than \oiii\ are taken by propagating errors in Equation \ref{ew_stack_eq}.   For $W_{\rm stack}([\rm{O~III]})$  and $(F_{\lambda}(5007) /F_{\lambda}(4341) )_{\rm stack} $ we take the error on the mean (the RMS of the individual measurements divided by the square root of the sample size in the stack).    The uncertainty on the relative flux ratio $(F(H \gamma) / F([\rm{O~III}]))_{\rm stack}$  is measured directly from bootstrap simulations, as described in \S \ref{stack_sec}.  Overall, this approach provides estimates for the equivalent widths of the Balmer lines in the stacked spectra,  as well as their uncertainties, so that we can correct the stack flux ratios for stellar absorption in our measurements of dust and metallicity. 

\section{Dust Extinction Correction Before or After Stacking Spectra?}  
\label{dust_stack_app} 

In \S \ref{stack_sec}, we described our methodology for stacking. In short,  spectra are continuum subtracted, and then normalized by the \oiii\ line flux to avoid weighting towards the sources with the strongest emission lines.   In this approach, the spectra are not corrected for dust extinction before stacking. Rather, an average extinction correction is applied to each stack, based on lines that we measure in the stack (and in some cases a prior based on lower redshift data; see Appendix \ref{bayes_met}).  For most galaxies in the sample, poor knowledge of the individual extinction corrections necessitate this strategy.    However, if the galaxies in a given stack exhibit a range in their dust extinction, then normalizing by \oiii\ flux can bias the relative \oii\ flux to higher values from systems with little dust.   A similar effect is present for \ha, for redshifts where it is observed. In this case, the \oiii\ normalization creates a bias towards strong \ha\ emission from galaxies with more dust.   The consequence of these biases is that dust extinction from \ha/\hb\ could be systematically high, while \oiii/\oii\ ratios are biased towards galaxies with low dust.  Hence, correcting the stack-measured \oii\ using the dust extinction also measured from the stack could underestimate the \oiii/\oii\ ratio from the combination of these two effects.  

We quantified this possible bias by correcting individual spectra for dust extinction prior to stacking, using the subset of galaxies where this is possible.    We take the 49 galaxies with \ha\ SNR  $> 10$ presented in \S \ref{ind_spec}, and divide this subsample into two mass bins with log $M/M_{\sun} \le 9.49$ and  log $M/M_{\sun} > 9.49$ (25 and 24 objects, respectively).   We made stacks of these same galaxies using our normal method, where extinction correction is done after stacking.   We find 12 + log (O/H) = 8.25 $\pm 0.02$ and 8.39 $\pm 0.01$  for the low and high mass bins, respectively.  On the other hand, when we apply the dust extinction that we derive for these individual galaxies before stacking, we find 12 + log (O/H) = 8.27 $\pm 0.02$ and 8.39 $\pm 0.01$.     For comparison, the mean of the individual metallicity measurements for these two mass bins are 12 + log (O/H) = 8.25 and 8.40, in excellent agreement with the stack results from either method.     Therefore, we conclude that our stacking method is robust to systematic biases due to variations in dust extinction among the samples.

\section{Bayesian Inference of Metallicity and Dust Extinction} 
\label{bayes_met} 
In order to calculate metallicities for both the stacked samples and individual galaxies, 
the spectra must be corrected for dust extinction and Balmer line stellar absorption.    Assessing these corrections, and how their uncertainties translate to uncertainties in metallicity is best addressed through a Bayesian analysis.  In this way, we can take advantage of \hg\ and \hd\ detections, or upper limits, while accounting for (or marginalizing over) stellar absorption and the contamination of \ha\ by \nii\ and \hg\ by \oiii\ $\lambda$4363.   Our method is similar to the approaches used previously for grism spectra by \cite{Jones15b} and \cite{Wang17, XWang19, XWang20}.  

We use Bayes' theorem to  write the posterior probability distribution of our model: 
\begin{equation} 
p (\theta | R)  = {{ p( R | \theta) p( \theta) } \over { p(R )}}, 
\label{bayes_th} 
\end{equation} 
where $\theta$ is the set of four physical parameters that model our data:  metallicity, E(B-V)$_{gas}$, the equivalent width of \hb\ stellar absorption, and the ratio of \oiii\ $\lambda$4363/\hg\ fluxes.   Likewise, $R$  represents the set of line flux ratios that our model must reproduce: (\ha + \nii)/\hb, (\hg\ + \oiii $\lambda$4363)/\hb, \hd/\hb, as well as $R_2  \equiv$ log(\oii/\hb), $R_3 \equiv$ log(\oiii/\hb),  and $O_{32} \equiv$ log(\oiii/\oii).  Then, $p(\theta)$ is the prior distribution, and $p(R)$ is a constant that normalizes the posterior probability distribution.   In most cases, we adopt flat priors, $p(\theta)$, that are bounded by reasonable parameter space.     Details regarding the model and priors are as follows:

\paragraph{Metallicity} 
We consider metallicities over the range of $ 7.6 <  {\rm 12+log(O/H)}  < 8.9$   where the \cite{Curti17} calibration is defined.  To produce a model of our data, 
we calculate line ratios from their calibration, considering  $R_2$, $R_3$, and $O_{32}$.    
We also model \nii/\ha, in order to account for the contamination of our \ha\ flux by (generally weak) \nii\ emission. This factors into our calculation of dust-extinction.
 However, as we discussed in \S \ref{met_cal},  unlike the oxygen-based line ratios, the relationship between \nii/\ha\ and metallicity likely does evolve to high-redshifts.   Therefore, we adopt the \nii/\ha\ calibration from \cite{Strom18}:  12 + log(O/H) = 8.77 + 0.34 $\times$ log(\nii/\ha).  Relative to local \nii/\ha\ calibrations, this relation implies stronger \nii\ at fixed metallicity.   Since \nii\ is still generally weak, the precise form of this correction does not strongly influence our metallicity inferences; however, the stronger \nii\ contamination implied by this choice often results in better agreement between the dust-extinction implied from  \ha/\hb\ and \hg/\hb.    
 
\begin{figure} 
\plotone{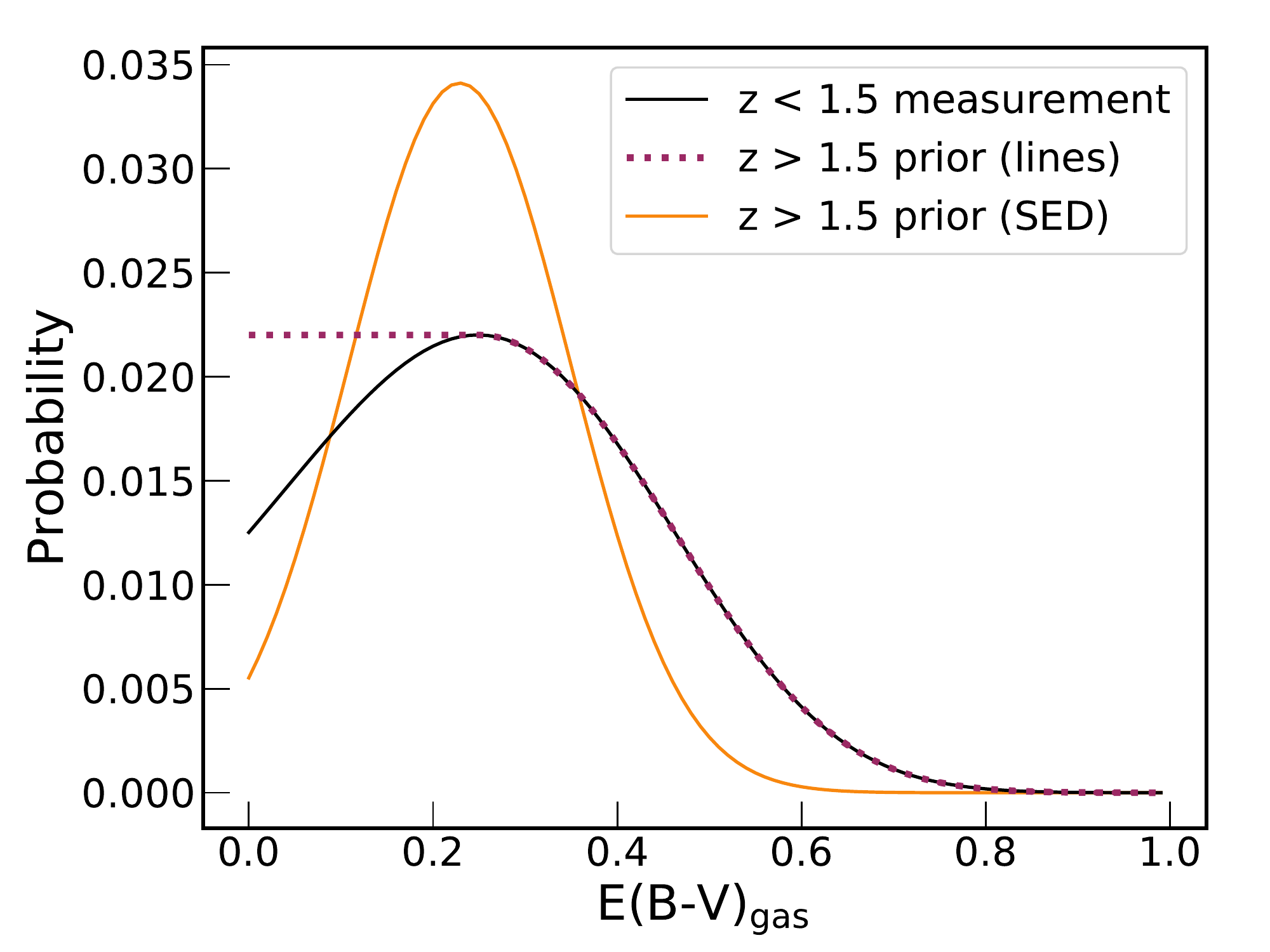} 
\caption{ For stacked spectra and individual galaxies at $z>1.5$, we adopt priors on dust extinction, since the constraints from the weaker Balmer lines are usually poor.  Here, we show two possibilities that we considered.  First, we derived priors from the posterior probability distributions for stacks at  $z<1.5$.   This example highlights the posterior probability distribution for E(B-V) for the stack of galaxies with $9.5 < {\rm log~} M/M_{\sun}  < 10.0$ (black curve).   The prior that we use for the higher redshift galaxies has the same high E(B-V) tail, but is flat to lower E(B-V), as illustrated by the magenta dotted curve.    Alternatively, we tested a prior based on the extinction derived from SED fits, after a conversion to nebular extinction following \cite{Shivaei20}, shown in the orange curve.  Priors are derived for for each of the five mass bins that we use in this paper. 
 \label{dustprior} }
\end{figure}

\paragraph{Dust Extinction} The line ratios that we evaluate for the metallicity calculations are next reddened, according to
\cite{Calzetti00}.  We also consider the Balmer line ratios that are not used in the metallicity calibrations described above.   We adopt intrinsic Balmer line ratios for $T_{e}$ = 10,000K:  \ha/\hb = 2.86, \hg/\hb = 0.468, \hd/\hb = 0.259.  A hotter $T_e$ does not have a significant impact on our dust extinction and subsequent metallicity estimates.  When \ha\ is included in the spectrum ($z< 1.5$), we use a flat dust prior that extends from  $0 <  E(B-V)_{gas} < 1.0$.   When \ha\ is not included, we adopt a prior, which is used in addition to any constraints that come from the \hg/\hb\ and \hd/\hb\ ratios.  We tested two methods for deriving the prior, which we illustrate in Figure \ref{dustprior} and quantify in Table \ref{dust_table}.  First, we derive a prior based on stacks of our  $z< 1.5$ sample, using galaxies in the same mass range. Extinction equivalent to or lower than at $1.3 < z<1.5$ is preferred for higher redshifts, but higher values are still allowed.  This prior is illustrated by the dashed magenta curve in Figure \ref{dustprior}.   Second, we used the mean of the extinction derived from the SED fits of the galaxies,  assuming $A_{V, nebular} / A_{V, stellar} = 2.1$ \citep{Shivaei20}, and a 50\% systematic uncertainty due to the stellar-to-nebular extinction conversion.   An example of this prior is shown as the orange curves in Figure \ref{dustprior}.   As is evident from Table \ref{dust_table},  the nebular extinction inferred from SED fits for galaxies in our entire redshift range ($1.3 < z < 2.3$) is very similar to the dust reddening derived from stacked sample of galaxies at $z< 1.5$.  
Ultimately, the method of estimating the prior has a negligible impact on metallicity that we derive (0.01 - 0.02 dex)  We therefore choose to use the prior based on Balmer lines in the $z< 1.5$ stacks, so that systematic uncertainties on the stellar-to-nebular extinction conversion can be avoided.

\begin{deluxetable}{lccc}
\tablecolumns{4} 
\tablewidth{0pt}
\tablecaption{Dust extinction in stacked spectra} 
\tablehead{
\colhead{ log(M/M$_{\sun}$)} & \colhead{N$_{z<1.5}$} &  \colhead{E(B-V)$_{z<1.5}$ } &  \colhead{E(B-V)$_{all, est}$}  \\
\colhead{(1)}   & \colhead{(2)} & \colhead{(3)} & \colhead{(4)} 
}
\startdata 
$< 8.5$      & 18    & $0.07_{-0.07}^{+0.22}$  & $0.12 \pm 0.06$ \\  [5pt]
8.5  - 9.0   &  40  &  $0.00_{-0.00}^{+0.16}$   &  $0.11 \pm 0.05$ \\ [5pt]
9.0  - 9.5    &  57  & $0.13_{-0.08}^{+0.10}$   &   $0.14 \pm 0.07$ \\   [5pt]
9.5 - 10.0   &  27   & $0.25_{-0.18}^{+0.16}$ &     $0.23 \pm 0.12$ \\ [5pt]
$>10.0$     &   11  &  $0.55_{-0.09}^{+0.09}$ &  $0.57 \pm 0.29$\\  [5pt]
\enddata 
\label{dust_table} 
\tablecomments{Two different estimates of extinction were considered for determining priors on dust.  {\bf (1):} The stellar mass bins used for these estimates are the same five that have been presented throughout this paper.    {\bf (2):}  The number of galaxies in each mass bin at $z < 1.5$, which we stack to obtain constraints on the dust extinction.  { \bf (3): } Dust extinction derived from the Balmer decrement measured in stacks of galaxies at $z< 1.5$.  While the measurements (or limits) for H$\gamma$/H$\beta$ and H$\delta$/H$\beta$ are included, the most robust constraints come from \ha/\hb.   {\bf (4):}   An estimate of the mean {\it nebular} extinction for galaxies in the full redshift range of our sample, determined from their SED fit, and multiplied by 2.1 following \cite{Shivaei20}.    In this case, we adopt a 50\% systematic uncertainty on the stellar-to-nebular extinction correction. } 
\end{deluxetable}

\paragraph{Balmer Line Stellar Absorption}  All line ratios involving Balmer lines are corrected for stellar absorption, reducing the modeled flux of individual lines to match the observed data.    In order to choose a range of stellar absorption for our model, we examine spectra from both continuous and instantaneous burst models in Starburst99 \citep{Leitherer99}, measuring the absorption equivalent widths in \ha, \hb, \hg, and \hd.  For populations with young stars, a reasonable range of \hb\ stellar absorption equivalent widths is 0  to 6 \AA, with the other Balmer lines tracking \hb.  Therefore,  we adopt stellar absorption equivalent widths in \hg\ and \hd\ that are equal to that of \hb, and \ha\ stellar absorption equivalent width that is 2/3 of \hb.      Then, the modeled line fluxes (and ratios) are corrected in proportion to their equivalent widths.  For example: 
\begin{equation} 
f (H\beta)_{mod}^{abs} = f(H\beta)_{mod}   { | W(H\beta)^{em} | \over  { (|W(H\beta)^{em} |  + |W(H\beta)^{*}|) } }, 
\end{equation}
where $f (H\beta)_{mod}^{abs}$ is the modeled flux that is reduced to account for stellar absorption and $f(H\beta)_{mod}$ is the modeled flux, prior to this correction. In this calculation,  $W(H\beta)^{em}$ 
and $W(H\beta)^{*}$ are the equivalent widths of the emission and the stellar absorption, respectively.    The emission line equivalent widths are estimated from broad-band photometry; we describe these estimates and assess their accuracy in Appendix \ref{appendix_ew}. 

\begin{figure*}[!t]
\begin{center} 
\includegraphics[trim = 50 20 50 70, scale=0.6]{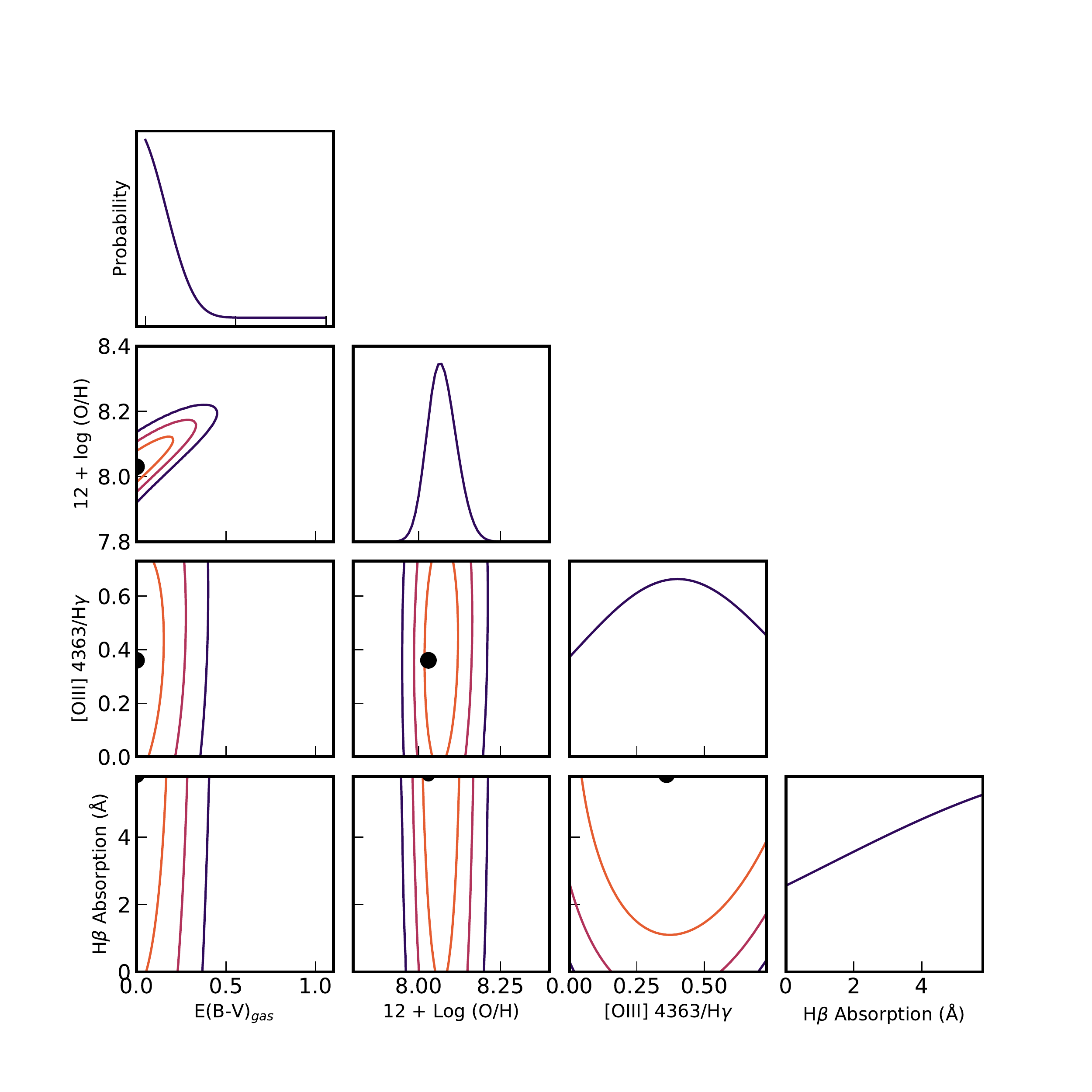} 
\caption{Contours show the 68, 95, 99.7\% confidence intervals for the four parameters that describe our model.   The filled points show the most likely solution for this stacked spectrum.   The one dimensional posterior probability distribution for each parameter is shown in the panels on the diagonal.   This example is for the 223 galaxies with masses between $10^{8.5}$ and $10^{9.0}$ M$_{\sun}$, covering the full redshift range of our sample ($1.3 < z < 2.3$).   \label{mle_conf} } 
\end{center} 
\end{figure*} 

\paragraph{Contamination of \hg\ by \oiii\ $\lambda$4363} 
If we use \hg\ to obtain constraints on dust extinction, we must correct for contamination by emission from \oiii $\lambda$4363.  This feature is strongest at the low-metallicities that we expect for our sample.
Therefore, we allow a contribution from this line, spanning from 1/40 to 1/500 of  \oiii\ $\lambda \lambda$4959,5007 for high and low electron temperatures \citep{Osterbrock} .   Although this emission line contribution should correlate with metallicity, given the systematic uncertainties discussed in \S \ref{met_cal},  we instead choose to model it independently.  

Using these ingredients to model the line ratios that we observe, $R_i$, we evaluate the likelihood function: 
\begin{equation}
p(R | \theta)  =  \prod_i e^{-( R_i - R_i^{mod}(\theta))^2/ (2 \sigma_i^2) }.  
\end{equation} 
Here, $R_i^{mod}$ describes the modeled line ratios, which are a function of $\theta$ (metallicity, stellar absorption equivalent width, dust extinction, and \oiii\ $\lambda$4363 contamination).  Likewise,  $\sigma_i$ represents the statistical error on the observed line flux ratios. For ratios involving Balmer lines, we add in quadrature an additional error to account for the uncertainty on the correction for the stellar absorption. This error arises from the uncertainty on the measured emission line equivalent widths  (see Appendix \ref{appendix_ew}).

Figure \ref{mle_conf}  shows an example of our calculation, highlighting that metallicity is the best-constrained parameter.  The contour plots and the one dimensional posterior probability distributions (on the diagonal panels) show that the \hb\ stellar absorption equivalent width and the \oiii $\lambda4363$/\hg\ flux ratio are not well-constrained by our data.  This characteristic is shared by all of the stacks that we consider.  However, the inclusion of these nuisance parameters in our model is important, as we marginalize over them to arrive at realistic uncertainties on metallicity and dust extinction.    Curiously, Figure \ref{mle_conf} shows that the metallicity does not depend strongly on the dust extinction.  This weak dependence is a consequence of the fact that our metallicity inferences require only a {\it relative} dust correction between \oiii\ and \oii, rather than an accurate absolute correction to model the reddened line fluxes.


\begin{thebibliography} 

\bibitem[Alexandroff et al.(2015)]{Alexandroff15}
 Alexandroff, R.~M., Heckman, T.~M., Borthakur, S., et al.\ 2015, \apj, 810, 104

\bibitem[Andrews \& Martini(2013)]{AM13}
 Andrews, B.~H., \& Martini, P.\ 2013, \apj, 765, 140

\bibitem[Asplund et al.(2009)]{Asplund09}
Asplund, M., Grevesse, N., Sauval, A.~J., \& Scott, P. 2009,\araa, 47, 481 

\bibitem[Astropy Collaboration et al.(2013)]{astropy13} 
Astropy Collaboration, Robitaille, T.~P., Tollerud, E.~J., et al.\ 2013, \aap, 558, A33

\bibitem[Astropy Collaboration et al.(2018)]{astropy18} 
Astropy Collaboration, Price-Whelan, A.~M., Sip{\H{o}}cz, B.~M., et al.\ 2018, \aj, 156, 123


\bibitem[Atek et al.(2010)]{Atek10} 
Atek, H., Malkan, M., McCarthy, P., et al.\ 2010, \apj, 723, 104

\bibitem[Atek et al.(2011)]{Atek11} 
Atek, H., Siana, B., Scarlata, C., et al.\ 2011, \apj, 743, 121


\bibitem[Baldwin et al.(1981)]{BPT} 
Baldwin, J.~A., Phillips, M.~M., \& Terlevich, R.\ 1981, \pasp, 93, 5

\bibitem[Bagley et al.(2020)]{Bagley20} 
Bagley, M.~B., Scarlata, C., Mehta, V., et al.\ 2020, \apj, 897, 98


\bibitem[Baronchelli et al.(2020)]{Baronchelli20} 
Baronchelli, I., Scarlata, C.~M., Rodighiero, G., et al.\ 2020, \apjs, 249, 12

\bibitem[Barro et al.(2019)]{Barro19} 
Barro, G., P{\'e}rez-Gonz{\'a}lez, P.~G., Cava, A., et al.\ 2019, \apjs, 243, 22



\bibitem[Belli et al.(2013)]{Belli13} 
Belli, S., Jones, T., Ellis, R.~S., et al.\ 2013, \apj, 772, 141


\bibitem[Berg et al.(2012)]{Berg12} 
Berg, D.~A., Skillman, E.~D., Marble, A.~R., et al.\ 2012, \apj, 754, 98

\bibitem[Bertin \& Arnouts(1996)]{SE} 
Bertin, E. \& Arnouts, S.\ 1996, \aaps, 117, 393


\bibitem[Binette et al.(2012)]{Binette12} 
Binette, L., Matadamas, R., H{\"a}gele, G.~F., et al.\ 2012, \aap, 547, A29

\bibitem[Bothwell et al.(2013)]{Bothwell13}
 Bothwell, M.~S., Maiolino, R., Kennicutt, R., et al.\ 2013, \mnras, 433, 1425


\bibitem[Bradley et al.(2021)]{Bradley21} 
Bradley, L.,  Sip\H{o}cz,  B., Robitaille, T., et al. 2021 (Zenodo)



\bibitem[Brammer et al.(2008)]{Brammer08} 
Brammer, G.~B., van Dokkum, P.~G., \& Coppi, P.\ 2008, \apj, 686, 1503


\bibitem[Bresolin(2007)]{Bresolin07} 
Bresolin, F.\ 2007, \apj, 656, 186

\bibitem[Bresolin et al.(2016)]{Bresolin16} 
Bresolin, F., Kudritzki, R.-P., Urbaneja, M.~A., et al.\ 2016, \apj, 830, 64

\bibitem[Brooks et al.(2007)]{Brooks07} 
Brooks, A.~M., Governato, F., Booth, C.~M., et al.\ 2007, \apjl, 655, L17

\bibitem[Brown et al.(2016)]{Brown16}
 Brown, J.~S., Martini, P., \& Andrews, B.~H.\ 2016, \mnras, 458, 1529

\bibitem[Bruzual \& Charlot(2003)]{BC03} 
Bruzual, G., \& Charlot, S.\ 2003, \mnras, 344, 1000


\bibitem[Calzetti et al.(2000)]{Calzetti00} 
Calzetti, D. Armus, L., Bohlin, R.~C., et al. 2000, \apj, 533, 682 

\bibitem[Chabrier(2003)]{Chabrier2003} 
Chabrier, G.\ 2003, \pasp, 115, 763

\bibitem[Charlot \& Fall(2000)]{CharlotFall00} 
Charlot, S. \& Fall, S.~M.\ 2000, \apj, 539, 718


\bibitem[Chisholm et al.(2018)]{Chisholm18} 
Chisholm, J., Tremonti, C., \& Leitherer, C.\ 2018, \mnras, 481, 1690


\bibitem[Coil et al.(2015)]{Coil15} 
Coil, A.~L., Aird, J., Reddy, N., et al.\ 2015, \apj, 801, 35

\bibitem[Colbert et al.(2013)]{Colbert13} 
Colbert, J.~W., Teplitz, H., Atek, H., et al.\ 2013, \apj, 779, 34


\bibitem[Cresci et al.(2012)]{Cresci12} 
Cresci, G., Mannucci, F., Sommariva, V., et al.\ 2012, \mnras, 421, 262

\bibitem[Croton et al.(2016)]{Croton16} 
Croton, D.~J., Stevens, A.~R.~H., Tonini, C., et al.\ 2016, \apjs, 222, 22

\bibitem[Curti et al.(2017)]{Curti17} 
Curti, M., Cresci, G., Mannucci, F., et al.\ 2017, \mnras, 465, 1384

\bibitem[Curti et al.(2020)]{Curti20} 
Curti, M., Mannucci, F., Cresci, G., et al.\ 2020, \mnras, 491, 944

\bibitem[Dalcanton(2007)]{Dalcanton07} 
Dalcanton, J.~J.\ 2007, \apj, 658, 941

\bibitem[Dav{\'e} et al.(2012)]{Dave12} 
Dav{\'e}, R., Finlator, K., \& Oppenheimer, B.~D.\ 2012, \mnras, 421, 98


\bibitem[Dav{\'e} et al.(2017)]{Dave17}
Dav{\'e}, R., Rafieferantsoa, M.~H., Thompson, R.~J., et al.\ 2017, \mnras, 467, 115

\bibitem[Dav{\'e} et al.(2019)]{Dave19} 
Dav{\'e}, R., Angl{\'e}s-Alc{\'a}zar, D., Narayanan, D., et al.\ 2019, \mnras, 486, 2827

\bibitem[Davies et al.(2017)]{Davies17} 
Davies, B., Kudritzki, R.-P., Lardo, C., et al.\ 2017, \apj, 847, 112

\bibitem[Dayal et al.(2013)]{Dayal13} 
Dayal, P., Ferrara, A., \& Dunlop, J.~S.\ 2013, \mnras, 430, 2891

\bibitem[De Lucia et al.(2020)]{DeLucia20}
De Lucia, G., Xie, L., Fontanot, F., Hirschmann, M.\ 2020, arXiv:2008.09127

\bibitem[De Rossi et al.(2017)]{DeRossi17} 
De Rossi, M.~E., Bower, R.~G., Font, A.~S., et al.\ 2017, \mnras, 472, 3354

\bibitem[Dom{\'\i}nguez et al.(2013)]{Dominguez13} 
Dom{\'\i}nguez, A., Siana, B., Henry, A.~L., et al.\ 2013, \apj, 763, 145

\bibitem[Donley et al.(2012)]{Donley12} 
Donley, J.~L., Koekemoer, A.~M., Brusa, M., et al.\ 2012, \apj, 748, 142

\bibitem[Dopita et al.(2013)]{Dopita13} 
Dopita, M.~A., Sutherland, R.~S., Nicholls, D.~C., et al.\ 2013, \apjs, 208, 10

\bibitem[Eddington(1913)]{Eddington13} 
Eddington, A.~S.\ 1913, \mnras, 73, 359


\bibitem[Eldridge \& Stanway(2016)]{Eldridge16} 
Eldridge, J.~J., \& Stanway, E.~R.\ 2016, \mnras, 462, 3302 

\bibitem[Eldridge et al.(2017)]{BPASS}
 Eldridge, J.~J., Stanway, E.~R., Xiao, L., et al.\ 2017, PASA, 34, e058

\bibitem[Ellison et al.(2008)]{Ellison08} 
Ellison, S.~L., Patton, D.~R., Simard, L., et al.\ 2008, \apjl, 672, L107

\bibitem[Emami et al.(2019)]{Emami19} 
Emami, N., Siana, B., Weisz, D.~R., et al.\ 2019, \apj, 881, 71


\bibitem[Erb et al.(2006)]{Erb06} 
Erb, D.~K., Shapley, A.~E., Pettini, M., et al.\ 2006, \apj, 644, 813

\bibitem[Erb(2008)]{Erb08} 
Erb, D.~K.\ 2008, \apj, 674, 151


\bibitem[Estrada-Carpenter et al.(2019)]{clear_paper1}
 Estrada-Carpenter, V., Papovich, C., Momcheva, I., et al.\ 2019, \apj, 870, 133
 
\bibitem[Fazio et al.(2004)]{Fazio04} 
Fazio, G.~G., Hora, J.~L., Allen, L.~E., et al.\ 2004, \apjs, 154, 10

\bibitem[Ferland et al.(2017)]{Ferland17} 
Ferland, G.~J., Chatzikos, M., Guzm{\'a}n, F., et al.\ 2017, RMxAA, 53, 385

\bibitem[Finlator \& Dav{\'e}(2008)]{Finlator08} 
Finlator, K. \& Dav{\'e}, R.\ 2008, \mnras, 385, 2181

\bibitem[Galametz et al.(2013)]{Galametz13} 
Galametz, A., Grazian, A., Fontana, A., et al.\ 2013, \apjs, 206, 10


\bibitem[Garc{\'\i}a-Rojas \& Esteban(2007)]{GarciaRojas07}
 Garc{\'\i}a-Rojas, J., \& Esteban, C.\ 2007, \apj, 670, 457

\bibitem[Garn \& Best(2010)]{Garn10} 
Garn, T. \& Best, P.~N.\ 2010, \mnras, 409, 421


\bibitem[Gburek et al.(2019)]{Gburek19} 
Gburek, T., Siana, B., Alavi, A., et al.\ 2019, \apj, 887, 168

\bibitem[Gillman et al.(2021)]{Gillman21} 
Gillman, S., Tiley, A.~L., Swinbank, A.~M., et al.\ 2021, \mnras, 500, 4229

\bibitem[Grasshorn Gebhardt et al.(2016)]{GG16} 
Grasshorn Gebhardt, H.~S., Zeimann, G.~R., Ciardullo, R., et al.\ 2016, \apj, 817, 10


\bibitem[Grogin et al.(2011)]{Grogin11}
 Grogin, N.~A., Kocevski, D.~D., Faber, S.~M., et al.\ 2011, \apjs, 197, 35 


\bibitem[Guo et al.(2013)]{Guo13}
 Guo, Y., Ferguson, H.~C., Giavalisco, M., et al.\ 2013, \apjs, 207, 24


\bibitem[Guo et al.(2016a)]{Guo16a} 
Guo, Y., Koo, D.~C., Lu, Y., et al.\ 2016a, \apj, 822, 103

\bibitem[Guo et al.(2016b)]{Guo16b} 
Guo, Y., Rafelski, M., Faber, S.~M., et al.\ 2016b, \apj, 833, 37


\bibitem[Henry et al.(2013a)]{Henry2013_clm} 
Henry, A., Martin, C.~L., Finlator, K., et al.\ 2013a, \apj, 769, 148

\bibitem[Henry et al.(2013b)]{Henry2013_wisp} 
Henry, A., Scarlata, C., Dom{\'\i}nguez, A., et al.\ 2013b, \apj, 776, L27

\bibitem[Henry et al.(2018)]{Henry18} 
Henry, A., Berg, D.~A., Scarlata, C., et al.\ 2018, \apj, 855, 96

\bibitem[Hirschauer et al.(2018)]{Hirschauer18} 
Hirschauer, A.~S., Salzer, J.~J., Janowiecki, S., et al.\ 2018, \aj, 155, 82


 
 \bibitem[Horne(1986)]{Horne86} 
 Horne, K.\ 1986, \pasp, 98, 609

 
 \bibitem[Izotov et al.(2012)]{Izotov12} 
 Izotov, Y.~I., Thuan, T.~X., \& Guseva, N.~G.\ 2012, \aap, 546, A122


 \bibitem[Jones et al.(2015a)]{Jones15a} 
 Jones, T., Martin, C., \& Cooper, M.~C.\ 2015a, \apj, 813, 126

\bibitem[Jones et al.(2015b)]{Jones15b}
 Jones, T., Wang, X., Schmidt, K.~B., et al.\ 2015b, \aj, 149, 107
 
 

\bibitem[Juneau et al.(2011)]{Juneau11} 
Juneau, S., Dickinson, M., Alexander, D.~M., et al.\ 2011, \apj, 736, 104

\bibitem[Juneau et al.(2014)]{Juneau14} 
Juneau, S., Bournaud, F., Charlot, S., et al.\ 2014, \apj, 788, 88

\bibitem[Kacprzak et al.(2016)]{Kacprzak16}
 Kacprzak, G.~G., van de Voort, F., Glazebrook, K., et al.\ 2016, \apjl, 826, L11

\bibitem[Kashino et al.(2017)]{Kashino17} 
Kashino, D., Silverman, J.~D., Sanders, D., et al.\ 2017, \apj, 835, 88

\bibitem[Kashino et al.(2019)]{Kashino19}
 Kashino, D., Silverman, J.~D., Sanders, D., et al.\ 2019, \apjs, 241, 10

\bibitem[Kennicutt(1998)]{Kennicutt98} 
Kennicutt, R.~C.\ 1998, \araa, 36, 189

\bibitem[Kewley \& Dopita(2002)]{KD02} 
Kewley, L.~J., \& Dopita, M.~A.\ 2002, \apjs, 142, 35 

\bibitem[Kewley \& Ellison(2008)]{KE08} 
Kewley, L.~J., \& Ellison, S.~L.\ 2008, \apj, 681, 1183

\bibitem[Kewley et al.(2013)]{Kewley13} 
Kewley, L.~J., Dopita, M.~A., Leitherer, C., et al.\ 2013, \apj, 774, 100

\bibitem[Kewley et al.(2015)]{Kewley15} 
Kewley, L.~J., Zahid, H.~J., Geller, M.~J., et al.\ 2015, \apj, 812, L20

\bibitem[Kobulnicky \& Kewley(2004)]{KK04} 
Kobulnicky, H.~A., \& Kewley, L.~J.\ 2004, \apj, 617, 240

\bibitem[Kocevski et al.(2018)]{Kocevski18} 
Kocevski, D.~D., Hasinger, G., Brightman, M., et al.\ 2018, \apjs, 236, 48

\bibitem[Koekemoer et al.(2011)]{Koekemoer11} 
Koekemoer, A. M., Faber, S., Ferguson, H. et al. 2011, ApJS, 197, 36

\bibitem[Kriek et al.(2015)]{Kriek15} 
Kriek, M., Shapley, A.~E., Reddy, N.~A., et al.\ 2015, \apjs, 218, 15

\bibitem[Kudritzki et al.(2016)]{Kudritzki16} 
Kudritzki, R.~P., Castro, N., Urbaneja, M.~A., et al.\ 2016, \apj, 829, 70

\bibitem[K{\"u}mmel et al.(2009)]{aXe} 
K{\"u}mmel, M., Walsh, J.~R., Pirzkal, N., et al.\ 2009, \pasp, 121, 59



\bibitem[Lara-L{\'o}pez et al.(2010)]{LaraLopez10} 
Lara-L{\'o}pez, M.~A., Cepa, J., Bongiovanni, A., et al.\ 2010, \aap, 521, L53

\bibitem[Lara-L{\'o}pez et al.(2013)]{LaraLopez13} 
Lara-L{\'o}pez, M.~A., Hopkins, A.~M., Lopez-Sanchez, A.~R., et al.\ 2013, \mnras, 433, L35

\bibitem[Lee et al.(2006)]{Lee06} 
Lee, H., Skillman, E.~D., Cannon, J.~M., et al.\ 2006, \apj, 647, 970

\bibitem[Lee et al.(2011)]{Lee2011} 
Lee, J.~C., Gil de Paz, A., Kennicutt, R.~C., et al.\ 2011, \apjs, 192, 6

\bibitem[Leitherer et al.(1999)]{Leitherer99}
Leitherer, C., Schaerer, D., Goldader, J., et al. 1999, \apjs, 123, 3 

\bibitem[Lilly et al.(2013)]{Lilly13} 
Lilly, S.~J., Carollo, C.~M., Pipino, A., et al.\ 2013, \apj, 772, 119

\bibitem[Lower et al.(2020)]{Lower20} 
Lower, S., Narayanan, D., Leja, J., et al.\ 2020, \apj, 904, 33

\bibitem[Lu et al.(2014)]{Lu14} 
Lu, Y., Wechsler, R.~H., Somerville, R.~S., et al.\ 2014, \apj, 795, 123

\bibitem[Ly et al.(2016)]{Ly16} 
Ly, C., Malkan, M.~A., Rigby, J.~R., et al.\ 2016, \apj, 828, 67


\bibitem[Ma et al.(2016)]{Ma16} 
Ma, X., Hopkins, P.~F., Faucher-Gigu{\`e}re, C.-A., et al.\ 2016, \mnras, 456, 2140

\bibitem[MacKenty et al.(2010)]{wfc3} 
MacKenty, J.~W., Kimble, R.~A., O'Connell, R.~W., et al.\ 2010, \procspie, 7731, 77310Z


\bibitem[Marinacci et al.(2018)]{Marinacci18} 
Marinacci, F., Vogelsberger, M., Pakmor, R., et al.\ 2018, \mnras, 480, 5113


\bibitem[Masters et al.(2014)]{Masters14} 
Masters, D., McCarthy, P., Siana, B., et al.\ 2014, \apj, 785, 153

\bibitem[Masters et al.(2016)]{Masters16} 
Masters, D., Faisst, A., \& Capak, P.\ 2016, \apj, 828, 18

\bibitem[Maier et al.(2015)]{Maier15} 
Maier, C., Ziegler, B.~L., Lilly, S.~J., et al.\ 2015, \aap, 577, A14


\bibitem[Maiolino et al.(2008)]{Maiolino08}
 Maiolino, R., Nagao, T., Grazian, A., et al.\ 2008, \aap, 488, 463


\bibitem[Mannucci et al.(2010)]{Mannucci10} 
Mannucci, F., Cresci, G., Maiolino, R., et al.\ 2010, \mnras, 408, 2115

\bibitem[Mannucci et al.(2011)]{Mannucci11} 
Mannucci, F., Salvaterra, R., \& Campisi, M.~A.\ 2011, \mnras, 414, 1263

\bibitem[Mehta et al.(2017)]{Mehta17} 
Mehta, V., Scarlata, C., Rafelski, M., et al.\ 2017, \apj, 838, 29

\bibitem[Merlin et al.(2015)]{Merlin15} 
Merlin, E., Fontana, A., Ferguson, H.~C., et al.\ 2015, \aap, 582, A15

\bibitem[Merlin et al.(2016)]{Merlin16}
 Merlin, E., Bourne, N., Castellano, M., et al.\ 2016, \aap, 595, A97
 
 
 \bibitem[Mobasher et al.(2015)]{Mobasher15}  
 Mobasher, B., Dahlen, T., Ferguson, H.~C., et al.\ 2015, \apj, 808, 101


\bibitem[Momcheva et al.(2016)]{Momcheva16} 
Momcheva, I.~G., Brammer, G.~B., van Dokkum, P.~G., et al.\ 2016, \apjs, 225, 27

\bibitem[Naiman et al.(2018)]{Naiman18} 
Naiman, J.~P., Pillepich, A., Springel, V., et al.\ 2018, \mnras, 477, 1206

\bibitem[Narayanan et al.(2018)]{Narayanan18}
Narayanan, D., Conroy, C., Dav{\'e}, R., et al.\ 2018, \apj, 869, 70


\bibitem[Nayyeri et al.(2017)]{Nayyeri17} 
Nayyeri, H., Hemmati, S., Mobasher, B., et al.\ 2017, \apjs, 228, 7


\bibitem[Nelson et al.(2018)]{Nelson18} 
Nelson, D., Pillepich, A., Springel, V., et al.\ 2018, \mnras, 475, 624

\bibitem[Nicholls et al.(2012)]{Nicholls12} 
Nicholls, D.~C., Dopita, M.~A., \& Sutherland, R.~S.\ 2012, \apj, 752, 148

\bibitem[Nomoto et al.(2013)]{Nomoto13} 
Nomoto, K., Kobayashi, C., \& Tominaga, N.\ 2013, \araa, 51, 457






\bibitem[Osterbrock(1989)]{Osterbrock} 
Osterbrock, D.~E.\ 1989, Astrophysics of Gaseous Nebulae and Active Galactic Nuclei


\bibitem[Pacifici et al.(2012)]{Pacifici12}
 Pacifici, C., Charlot, S., Blaizot, J., et al.\ 2012, \mnras, 421, 2002

\bibitem[Pacifici et al.(2016)]{Pacifici16} 
Pacifici, C., Kassin, S.~A., Weiner, B.~J., et al.\ 2016, \apj, 832, 79

\bibitem[Pirzkal et al.(2017)]{Pirzkal17} 
Pirzkal, N., Malhotra, S., Ryan, R.~E., et al.\ 2017, \apj, 846, 84


\bibitem[Peeples \& Shankar(2011)]{Peeples11}
 Peeples, M.~S. \& Shankar, F.\ 2011, \mnras, 417, 2962


\bibitem[Peimbert et al.(2007)]{Peimbert07} 
Peimbert, M., Peimbert, A., Esteban, C., et al.\ 2007, Revista Mexicana De Astronomia Y Astrofisica Conference Series, 72

\bibitem[Pettini \& Pagel(2004)]{PP04} 
Pettini, M., \& Pagel, B.~E.~J.\ 2004, \mnras, 348, L59

\bibitem[Pillepich et al.(2018)]{Pillepich18} 
Pillepich, A., Nelson, D., Hernquist, L., et al.\ 2018, \mnras, 475, 648

\bibitem[Pilyugin \& Grebel(2016)]{Pil16}
 Pilyugin, L.~S., \& Grebel, E.~K.\ 2016, \mnras, 457, 3678
 
\bibitem[Pilyugin et al.(2012)]{Pil12} 
Pilyugin, L.~S., Grebel, E.~K., \& Mattsson, L.\ 2012, \mnras, 424, 2316

 \bibitem[Pilyugin \& Thuan(2005)]{Pil05} 
 Pilyugin, L.~S., \& Thuan, T.~X.\ 2005, \apj, 631, 231

\bibitem[Planck Collaboration et al.(2016)]{Planck15} 
Planck Collaboration, Ade, P.~A.~R., Aghanim, N., et al.\ 2016, \aap, 594, A13

 
 \bibitem[Porter et al.(2012)]{Porter12} 
Porter, R.~L., Ferland, G.~J., Storey, P.~J., et al.\ 2012, \mnras, 425, L28

\bibitem[Rutkowski et al.(2016)]{Rutkowski16} 
Rutkowski, M.~J., Scarlata, C., Haardt, F., et al.\ 2016, \apj, 819, 81

\bibitem[Rafelski et al.(2015)]{Rafelski15} 
Rafelski, M., Teplitz, H.~I., Gardner, J.~P., et al.\ 2015, \aj, 150, 31


\bibitem[Salim et al.(2014)]{Salim14} 
Salim, S., Lee, J.~C., Ly, C., et al.\ 2014, \apj, 797, 126

\bibitem[Salim et al.(2015)]{Salim15} 
Salim, S., Lee, J.~C., Dav{\'e}, R., et al.\ 2015, \apj, 808, 25

\bibitem[Salpeter(1955)]{Salpeter}
 Salpeter, E.~E.\ 1955, \apj, 121, 161

\bibitem[S{\'a}nchez Almeida et al.(2014)]{SA14}
S{\'a}nchez Almeida, J., Elmegreen, B.~G., Mu{\~n}oz-Tu{\~n}{\'o}n, C., et al.\ 2014, \aapr, 22, 71

\bibitem[Sanders et al.(2015)]{Sanders15} 
Sanders, R.~L., Shapley, A.~E., Kriek, M., et al.\ 2015, \apj, 799, 138

\bibitem[Sanders et al.(2016)]{Sanders16} 
Sanders, R.~L., Shapley, A.~E., Kriek, M., et al.\ 2016, \apj, 816, 23

\bibitem[Sanders et al.(2017)]{Sanders17} 
Sanders, R.~L., Shapley, A.~E., Zhang, K., et al.\ 2017, \apj, 850, 136


\bibitem[Sanders et al.(2018)]{Sanders18} 
Sanders, R.~L., Shapley, A.~E., Kriek, M., et al.\ 2018, \apj, 858, 99

\bibitem[Sanders et al.(2020)]{Sanders19} 
Sanders, R.~L., Shapley, A.~E., Reddy, N.~A., et al.\ 2020, \mnras, 491, 1427


\bibitem[Shapley et al.(2015)]{Shapley15} 
Shapley, A.~E., Reddy, N.~A., Kriek, M., et al.\ 2015, \apj, 801, 88

\bibitem[Shirazi et al.(2014)]{Shirazi14}
 Shirazi, M., Brinchmann, J., \& Rahmati, A.\ 2014, \apj, 787, 120
 
 \bibitem[Shivaei et al.(2020)]{Shivaei20} 
 Shivaei, I., Reddy, N., Rieke, G., et al.\ 2020, \apj, 899, 117


\bibitem[Skelton et al.(2014)]{Skelton14} 
Skelton, R.~E., Whitaker, K.~E., Momcheva, I.~G., et al.\ 2014, \apjs, 214, 24

\bibitem[Somerville \& Dav{\'e}(2015)]{SD15} 
Somerville, R.~S. \& Dav{\'e}, R.\ 2015, \araa, 53, 51

\bibitem[Springel et al.(2018)]{Springel18}
Springel, V., Pakmor, R., Pillepich, A., et al.\ 2018, \mnras, 475, 676


\bibitem[Stasi{\'n}ska(2002)]{Stasinska02}
 Stasi{\'n}ska, G.\ 2002, Revista Mexicana De Astronomia Y Astrofisica Conference Series, 62
 
 \bibitem[Stefanon et al.(2017)]{Stefanon17} 
 Stefanon, M., Yan, H., Mobasher, B., et al.\ 2017, \apjs, 229, 32


\bibitem[Steidel et al.(2014)]{Steidel14} 
Steidel, C.~C., Rudie, G.~C., Strom, A.~L., et al.\ 2014, \apj, 795, 165 

\bibitem[Steidel et al.(2016)]{Steidel16} 
Steidel, C.~C., Strom, A.~L., Pettini, M., et al.\ 2016, \apj, 826, 159



\bibitem[Strom et al.(2017)]{Strom17} 
Strom, A.~L., Steidel, C.~C., Rudie, G.~C., et al.\ 2017, \apj, 836, 164

\bibitem[Strom et al.(2018)]{Strom18} 
Strom, A.~L., Steidel, C.~C., Rudie, G.~C., et al.\ 2018, \apj, 868, 117


\bibitem[Theios et al.(2019)]{Theios19} 
Theios, R.~L., Steidel, C.~C., Strom, A.~L., et al.\ 2019, \apj, 871, 128


\bibitem[Torrey et al.(2018)]{Torrey18} 
Torrey, P., Vogelsberger, M., Hernquist, L., et al.\ 2018, \mnras, 477, L16


\bibitem[Torrey et al.(2019)]{Torrey19} 
Torrey, P., Vogelsberger, M., Marinacci, F., et al.\ 2019, \mnras, 484, 5587

\bibitem[Tremonti et al.(2004)]{Tremonti04} 
Tremonti, C.~A., Heckman, T.~M., Kauffmann, G., et al.\ 2004, \apj, 613, 898

\bibitem[Trump et al.(2011)]{Trump11}
 Trump, J.~R., Weiner, B.~J., Scarlata, C., et al.\ 2011, \apj, 743, 144
 
 \bibitem[Trump et al.(2013)]{Trump13} 
 Trump, J.~R., Konidaris, N.~P., Barro, G., et al.\ 2013, \apj, 763, L6

\bibitem[van Dokkum et al.(2013)]{pvd_udf} 
van Dokkum, P., Brammer, G., Momcheva, I., et al.\ 2013, arXiv e-prints, arXiv:1305.2140

\bibitem[Villforth et al.(2010)]{Villforth10} 
Villforth, C., Koekemoer, A.~M., \& Grogin, N.~A.\ 2010, \apj, 723, 737



\bibitem[B. Wang et al.(2019)]{Wang19} 
Wang, B., Heckman, T.~M., Leitherer, C., et al.\ 2019, \apj, 885, 57


\bibitem[Wang et al.(2017)]{Wang17} 
Wang, X., Jones, T.~A., Treu, T., et al.\ 2017, \apj, 837, 89

\bibitem[X. Wang et al.(2019)]{XWang19} 
Wang, X., Jones, T.~A., Treu, T., et al.\ 2019, \apj, 882, 94

\bibitem[Wang et al.(2020)]{XWang20} 
Wang, X., Jones, T.~A., Treu, T., et al.\ 2020, \apj, 900, 183


\bibitem[Whitaker et al.(2014)]{Whitaker14} 
Whitaker, K.~E., Franx, M., Leja, J., et al.\ 2014, \apj, 795, 104

\bibitem[Wisnioski et al.(2015)]{Wisnioski15} 
Wisnioski, E., F{\"o}rster Schreiber, N.~M., Wuyts, S., et al.\ 2015, \apj, 799, 209

\bibitem[Wright et al.(2009)]{Wright09}
 Wright, S.~A., Larkin, J.~E., Law, D.~R., et al.\ 2009, \apj, 699, 421

\bibitem[Wuyts et al.(2012)]{Wuyts12} 
Wuyts, E., Rigby, J.~R., Sharon, K., et al.\ 2012, \apj, 755, 73

\bibitem[Wuyts et al.(2016)]{Wuyts16}
 Wuyts, E., Wisnioski, E., Fossati, M., et al.\ 2016, \apj, 827, 74

\bibitem[Yabe et al.(2012)]{Yabe12} 
Yabe, K., Ohta, K., Iwamuro, F., et al.\ 2012, \pasj, 64, 60

\bibitem[Yabe et al.(2015a)]{Yabe15}
 Yabe, K., Ohta, K., Akiyama, M., et al.\ 2015a, \pasj, 67, 102
 
 \bibitem[Yabe et al.(2015b)]{Yabe15b} 
 Yabe, K., Ohta, K., Akiyama, M., et al.\ 2015b, \apj, 798, 45

 
 \bibitem[Yates et al.(2012)]{Yates12} 
 Yates, R.~M., Kauffmann, G., \& Guo, Q.\ 2012, \mnras, 422, 215

\bibitem[Zahid et al.(2011)]{Zahid11} 
Zahid, H.~J., Kewley, L.~J., \& Bresolin, F.\ 2011, \apj, 730, 137

\bibitem[Zahid et al.(2014)]{Zahid14} 
Zahid, H.~J., Kashino, D., Silverman, J.~D., et al.\ 2014, \apj, 792, 75





\end{thebibliography}
\end{document}